\documentclass[prd,twocolumn,showpacs,floatfix,amsmath,amssymb,floatfix,nofootinbib]{revtex4}
\usepackage{graphicx,color,dcolumn,booktabs,bm}
\usepackage{longtable,lscape,comment,braket}
\usepackage{txfonts}
\usepackage{overpic}
\usepackage{amssymb,tipa}
\usepackage{indentfirst}
\usepackage{feynmf}   
\usepackage{slashed}  
\usepackage{cases}
\usepackage{epstopdf}
\usepackage{psfrag}
\usepackage{subfigure}
\usepackage{color}
\usepackage{multirow}

\graphicspath{{Figures/}} %
\usepackage[colorlinks, citecolor=blue,anchorcolor=red,menucolor=red, linkcolor=red,filecolor=red,runcolor=red,urlcolor=blue,frenchlinks=red]{hyperref}

\usepackage{ulem}
\def\be{\begin{equation}}
\def\ee{\end{equation}}
\def\Tr{\textnormal{Tr}}

\begin{document}
\title{$\bar{B}^{(\ast)} \bar{B}^{(\ast)}$ interactions in chiral effective field theory}

\author{Bo Wang$^{1,2,3,4}$}\email{bo-wang@pku.edu.cn}
\author{Zhan-Wei Liu$^{1,2}$}\email{liuzhanwei@lzu.edu.cn}
\author{Xiang Liu$^{1,2}$}\email{xiangliu@lzu.edu.cn}
\affiliation{$^1$School of Physical Science and Technology, Lanzhou University,
Lanzhou 730000, China\\
$^2$Research Center for Hadron and CSR Physics,
Lanzhou University $\&$ Institute of Modern Physics of CAS,
Lanzhou 730000, China\\
$^3$ School of Physics and State Key Laboratory of Nuclear Physics and Technology, Peking University, Beijing 100871, China\\
$^4$Center of High Energy Physics, Peking University, Beijing 100871, China}
\begin{abstract}
In this work, the intermeson interactions of double-beauty $\bar{B}\bar{B}$, $\bar{B}\bar{B}^\ast$, and $\bar{B}^\ast\bar{B}^\ast$ systems have been studied with heavy meson chiral effective field theory. The effective potentials are calculated with Weinberg's scheme up to one-loop level. At the leading order, four-body contact interactions and one-pion exchange contributions are considered. In addition to two-pion exchange diagrams, we include the one-loop
chiral corrections to contact terms and one-pion exchange diagrams at the next-to-leading order. The behaviours of effective potentials both in momentum space and coordinate space are investigated and discussed extensively. We notice the contact terms play important roles in determining the characteristics of the total potentials. Only the potentials in $I(J^P)=0(1^+)$ $\bar{B}\bar{B}^\ast$ and $\bar{B}^\ast\bar{B}^\ast$ systems are attractive, and the corresponding binding energies in these two channels are solved to be $\Delta E_{\bar{B}\bar{B}^\ast}\simeq -12.6^{+9.2}_{-12.9}$ MeV and $\Delta E_{\bar{B}^\ast\bar{B}^\ast}\simeq -23.8^{+16.3}_{-21.5}$ MeV, respectively. The masses of $0(1^+)$ $\bar{B}\bar{B}^\ast$ and $\bar{B}^\ast\bar{B}^\ast$ states lie above the threshold of their electromagnetic decay modes $\bar{B}\bar{B}\gamma$ and $\bar{B}\bar{B}\gamma\gamma$, and thus they can be reconstructed via electromagnetic interactions. Our calculation not only provides some useful information to explore exotic doubly-bottomed molecular states for future experiments, but also is helpful for the extrapolations of Lattice QCD simulations.
\end{abstract}
 \maketitle

\section{introduction}\label{Sec1}
Hunting for exotic multiquark states beyond the conventional meson and baryon configurations is a long-standing problem of QCD \cite{GellMann:1964nj,Zweig:19xx,Jaffe:1976ig}.
After the pioneering exotic state $X(3872)$ was discovered in 2003 by the Belle Collaboration \cite{Choi:2003ue}, a new perspective of hadronic physics was opened, and plenty of
exotic hadrons called $XYZ$ states were observed at experiments \cite{Chen:2016qju,Guo:2017jvc}. Lots of experimental and theoretical efforts have been paid to understand
the nature of these states, but they still seem very elusive.

Most of these states cannot be assigned into the framework of conventional quark model. For example, if we treat $X(3872)$ as the $\chi_c^\prime(2P)$ charmonium, the mass of it would
be about 78 MeV lower than the prediction of relativized quark model  \cite{Godfrey:1985xj}. Furthermore, the isospin violation process $X(3872)\to J/\psi\rho$ with the substantial decay
width makes the situation more complicated and mysterious. Thus various pictures and explanations were put forward to probe the inner structure of $X(3872)$, such as molecular model \cite{Swanson:2003tb,Wong:2003xk,AlFiky:2005jd,Fleming:2007rp,Liu:2008fh,Liu:2008tn},
tetraquark state \cite{Maiani:2004vq,Chen:2010ze}, traditional axial vector charmonium \cite{Barnes:2003vb,Kong:2006ni}, Lattice QCD simulation \cite{Chiu:2006hd,Padmanath:2015era},
and other theoretical schemes \cite{Kang:2016jxw} (for a review, see Ref. \cite{Chen:2016qju}). Among them, the shallowly bound molecular picture, i.e., the deuteron-like configuration, is the most popular one, because
the mass of $X(3872)$ is very close to the threshold of $D^0\bar{D}^{\ast0}$. Like study the interaction of proton and neutron is a {\it sine qua non} for understanding the characteristics of deuteron and
nuclear force, if we want to comprehensively understand the nature of $X(3872)$ and other resonances in $XYZ$ family, to research the strong interactions between heavy mesons would be
an ineluctable key issue.

Unlike a plethora of $XYZ$ states have been observed in charmonium energy region \cite{Chen:2016qju,Guo:2017jvc,Esposito:2016noz}, only two $Z_b$ states were reported in bottomonium
spectrum \cite{Belle:2011aa}. There is also no unanimous conclusion on what the $Z_b(10610)$ and $Z_b(10650)$ really are \cite{Chen:2016qju}, but one promising interpretation about the two charged
bottomonium-like states is molecular conjecture which are composed of $B\bar{B}^\ast$ and $B^\ast\bar{B}^\ast$, respectively \cite{Bondar:2011ev,Sun:2011uh,Dias:2014pva,Ohkoda:2011vj,Kang:2016ezb}.
Stimulated by the continuously experimental observations of novelly hidden heavy flavor exotic states over the past decade, the interactions between heavy meson-heavy antimeson systems
have been intensively investigated. For example, before $Z_b(10610)$ was observed in 2011, Liu {\it et al} had predicted a loosely bound $S$-wave $B\bar{B}^\ast$ molecular state in the
framework of one boson exchange (OBE) model \cite{Liu:2008fh}, and later, the $B\bar{B}^\ast$ and $B^\ast\bar{B}^\ast$ systems were again studied with OBE model \cite{Sun:2011uh,Dias:2014pva,Ohkoda:2011vj,Sun:2012zzd,Zhao:2015mga}.
In Ref. \cite{Liu:2016kqx}, the pion-mediated interaction associated with coupled-channel effect is exploited to study the $B\bar{B}^\ast$ and $B^\ast\bar{B}^\ast$ systems.
And some other related works such as QCD sum rule \cite{Chen:2015ata}, Bethe-Salpeter approach \cite{He:2014nya}, coupled-channel model \cite{Coito:2016ads}, and so on.

As mentioned above, up to now, the observed exotic states are mainly composed of $Q\bar{Q}q\bar{q}$ or $Q\bar{Q}qqq$ (where $Q$ denotes $b$ or $c$ quark, and $q$ stands for $u$ or $d$ quark) \cite{Chen:2016qju}.
Whereas, nowadays, the exotic clusters with double charm or bottom still conceal themselves from the field of our view. Fortunately,
LHCb Collaboration reported the observation of doubly charmed baryon $\Xi_{cc}^{++}$ very recently \cite{Aaij:2017ueg}, which triggered many discussions on whether the stable $QQ\bar{q}\bar{q}$
tetraquark states can exist in nature. For instance, in Ref. \cite{Karliner:2017qjm}, Karliner {\it et al} predicted a doubly bottomed tetraquark $bb\bar{u}\bar{d}$ with $J^P=1^+$ at $10389\pm12$ MeV,
which lies far below the $B^-\bar{B}^\ast$ and electromagnetic decay $B^-\bar{B}^0\gamma$ thresholds. Eichten {\it et al} also found the double-beauty state with constituent quarks $bb\bar{u}\bar{d}$ is
stable against strong interactions \cite{Eichten:2017ffp}. Based on the quark-diquark symmetry and the input of the mass of $\Xi_{cc}^{++}$, Ref. \cite{Mehen:2017nrh} indicated the existence of a stable
double-bottom tetraquark state in $I=0$ channel.

Actually, the exploration on the existence of stable $QQ\bar{q}\bar{q}$ tetraquark states is always an intriguing topic. Many meaningful works, including quark model potential, QCD sum rule, color-magnetic interaction, and Lattice QCD, have been utilized to study $bb\bar{u}\bar{d}$ systems. Date back to 1985, in the framework of non-relativistic potential model, Ref. \cite{Zouzou:1986qh} found the state with quark contents $bb\bar{u}\bar{d}$ to be bound. Later, Barnes {\it et al} \cite{Barnes:1999hs} investigated $\bar{B}\bar{B}$, $\bar{B}\bar{B}^\ast$, and $\bar{B}^\ast \bar{B}^\ast$ intermeson interactions in quark model potential, and after solving two-meson Schr\"odinger equations they found $I=0$ $BB^\ast$ ($[b\bar{q}][b\bar{q}]$) channel is attractive (for other related works, see Refs. \cite{Vijande:2009kj,Ebert:2007rn,Zhang:2007mu,Du:2012wp,Luo:2017eub}). In Refs. \cite{Michael:1999nq,Detmold:2007wk,Bicudo:2012qt,Brown:2012tm,Bicudo:2015vta,Bicudo:2015kna,Francis:2016hui,Bicudo:2016ooe}, the potential between two $B$ mesons as a function of two static $b$ quark distance is
calculated in quenched Lattice QCD (for a review, see Ref. \cite{Beane:2008dv}), and especially in Refs. \cite{Bicudo:2015kna,Francis:2016hui,Bicudo:2016ooe} the binding energy region of $bb\bar{u}\bar{d}$ system is
evaluated to be $20-200$ MeV, which is compatible with phenomenological models' predictions \cite{Karliner:2017qjm,Eichten:2017ffp}.

Inspired by the observation of doubly charmed baryon $\Xi_{cc}^{++}$ \cite{Aaij:2017ueg} and many theoretical works as mentioned before, in this work we use chiral effective field theory ($\chi$EFT) to study the
 physically allowed $\bar{B}^{(\ast)} \bar{B}^{(\ast)} ([b\bar{q}][b\bar{q}])$ (see Tab. \ref{allowedBB}), i.e., $\bar{B}\bar{B}$, $\bar{B}\bar{B}^\ast$, and $\bar{B}^\ast \bar{B}^\ast$ intermeson interactions, to see in which channel
 the potential is attractive and to search for the possible bound states.
\begin{table}
\renewcommand{\arraystretch}{1.1}
 \tabcolsep=1.5pt
\caption{The physically allowed $\bar{B}\bar{B}$, $\bar{B}\bar{B}^\ast$, and $\bar{B}^\ast \bar{B}^\ast$ states \cite{Barnes:1999hs}, where $I_{\text{tot}}$ and $S_{\text{tot}}$ designates the total isospin and total
spin of two $B$ mesons, respectively. Due to the constraints of symmetry, the quantum numbers of $\bar{B}\bar{B}$ and $\bar{B}^\ast \bar{B}^\ast$ systems must satisfy the the selection rule $L+S_{\text{tot}}+I_{\text{tot}}+2i=Even\ number$, where $L$ is the orbital angular momentum between two $B$ mesons, and $i$ denotes the isospin of one $B$ meson, which is $1/2$.}\label{allowedBB}
\setlength{\tabcolsep}{3mm}
{
\begin{tabular}{c|cccc}
\hline\hline
\multirow{2}{*}{System}&Total isospin&\multicolumn{3}{c}{Total spin}\\
&$I_{\text{tot}}$&$S_{\text{tot}}=0$&$S_{\text{tot}}=1$&$S_{\text{tot}}=2$\\
\hline
\multirow{2}{*}{$\bar{B}\bar{B}$}&          1&            even $L$&               &   \\
                     &          0&            odd  $L$&               &   \\
\hline
\multirow{2}{*}{$\bar{B}\bar{B}^\ast$}&     1&            &        all $L$       &   \\
                     &          0&            &        all  $L$       &   \\
\hline
\multirow{2}{*}{$\bar{B}^\ast \bar{B}^\ast$}&     1&   even $L$         &        odd $L$       &even $L$   \\
                     &          0&odd  $L$            &        even $L$       &odd $L$   \\
\hline\hline
\end{tabular}
}
\end{table}

$\chi$EFT has been widely employed to study the interactions of nucleon-nucleon ($N$-$N$) and meson-meson ($M$-$M$) systems with flying colours. For the successful application of $\chi$EFT in $N$-$N$ systems, one can
see some important works in Refs. \cite{Bernard:1995dp,Epelbaum:2008ga,Machleidt:2011zz,Meissner:2015wva} and the references therein, or see Ref. \cite{Scherer:2002tk} for an introduction to $\chi$EFT. Here, we give a short review about the use of $\chi$EFT in $M$-$M$ systems, mainly focus on heavy meson sectors. In Ref. \cite{AlFiky:2005jd}, AlFiky {\it et al} proposed an approach with respecting the heavy quark symmetry and chiral
symmetry to deal with $X(3872)$ as a molecular state of $D^0\bar{D}^{\ast0}$. Similarly, there are also an abundance of investigations on $D\bar{D}^\ast(B\bar{B}^\ast)$ systems with $\chi$EFT, concerning the molecular assumptions
of $X(3872)$ and two $Z_b$ states \cite{Fleming:2008yn,Baru:2011rs,Valderrama:2012jv,Nieves:2012tt,Meng:2014ota,Baru:2015tfa,Wang:2013kva,Baru:2015nea,Jansen:2015lha,Braaten:2015tga,Baru:2016iwj} (or see Refs. \cite{Chen:2016qju,Guo:2017jvc} for a review). In addition to utilizing $\chi$EFT to study heavy meson-heavy antimeson systems, applying $\chi$EFT to heavy meson-heavy meson systems (such as $D^{(\ast)}D^{(\ast)}$, $D^{(\ast)}\bar{B}^{(\ast)}$ and $B^{(\ast)}B^{(\ast)}$) is also an interesting topic. In our two previous works, first in Ref. \cite{Liu:2012vd}, we gave a tentative usage of $\chi$EFT in $\bar{B}\bar{B}$ system, and obtained the strong interaction potentials in momentum space. Then in Ref. \cite{Xu:2017tsr}, we calculated $S$-wave $DD^{\ast}$ potential at one-loop level, noticed in the $0(1^+)$ channel it is attractive, and obtained a bound state with the binding energy $\Delta E_{DD^{\ast}}=-15.6^{+14.7}_{-19.2}$ MeV. We also notice that in Ref. \cite{Abreu:2015jma}, both open charm and bottom states ($D^{(\ast)} \bar{B}^{(\ast)}$) are studied with heavy meson chiral effective field theory (HM$\chi$EFT).

Based on Refs. \cite{Liu:2012vd,Xu:2017tsr}, we naturally extend our study to $\bar{B}^{(\ast)}\bar{B}^{(\ast)}$ systems, and for these kinds of states their typical quark configuration is $[b\bar{q}][b\bar{q}]$. To inquiry whether there exist such heavy flavor molecular states with double bottom or not, in our work we consider the $S$-wave ($L=0$) interactions of different $\bar{B}^{(\ast)}\bar{B}^{(\ast)}$ systems. According to the selection rule listed in Tab. \ref{allowedBB}, the $I(J^P)$ numbers for the physically allowed states are: $1(0^+)$ for $\bar{B}\bar{B}$; $1(1^+)$ and $0(1^+)$ for $\bar{B}\bar{B}^\ast$; $1(0^+)$, $1(2^+)$, and $0(1^+)$ for $\bar{B}^\ast \bar{B}^\ast$. In order to get the interaction potentials for these different channels in coordinate space, we firstly calculate the amplitudes in momentum space with $SU(2)$ HM$\chi$EFT up to one-loop level, where the amplitudes include the contributions from four-body contact interaction (FBCI), one-loop corrections to four-body contact interaction, one-pion exchange (OPE) contributions, one-loop corrections to one-pion exchange parts, and two-pion exchange (TPE) contributions. After subtracting the two-particle reducible (2PR) contributions in some typical Feynman diagrams, we make the Fourier transformation on the potentials in momentum space, and thus the potentials in coordinate space can be obtained. Finally, by solving the nonperturbative equations such as Schr\"odinger equation, Lippmann-Schwinger equation, and so on (in this paper, the Schr\"odinger equation is solved numerically), one can not only recover the 2PR contributions, but also get the binding energy $\Delta E$ in the attractive channels. In this way, we can predict the possible molecular states in $\bar{B}^{(\ast)}\bar{B}^{(\ast)}$ systems and make a comparison with other phenomenological models \cite{Barnes:1999hs,Li:2012ss}.

This paper is organized as follows. After the introduction, we show the effective Lagrangians used in this work and the Weinberg's formalism in Sec. \ref{Sec2}. The calculations of effective potentials and the analyses are represented in Sec. \ref{Sec3}, the contributions of the low energy constants at the next-to-leading order are estimated in Sec. \ref{Two_Order_LECs}, the results are summarized in Sec. \ref{Sec4}, and some needful formulas are given in the Appendix.

\section{effective lagrangians and Weinberg's formalism}\label{Sec2}
\subsection{Effective Lagrangians}
In the framework of HM$\chi$EFT, the scattering amplitudes can be expanded order by order with a small parameter $\epsilon=q/\Lambda_\chi$, where $q$ is either the momentum of Goldstone bosons or the residual momentum of heavy mesons, and $\Lambda_\chi$ represents either the chiral breaking scale or the mass of a heavy meson. Except for $\bar{B}\bar{B}$, the scattering amplitudes of both $\bar{B}\bar{B}^{\ast}$ and $\bar{B}^{\ast}\bar{B}^{\ast}$ channels at leading order $\mathcal{O}(\epsilon^0)$ have the contributions from FBCI and OPE, which are described by the leading effective Lagrangians in Eq. (\ref{4H}) \cite{AlFiky:2005jd,Valderrama:2012jv,Liu:2012vd} and Eq. (\ref{Hphi}) \cite{Burdman:1992gh,Wise:1992hn,Yan:1992gz}, respectively.
\begin{eqnarray}\label{4H}
\mathcal{L}_{4H}^{(0)}&=&D_a \Tr\left[H\gamma_\mu\bar{H}\right]\Tr\left[H\gamma^\mu\bar{H}\right]\nonumber\\
&&+D_b \Tr\left[H\gamma_\mu\gamma_5\bar{H}\right]\Tr\left[H\gamma^\mu\gamma_5\bar{H}\right]\nonumber\\
&&+E_a \Tr\left[H\gamma_\mu\tau^a\bar{H}\right]\Tr\left[H\gamma^\mu\tau_a\bar{H}\right]\nonumber\\
&&+E_b \Tr\left[H\gamma_\mu\gamma_5\tau^a\bar{H}\right]\Tr\left[H\gamma^\mu\gamma_5\tau_a\bar{H}\right],
\end{eqnarray}
where $D_a$, $D_b$, $E_a$, and $E_b$ are four independent low energy constants (LECs), which are determined later. $\tau^a$ represents the pauli matrix.
\begin{eqnarray}\label{Hphi}
\mathcal{L}_{H\phi}^{(1)}&=&-\left\langle\left(iv\cdot\partial H\right)\bar{H}\right\rangle+\left\langle Hv\cdot\Gamma\bar{H}\right\rangle+g\left\langle H\slashed{u}\gamma_5\bar{H}\right\rangle\nonumber\\
&&-\frac{1}{8}\Delta\left\langle H\sigma^{\mu\nu}\bar{H}\sigma_{\mu\nu}\right\rangle,
\end{eqnarray}
where $v=(1,\vec{0})$ is the four-velocity of heavy mesons, and $H$ denotes the degenerated $B$ and $B^\ast$ doublet in heavy quark limit, which can be expressed as:
\begin{eqnarray}
H&=&\frac{1+\slashed{v}}{2}\left(P_\mu^\ast\gamma^\mu+iP\gamma_5\right),\nonumber\\
\bar{H}&=&\gamma^0H^\dag\gamma^0=\left(P_\mu^{\ast\dag}\gamma^\mu+iP^\dag\gamma_5\right)\frac{1+\slashed{v}}{2}.
\end{eqnarray}
\begin{eqnarray}
P=\left(B^-,\bar{B}^0\right),\quad\quad\quad P_\mu^\ast=\left(B^{\ast-},\bar{B}^{\ast0}\right).
\end{eqnarray}
The last term in Eq. (\ref{Hphi}) accounts for the mass shift of $B$ and $B^\ast$, which will not vanish in chiral limit, and $\Delta$ is the mass difference of $(B,B^\ast)$ doublet. The tensor $\sigma^{\mu\nu}$ is defined as
$i[\gamma_\mu,\gamma_\nu]/2$. Besides, the chiral connection $\Gamma_\mu$ and axial vector current $u_\mu$ are illustrated as follows,
\begin{eqnarray}
\Gamma_\mu=\frac{i}{2}\left[\xi^\dag,\partial_\mu\xi\right],\quad\quad\quad u_\mu=\frac{i}{2}\left\{\xi^\dag,\partial_\mu\xi\right\},
\end{eqnarray}
where $\xi=\exp(i\phi/2f)$. $f$ is the pion decay constant, and the triplet pion field $\phi$ is defined as
\begin{displaymath}
\phi =\sqrt{2}
\left( \begin{array}{cc}
\frac{\pi^0}{\sqrt{2}} & \pi^+\\
\pi^- & -\frac{\pi^0}{\sqrt{2}}
\end{array} \right).
\end{displaymath}

As indicated in Ref. \cite{Liu:2012vd}, other forms of FBCI at the leading order with different Lorentz structures are not independent, such as $\Tr\left[H\gamma_\mu\bar{H}H\gamma_\mu\bar{H}\right]$, $\Tr\left[H\bar{H}\right]\Tr\left[H\bar{H}\right]$, and $\Tr\left[H\sigma_{\mu\nu}\bar{H}\right]\Tr\left[H\sigma^{\mu\nu}\bar{H}\right]$, which can be expressed as the linear combinations of the terms involved in Eq. (\ref{4H}). Some terms like $\Tr\left[H\gamma_5\bar{H}\right]\Tr\left[H\gamma_5\bar{H}\right]$ and $\Tr\left[H\gamma_5\tau^a\bar{H}\right]\Tr\left[H\gamma_5\tau_a\bar{H}\right]$ vanish in heavy quark limit.

At the next-to-leading order, i.e., $\mathcal{O}(\epsilon^2)$, the scattering amplitudes can be decomposed into four parts, one-loop corrections to FBCI and OPE, TPE, and the tree diagrams governed by $\mathcal{O}(\epsilon^2)$ Lagrangians. In our calculations, we introduce the $\mathcal{O}(\epsilon^2)$ FBCI Lagrangians to renormalize the $\mathcal{O}(\epsilon^2)$ loop diagrams, which read \cite{Liu:2012vd,Xu:2017tsr},
\begin{widetext}
 \begin{eqnarray}\label{4H2h}
 \mathcal L^{(2,h)}_{4H}&=& D_{a}^h\Tr\left[H \gamma_\mu \bar H \right]\Tr\left[ H
 \gamma^\mu\bar H\right] \Tr(\chi_+)+D_{b}^h\Tr\left[H \gamma_\mu\gamma_5
 \bar H \right]\Tr\left[ H \gamma^\mu\gamma_5\bar H\right]\Tr(\chi_+)\nonumber \\
 &&+E_{a}^h \Tr\left[H \gamma_\mu\tau^a \bar H \right]\Tr\left[ H
 \gamma^\mu\tau_a\bar H\right]\Tr(\chi_+)+E_{b}^h \Tr\left[H
 \gamma_\mu\gamma_5\tau^a \bar H \right]\Tr\left[ H
 \gamma^\mu\gamma_5\tau_a\bar H\right]\Tr(\chi_+),
 \end{eqnarray}
  \begin{eqnarray}\label{4H2v}
 \mathcal L^{(2,v)}_{4H}&=& \bigg\{D_{a1}^v\Tr\left[(v\cdot D H) \gamma_\mu
 (v\cdot D \bar H) \right]\Tr\left[ H \gamma^\mu\bar H\right]+D_{a2}^v\Tr\left[(v\cdot D
 H) \gamma_\mu \bar H \right]\Tr\left[ (v\cdot D H) \gamma^\mu\bar H\right]
 \nonumber\\&& +D_{a3}^v\Tr\left[(v\cdot D H) \gamma_\mu \bar H \right]\Tr\left[  H
 \gamma^\mu(v\cdot D \bar H)\right] +D_{a4}^v\Tr\left[\left((v\cdot D)^2 H\right)
 \gamma_\mu \bar H \right]\Tr\left[  H \gamma^\mu\bar H \right] \nonumber\\&&
 +D_{b1}^v\Tr\left[(v\cdot D H) \gamma_\mu\gamma_5 (v\cdot D \bar H)
 \right]\Tr\left[ H \gamma^\mu\gamma_5\bar H\right]+... +E_{a1}^v \Tr\left[(v\cdot D H)
 \gamma_\mu\tau^a (v\cdot D \bar H) \right]\Tr\left[ H
 \gamma^\mu\tau_a\bar H\right]+... \nonumber\\&& +E_{b1}^v\Tr\left[(v\cdot
 D H) \gamma_\mu\gamma_5\tau^a (v\cdot D \bar H) \right]\Tr\left[ H
 \gamma^\mu\gamma_5\tau_a \bar H\right] +...\bigg\} +\text{H.c.},
  \end{eqnarray}
 \begin{eqnarray}\label{4H2q}
 \mathcal L^{(2,q)}_{4H}&=& \bigg\{D_1^q\Tr\left[(D^\mu H) \gamma_\mu\gamma_5
 (D^\nu \bar H) \right]\Tr\left[ H \gamma_\nu\gamma_5\bar H\right] +D_2^q\Tr\left[(D^\mu
 H) \gamma_\mu\gamma_5 \bar H \right]\Tr\left[ (D^\nu H)
 \gamma_\nu\gamma_5\bar H\right] \nonumber\\&& +D_3^q\Tr\left[(D^\mu H)
 \gamma_\mu\gamma_5 \bar H \right]\Tr\left[ H \gamma_\nu\gamma_5(D^\nu \bar
 H)\right]+D_4^q\Tr\left[(D^\mu D^\nu H) \gamma_\mu\gamma_5 \bar H \right]\Tr\left[ H
 \gamma_\nu\gamma_5 \bar H\right] \nonumber\\&& +E_1^q\Tr\left[(D^\mu H)
 \gamma_\mu\gamma_5 \tau^a(D^\nu \bar H) \right]\Tr\left[ H
 \gamma_\nu\gamma_5\tau_a\bar H\right] +...\bigg\}+\text{H.c.},
  \cdots,
 \end{eqnarray}
 \end{widetext}
 where
\begin{eqnarray}
 \tilde\chi_\pm=\chi_\pm-\frac12\Tr[\chi_\pm],~~~~ \chi_\pm=\xi^\dagger\chi\xi^\dagger\pm\xi\chi\xi,~~~~ \chi=m_\pi^2.
\end{eqnarray}
The large amounts of LECs appeared in Eqs. (\ref{4H2h})-(\ref{4H2q}) contain both the finite and infinite parts, and the infinite parts can be used to cancel the divergences of the loop diagrams at $\mathcal{O}(\epsilon^2)$. Nevertheless, the finite parts can not be fitted due to the lack of enough data right now, and thus we neglect the contributions from $\mathcal{O}(\epsilon^2)$ tree diagrams that governed by the Lagrangians in Eqs. \eqref{4H2h}-\eqref{4H2q} in our computations for the moment. However, an estimation of the errors caused by the $\mathcal{O}(\epsilon^2)$ tree graphes is presented in Sec. \ref{Two_Order_LECs}.
\subsection{Weinberg's formalism}
Before formally carrying out our calculations, we give a brief introduction to Weinberg's power counting scheme \cite{Weinberg:1990rz,Weinberg:1991um} since it is the core of the theoretical calculations in this work.
\begin{figure}[hptb]
    \vspace{0.3cm}
	\scalebox{0.7}{\includegraphics[width=\columnwidth]{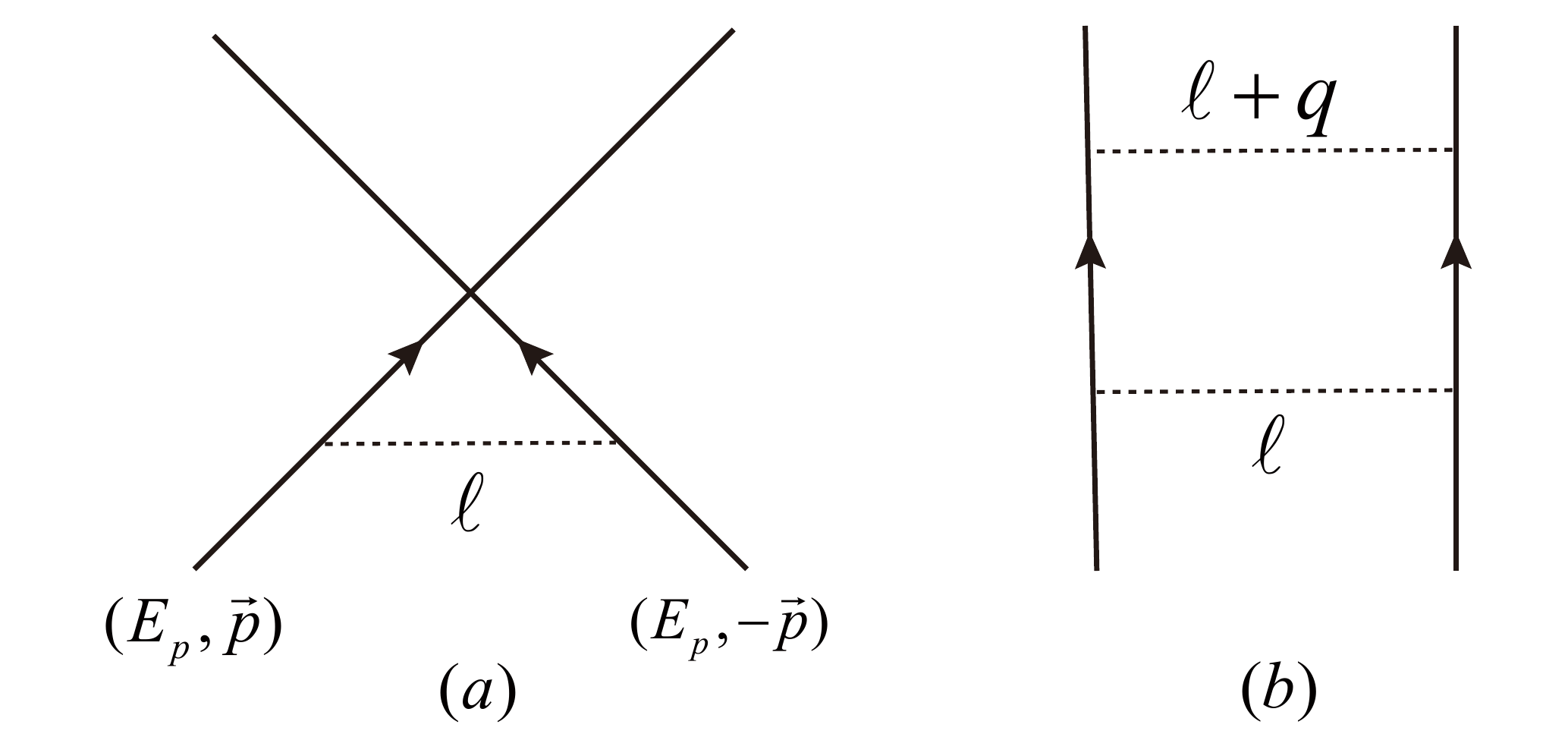}}
	\caption{Two types of 2PR Feynman diagrams of $N$-$N$ scattering, which will also appear in this work. \label{2PR_Diagram}}
\end{figure}

We take the two-nucleon scattering process as an example. Considering the one-loop Feynman diagrams displayed in Fig. \ref{2PR_Diagram}, in the non-relativistic limit the scattering amplitude of Fig. \ref{2PR_Diagram}(a) is proportional to
\begin{eqnarray}
i\int\frac{d^D\ell}{(2\pi)^D}\frac{1}{(-\ell_0+i\varepsilon)(\ell_0+i\varepsilon)(\ell^2-m_\pi^2+i\varepsilon)}.
\end{eqnarray}
Inspecting the above loop integral we can find that the integration over $\ell_0$ is ill-defined because it has poles above and below the real axis at $\ell_0=\pm i\varepsilon$ (which is always called as pinch singularity \cite{Weinberg:1991um,Machleidt:2011zz}). This problem can be cured by including the kinetic energy of the nucleon in the leading terms but not treating it as a perturbation. Then the pole positions of the two nucleons' propagators will be shifted to
\begin{eqnarray}
\ell_0=\pm\left(\mathcal{E}-\vec{\ell}^{2}/2M_N\right)\pm i\varepsilon,\nonumber
\end{eqnarray}
where $\mathcal{E}=\vec{p}^{2}/2M_N$, and $M_N$ is the mass of the nucleon. As is pointed out by Weinberg \cite{Weinberg:1990rz,Weinberg:1991um}, after the $\ell_0$ integration being performed with residue theorem the contribution from the pole of nucleon is enhanced by a large factor $M_N/|\vec{p}|$, and this strong enhancement explicitly breaks the naive power counting with which the $\ell_0$ integral should have been of $\mathcal{O}(1/|\vec{p}|)$.

With the formalism given in Ref. \cite{Weinberg:1991um}, we concentrate on the {\it effective potential} (i.e., the irreducible graphs) but not directly calculate the scattering amplitudes. Namely, removing the 2PR part contributions that originate from intermediate on-shell nucleon states (``infrared enhancement''), the irreducible diagrams that make up the effective potential can then be calculated perturbatively. The 2PR part will be automatically recovered when the effective potential (or kernel) is inserted into the Schr\"odinger equation (or Lippmann-Schwinger equation).

We will encounter the same problem as discussed above when studying the interactions of two $B$ mesons with HM$\chi$EFT. We follow the ideas in studying $N$-$N$ system \cite{Weinberg:1991um}, i.e., we calculate the effective potentials of two $B$ mesons first. One can find more details of calculations in Appendix \ref{LoopIntegral}.
\section{effective potentials of $\bar{B}^{(\ast)}\bar{B}^{(\ast)}$ systems}\label{Sec3}
\subsection{$\bar{B}\bar{B}$ system}
The $\bar{B}\bar{B}$ system has been studied in Ref. \cite{Liu:2012vd} with HM$\chi$EFT, but in Ref. \cite{Liu:2012vd} the four LECs were unevaluated and only the TPE contributions in momentum space were shown, and thus the behavior of the total potential is still ambiguous. Therefore, in this part, we revisit this system and give a more intuitional form of the potential in coordinate space to see whether a bound state can exist in $\bar{B}\bar{B}$ system.
\begin{figure}[hptb]
    \vspace{0.3cm}
	\scalebox{1.0}{\includegraphics[width=\columnwidth]{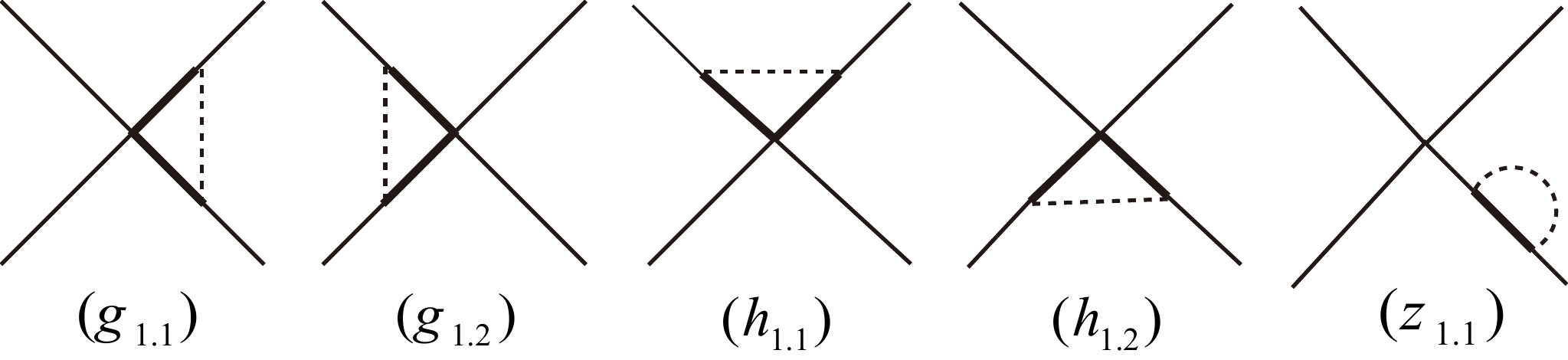}}
	\caption{One-loop corrections to the contact vertex at $\mathcal{O}(\epsilon^2)$. The thin solid, thick solid, and dashed lines represent $\bar{B}$ meson, $\bar{B}^\ast$ meson, and pion, respectively. \label{BB_Contact_Corrections}}
\end{figure}
\begin{figure}[hptb]
	\scalebox{1.0}{\includegraphics[width=\columnwidth]{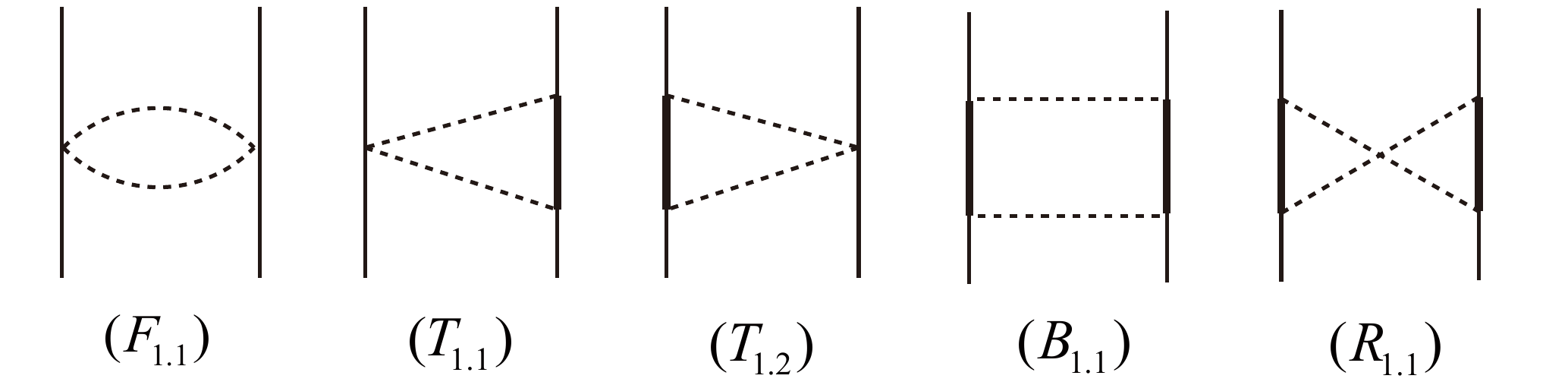}}
	\caption{Two pion exchange diagrams at $\mathcal{O}(\epsilon^2)$. There is one football diagram ($F_{1.1}$), two triangle diagrams ($T_{1.1}$ and $T_{1.2}$), one box diagram ($B_{1.1}$), and one crossed diagram ($R_{1.1}$). Notations same as in Fig. \ref{BB_Contact_Corrections}.\label{BB_TPE}}
\end{figure}

From Tab. \ref{allowedBB}, one can see that only $1(0^+)$ $\bar{B}\bar{B}$ state survives under $S$-wave interaction, which corresponds to the $B^-(p_1)B^-(p_2)\to B^-(p_3)B^-(p_4)$ elastic scattering process. At the leading order $\mathcal{O}(\epsilon^0)$, OPE is excluded and the amplitude only receive contribution from the FBCI, by expanding the Lagrangian in Eq. (\ref{4H}), one can easily get
 \begin{eqnarray}\label{BBZeroOrderFBCI}
\mathcal{Y}_{I=1}^{(0)\text{FBCI}}=8(D_a+E_a).
 \end{eqnarray}

At $\mathcal{O}(\epsilon^2)$,  the amplitudes receive both the one-loop diagrams for FBCI corrections and TPE diagrams which are illustrated in Fig. \ref{BB_Contact_Corrections} and Fig. \ref{BB_TPE}, respectively. The amplitudes correspond to the diagrams in Fig. \ref{BB_Contact_Corrections} are listed in Eqs. \eqref{AMP_BBTwoOrderFBCI_Yg11}-\eqref{AMP_BBTwoOrderFBCI_Yz11},
\begin{eqnarray}
\mathcal{Y}^{g_{1.1}}&=&\mathcal{C}^{g_{1.1}}\frac{g^2}{f^2}\left\{(D-1)\mathcal{J}_{22}^g\right\}_r(m_\pi,\mathcal{E}-\Delta,\mathcal{E}-\Delta),\label{AMP_BBTwoOrderFBCI_Yg11}
\end{eqnarray}
\begin{eqnarray}
\mathcal{Y}^{g_{1.2}}&=&\mathcal{C}^{g_{1.2}}\frac{g^2}{f^2}\left\{(D-1)\mathcal{J}_{22}^g\right\}_r(m_\pi,\mathcal{E}-\Delta,\mathcal{E}-\Delta),
\end{eqnarray}
\begin{eqnarray}
\mathcal{Y}^{h_{1.1}}&=&\mathcal{C}^{h_{1.1}}\frac{g^2}{f^2}\left\{(D-1)\mathcal{J}_{22}^h\right\}_r(m_\pi,\mathcal{E}-\Delta,\mathcal{E}-\Delta),
\end{eqnarray}
\begin{eqnarray}
\mathcal{Y}^{h_{1.2}}&=&\mathcal{C}^{h_{1.2}}\frac{g^2}{f^2}\left\{(D-1)\mathcal{J}_{22}^h\right\}_r(m_\pi,\mathcal{E}-\Delta,\mathcal{E}-\Delta),
\end{eqnarray}
\begin{eqnarray}
\mathcal{Y}^{z_{1.1}}&=&\mathcal{C}^{z_{1.1}}\frac{g^2}{f^2}\left\{(D-1)\frac{\partial}{\partial x}\mathcal{J}_{22}^a\right\}_r(m_\pi,x)\Big|_{x\to\mathcal{E}-\Delta},\label{AMP_BBTwoOrderFBCI_Yz11}
\end{eqnarray}
where the coefficients $\mathcal{C}^x$ are flavor and LECs dependent, their concrete values are:
\begin{eqnarray}
\mathcal{C}^{g_{1.1}}&=&\mathcal{C}^{g_{1.2}}=-4(3D_a-D_b-E_a-5E_b),\nonumber\\
\mathcal{C}^{h_{1.1}}&=&\mathcal{C}^{h_{1.2}}=4(D_b+E_b),\quad\mathcal{C}^{z_{1.1}}=-6(D_a+E_a).
\end{eqnarray}

Various scalar functions $\mathcal{J}_x^y$ are defined in Appendix \ref{LoopIntegral_1}, $m_\pi$ is the mass of pion, $D$ is the dimension where the loop integral is performed and approaches 4 at last. $\mathcal{E}$ represents the residual energy of $\bar{B}$ and $\bar{B}^\ast$ mesons, and is defined as $\mathcal{E}=E_{\bar{B}^{(\ast)}}-M_{\bar{B}^{(\ast)}}$. $\{X\}_r$ denotes the finite part of $X$ \cite{Liu:2012vd},
\begin{eqnarray}
\{X\}_r=\lim_{D\to4}\left(X-L\frac{\partial}{\partial L}X\right)+\frac{1}{16\pi^2}\lim_{D\to4}\left(\frac{\partial}{\partial D}\frac{\partial}{\partial L}X\right),
\end{eqnarray}
which is equivalent to make use of the modified minimal subtraction ($\mathrm{\overline{MS}}$) scheme.

The amplitudes of the diagrams $F_{1.1}$, $T_{1.1}$, $T_{1.2}$, $B_{1.1}$, and $R_{1.1}$ in Fig. \ref{BB_TPE} are illustrated in Eqs. \eqref{AMP_BBTPE_F11}-\eqref{AMP_BBTPE_R11}, respectively.
\begin{widetext}
\begin{eqnarray}
\mathcal{Y}^{F_{1.1}}&=&-\frac{1}{4f^4}\bigg\{q_0^2\left(\mathcal{J}_{0}^F+4\mathcal{J}_{11}^F+4\mathcal{J}_{21}^F\right)+4\mathcal{J}_{22}^F\bigg\}_r(m_\pi,q),\label{AMP_BBTPE_F11}
\end{eqnarray}
\begin{eqnarray}
\mathcal{Y}^{T_{1.1}}&=&-\frac{g^2}{2f^4}\Bigg\{(D-1)\bigg[q_0\left(\mathcal{J}_{21}^T+2\mathcal{J}_{31}^T\right)+2\mathcal{J}_{34}^T\bigg]-q_0\vec{q}^2\left(\mathcal{J}_{11}^T+3\mathcal{J}_{22}^T+2\mathcal{J}_{32}^T\right)-2\vec{q}^2\left(\mathcal{J}_{24}^T+\mathcal{J}_{33}^T\right)\Bigg\}_r(m_\pi,\mathcal{E}-\Delta,q),
\end{eqnarray}
\begin{eqnarray}
\mathcal{Y}^{T_{1.2}}&=&-\frac{g^2}{2f^4}\Bigg\{(D-1)\bigg[q_0\left(\mathcal{J}_{21}^T+2\mathcal{J}_{31}^T\right)+2\mathcal{J}_{34}^T\bigg]-q_0\vec{q}^2\left(\mathcal{J}_{11}^T+3\mathcal{J}_{22}^T+2\mathcal{J}_{32}^T\right)-2\vec{q}^2\left(\mathcal{J}_{24}^T+\mathcal{J}_{33}^T\right)\Bigg\}_r(m_\pi,\mathcal{E}-\Delta,q),
\end{eqnarray}
\begin{eqnarray}
\mathcal{Y}^{B_{1.1}}&=&-\frac{g^4}{4f^4}\Bigg\{(D^2-1)\mathcal{J}_{41}^B-\vec{q}^2\bigg[2(D+1)\left(\mathcal{J}_{31}^B+\mathcal{J}_{42}^B\right)+\mathcal{J}_{21}^B\bigg]+(\vec{q}^2)^2\left(\mathcal{J}_{22}^B+2\mathcal{J}_{32}^B+\mathcal{J}_{43}^B\right)\Bigg\}_r(m_\pi,\mathcal{E}-\Delta,\mathcal{E}-\Delta,q),
\end{eqnarray}
\begin{eqnarray}
\mathcal{Y}^{R_{1.1}}&=&-\frac{5g^4}{4f^4}\Bigg\{(D^2-1)\mathcal{J}_{41}^R-\vec{q}^2\bigg[2(D+1)\left(\mathcal{J}_{31}^R+\mathcal{J}_{42}^R\right)+\mathcal{J}_{21}^R\bigg]+(\vec{q}^2)^2\left(\mathcal{J}_{22}^R+2\mathcal{J}_{32}^R+\mathcal{J}_{43}^R\right)\Bigg\}_r(m_\pi,\mathcal{E}-\Delta,\mathcal{E}-\Delta,q).\label{AMP_BBTPE_R11}
\end{eqnarray}
\end{widetext}

Here we need to stress that in rigorous heavy quark limit, i.e., when the mass difference $\Delta=0$, the amplitudes induced by the diagrams $h_{1.1}$, $h_{1.2}$, and $B_{1.1}$ in Fig. \ref{BB_Contact_Corrections} and Fig. \ref{BB_TPE} would approach to infinity. Thus in order to get the correct potential, we employ Weinberg's formalism to subtract the 2PR contributions that stem from the double poles of the two intermediate heavy vector mesons. The contributions from the $B^\ast$ meson poles are enhanced in the $1/M_B$ expansion and repeat the results of the iterated OPE diagrams, and thus we only need to consider the contribution from pion poles. The related details are shown in Appendix \ref{LoopIntegral_2} and \ref{LoopIntegral_3}.

The effective potential $\mathcal{V}$ can be derived from the 2PI amplitude $\mathcal{Y}_{\text{2PI}}$ by the relation
\begin{eqnarray}\label{V_A}
\mathcal{V}=-\frac{1}{4}\mathcal{Y}_{\text{2PI}},
\end{eqnarray}
where the factor $-1/4$ comes from OBE model \cite{Sun:2012zzd,Li:2012ss}, which originates from Breit approximation.

The parameters used in the calculations are: $\Delta=0.045$ GeV, $m_\pi=0.139$ GeV, $f=0.092$ GeV, $g=0.52$ \cite{Patrignani:2016xqp,Ohki:2008py,Detmold:2012ge}, and the renormalization scale $\lambda=4\pi f$. For simplicity, as in the OBE model \cite{Liu:2008fh}, the transferred energy $q_0$ and residual energy $\mathcal{E}$ of heavy mesons are both set to be zero.

By calculating Eqs. \eqref{AMP_BBTwoOrderFBCI_Yg11}-\eqref{AMP_BBTwoOrderFBCI_Yz11}, we obtain the result of one-loop corrections to FBCI,
\begin{eqnarray}\label{BBTwoOrderFBCI}
\mathcal{Y}_{I=1}^{(2)\text{FBCI}}=4(-0.18D_b+0.1E_a-0.08E_b).
\end{eqnarray}
We can see the contribution of $D_a$ vanishes, while $D_b$ and $E_b$ emerge at $\mathcal{O}(\epsilon^2)$. In addition, the result in Eq. (\ref{BBTwoOrderFBCI}) is about one order of magnitude smaller than that of Eq. (\ref{BBZeroOrderFBCI}), which manifests the convergence of chiral correction is very good.

\begin{figure*}
\begin{minipage}[t]{0.45\linewidth}
\centering
\includegraphics[width=\columnwidth]{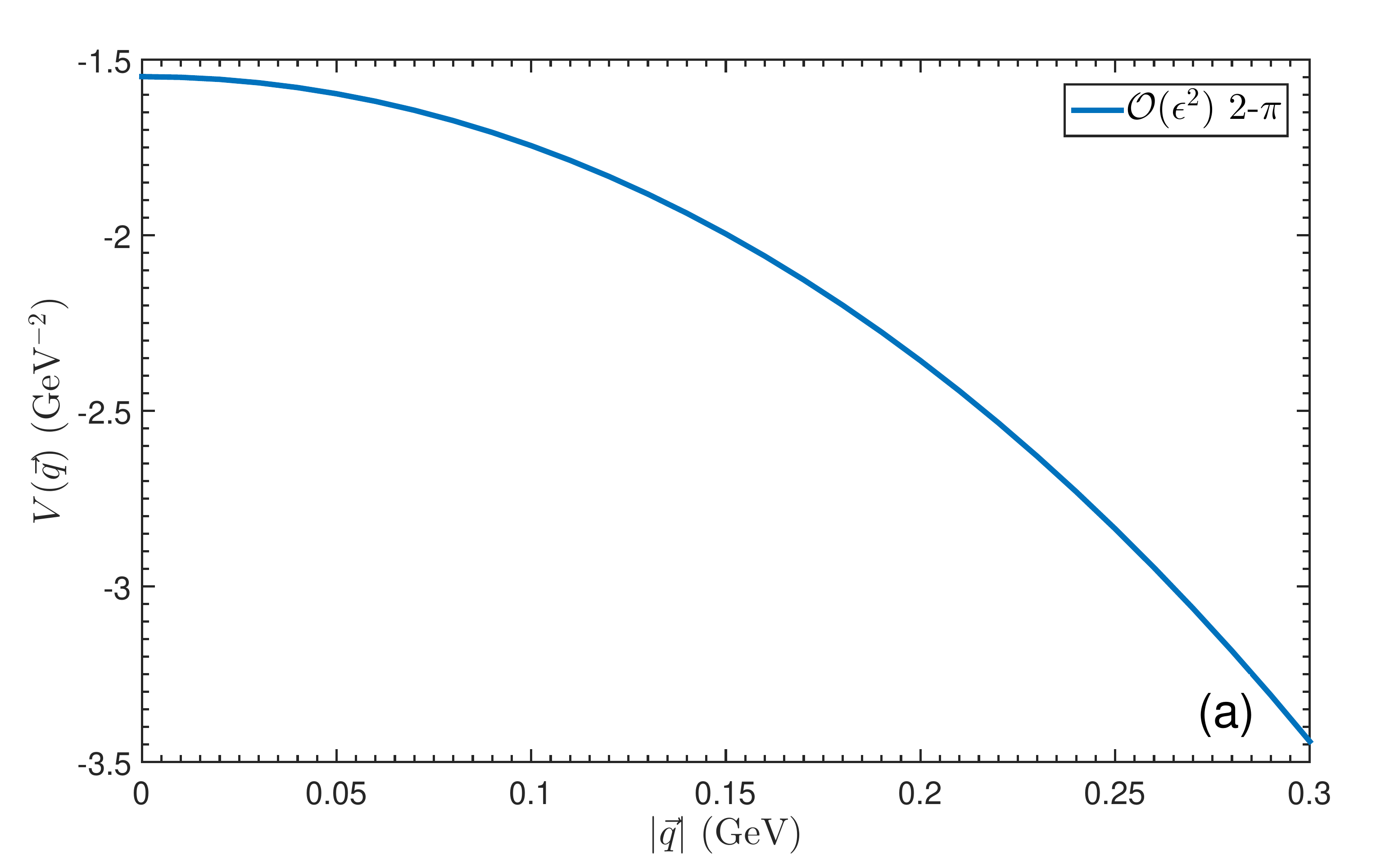}
\end{minipage}%
\begin{minipage}[t]{0.45\linewidth}
\centering
\includegraphics[width=\columnwidth]{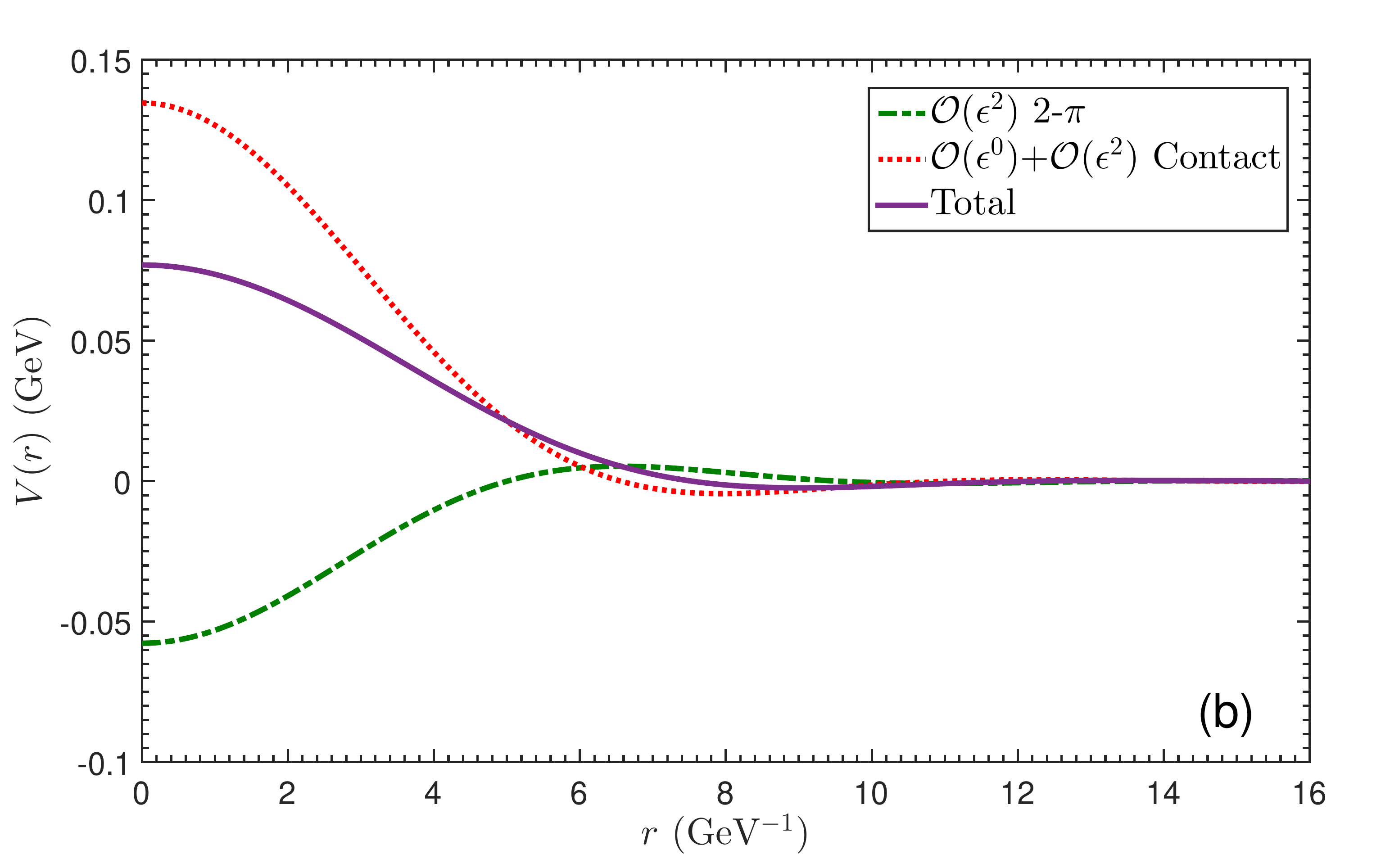}
\end{minipage}
\caption{Figure (a) represents the TPE potential of $1(0^+)$ $\bar{B}\bar{B}$ system in momentum space. The potentials in coordinate space are shown in figure (b), where the dot-dashed line describes the TPE potential, the dotted line denotes the sum of FBCI potentials at $\mathcal{O}(\epsilon^0)$ and $\mathcal{O}(\epsilon^2)$, and the solid line stands for the total potential. \label{BB_Momentum_Euclid}}
\end{figure*}

Because the FBCI and one-loop corrections to FBCI describe the short distance interactions, and they are independent of the transferred momentum $|\vec{q}|$. Thus the $|\vec{q}|$ dependent part can only come from TPE. The result of TPE in momentum space is displayed in Fig. \ref{BB_Momentum_Euclid}(a). We can see that the potential is negative when $|\vec{q}|$ is varied from $0$ to $300$ MeV, which indicates the TPE part supplies the attractive potential, and this potential would be deeper if $|\vec{q}|$ becomes larger.

Because the potential in coordinate space can give more intuitional information, so we make the Fourier transformation on the potential in momentum space by the following way,
\begin{eqnarray}\label{Four_Tansf}
\mathcal{V}(r)=\int\frac{d^3\bold{q}}{(2\pi)^3}e^{-i\bf{q\cdot r}}\mathcal{V}(\bf q)\mathcal{F}(\bf q).
\end{eqnarray}
We should note that, because the potential $\mathcal{V}(\bf q)$ is expanded as the power series of the transferred momentum $q$ in $\chi$EFT, so we need to regularize $\mathcal{V}(\bf q)$ to suppress the contributions from high momentum to avoid the ultraviolet divergence of the integral, which is the manifestation of the fact that $\chi$EFT is valid only for $q\ll\Lambda_\chi$. Different approaches in Refs. \cite{Fleming:1999ee,Phillips:1997xu,Frederico:1999ps,Birse:2005um,Beane:2001bc,PavonValderrama:2005gu,Nogga:2005hy,PavonValderrama:2005wv,PavonValderrama:2007nu,Entem:2007jg,Yang:2007hb,Beane:2008bt,Valderrama:2007ja,Cordon:2009pj,Gamermann:2009uq} were developed to regularize/renormalize $\mathcal{V}(\bf q)$ non-perturbatively. In this work, as in Refs. \cite{Xu:2017tsr,Ordonez:1995rz,Epelbaum:1999dj,Ren:2016jna}, we adopt a simple Gauss regulator $\mathcal{F}(\bold{q})=\exp(-\bold{q}^{2n}/\Lambda^{2n})$, and we set $n=2$ as in Ref. \cite{Ren:2016jna}. The value of the cutoff parameter $\Lambda$ is commonly below the mass of $\rho$ meson in the $N$-$N$ case \cite{Epelbaum:2014efa}, so we use a moderate value $\Lambda=0.7$ GeV as adopted in Ref. \cite{Xu:2017tsr} to give predictions.

If we want to get the numerical results of the effective potential in coordinate space, we have to know the concrete values of the four LECs (see Eq. \eqref{4H}). The values of the LECs have been determined in Ref. \cite{Xu:2017tsr} by exploiting the resonance saturation model \cite{Epelbaum:2001fm,Ecker:1988te,Du:2016tgp}, which read (all in units of GeV$^{-2}$),
\begin{eqnarray}\label{Values_LECs}
D_a&=&-6.62\pm0.15,~~~~~D_b=0\pm1.96,\nonumber\\
E_a&=&-5.74\pm0.45,~~~~~E_b=0\pm1.89.
\end{eqnarray}

With the preparations above, the effective potential of $1(0^+)$ $\bar{B}\bar{B}$ system in coordinate space is given in Fig. \ref{BB_Momentum_Euclid}(b). From the figure, we can read that although the TPE potential is attractive, the contribution from FBCI is dominantly repulsive, and thus the total potential is undoubtedly repulsive. This result indicates that it is impossible to find a bound state solution in $1(0^+)$ $\bar{B}\bar{B}$ system, which is in coincidence with the calculations of Lattice QCD \cite{Bicudo:2015kna}.

\subsection{$\bar{B}\bar{B}^\ast$ system}
\begin{figure}[hptb]
\vspace{0.3cm}
	\scalebox{0.65}{\includegraphics[width=\columnwidth]{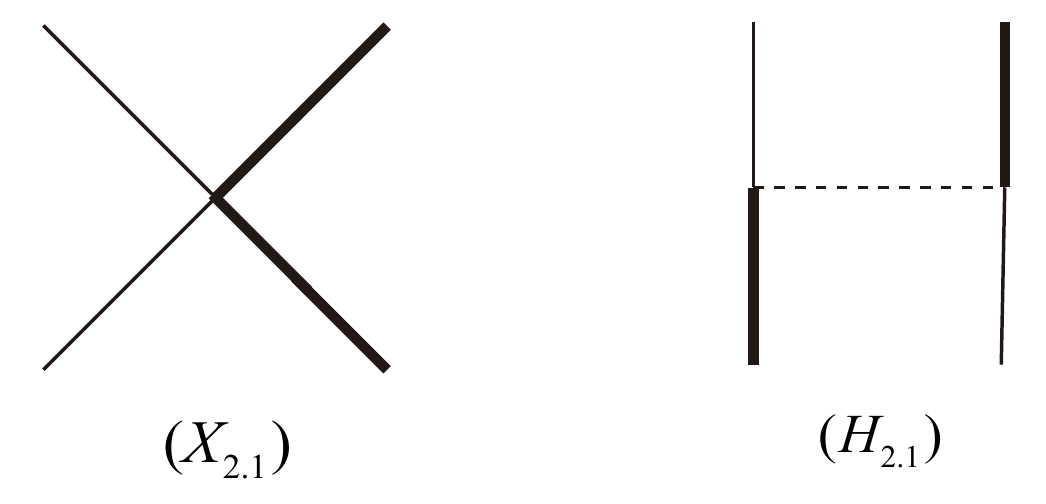}}
	\caption{The diagrams for $\bar{B}\bar{B}^\ast$ system at $\mathcal{O}(\epsilon^0)$, where $X_{2.1}$ depicts the contact interaction, and $H_{2.1}$ is OPE diagram. Notations same as in Fig. \ref{BB_Contact_Corrections}.\label{BBast_ZeroOrder}}
\end{figure}

\begin{figure*}[hptb]
	\scalebox{0.9}{\includegraphics[width=18cm]{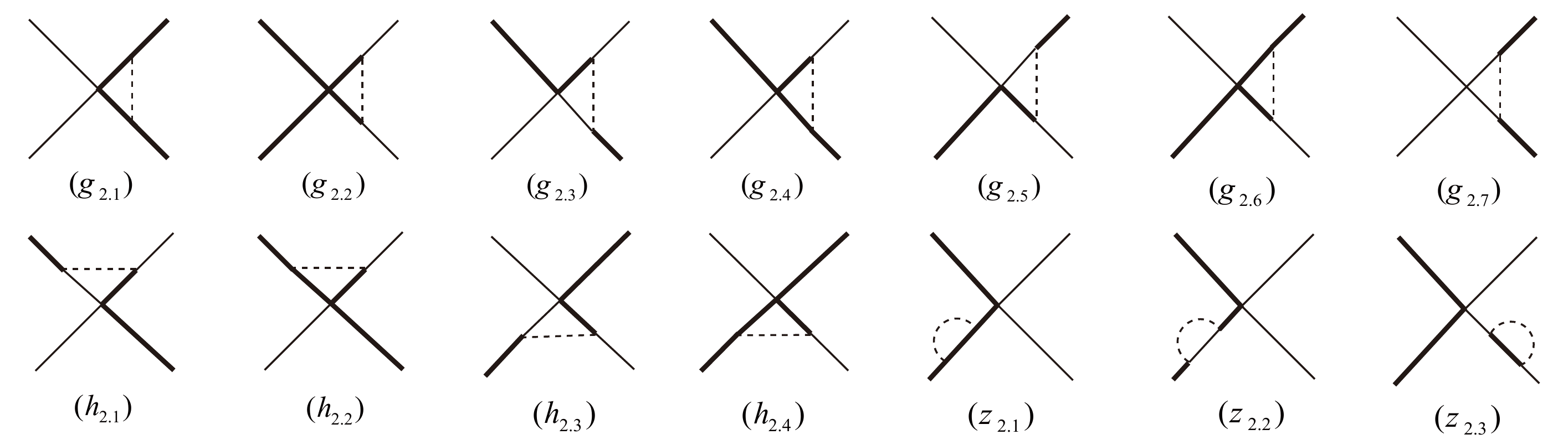}}
	\caption{One-loop corrections to the FBCI of $\bar{B}\bar{B}^\ast$ system at $\mathcal{O}(\epsilon^2)$, which include the corrections to the $\bar{B}\bar{B}^\ast\bar{B}\bar{B}^\ast$ contact vertex
 ($g_{2.1}$$\sim$$h_{2.4}$), and wave function renormalizations of external legs ($z_{2.1}$$\sim$$z_{2.3}$). Notations same as in Fig. \ref{BB_Contact_Corrections}.\label{BBast_Contact_Corrections}}
\end{figure*}

\begin{figure*}[hptb]
	\scalebox{0.825}{\includegraphics[width=18cm]{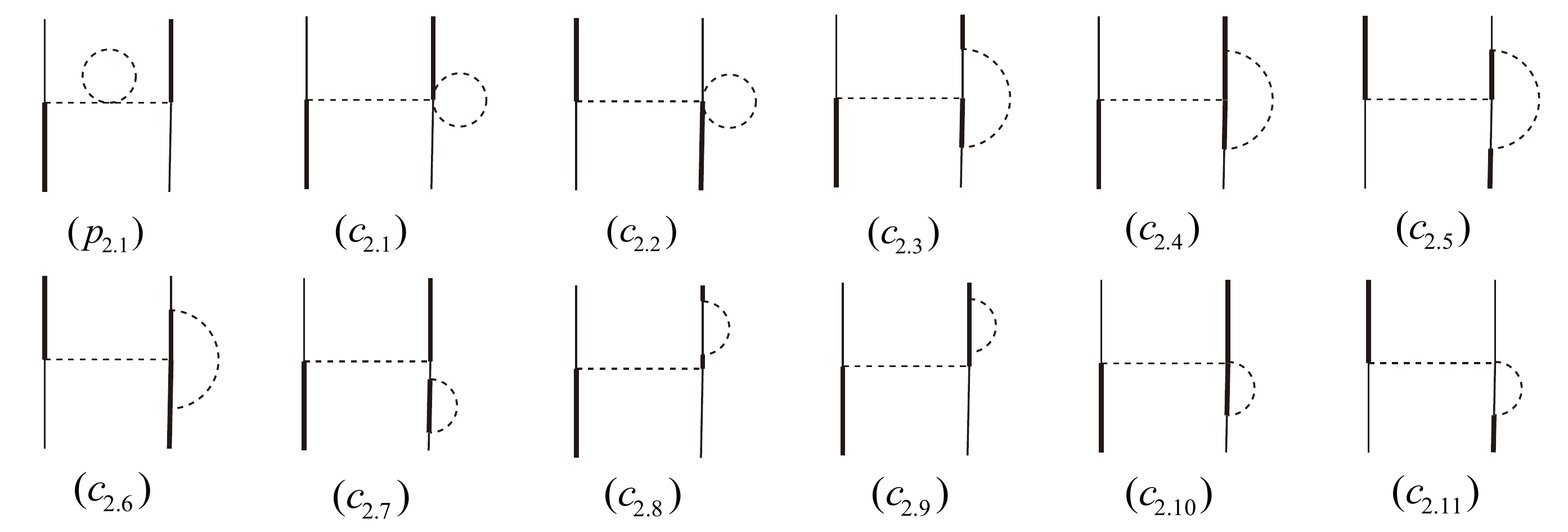}}
	\caption{One-loop corrections to the OPE diagram of $\bar{B}\bar{B}^\ast$ system at $\mathcal{O}(\epsilon^2)$, which include the renormalizations of pion line ($p_{2.1}$), $\bar{B}\bar{B}^\ast\pi$ vertex ($c_{2.1}$$\sim$$c_{2.6}$, $c_{2.10}$$\sim$$c_{2.11}$), and external legs ($c_{2.7}$$\sim$$c_{2.9}$). Notations same as in Fig. \ref{BB_Contact_Corrections}.\label{BBast_OPE_Corrections}}
\end{figure*}

\begin{figure*}[hptb]
	\scalebox{1.0}{\includegraphics[width=18cm]{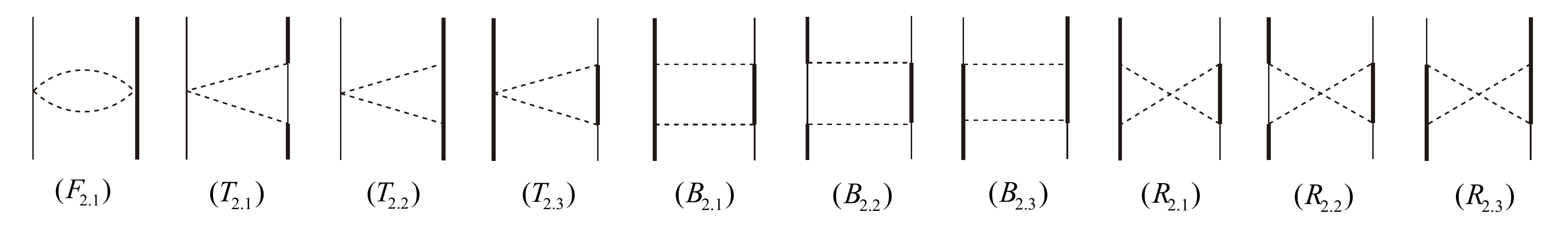}}
	\caption{Two-pion exchange diagrams of $\bar{B}\bar{B}^\ast$ system at $\mathcal{O}(\epsilon^2)$. Notations same as in Fig. \ref{BB_Contact_Corrections}.\label{BBast_TPE}}
\end{figure*}

For the $S$-wave $\bar{B}\bar{B}^\ast$ system, there are two different isospin states, i.e., $1(1^+)$ and $0(1^+)$ $\bar{B}\bar{B}^\ast$ states. At the leading order, both FBCI and OPE diagrams contribute to the scattering amplitudes, which are illustrated in Fig. \ref{BBast_ZeroOrder}, at the next-to-leading order, the amplitude is composed of one-loop corrections to FBCI and OPE diagrams, and TPE diagrams, which are shown in Fig. \ref{BBast_Contact_Corrections}, Fig. \ref{BBast_OPE_Corrections}, and Fig. \ref{BBast_TPE}, respectively.

According to the effective Lagrangians given in Eq. (\ref{4H}) and Eq. (\ref{Hphi}), we can easily get the amplitudes of the elastic scattering process $\bar{B}(p_1)\bar{B}^\ast(p_2)\to\bar{B}(p_3)\bar{B}^\ast(p_4)$ from the diagrams $H_{2.1}$ and $X_{2.1}$ in Fig. \ref{BBast_ZeroOrder},
\begin{eqnarray}\label{BBast_XI1}
\mathcal{Y}^{X_{2.1}}_{I=1}&=&8(-D_a+D_b-E_a+E_b)(\epsilon_2\cdot\epsilon_4^\ast),\\
\mathcal{Y}^{H_{2.1}}_{I=1}&=&-\frac{g^2}{f^2}\frac{(q\cdot\epsilon_2)(q\cdot\epsilon_4^\ast)}{q^2-m_\pi^2},
\end{eqnarray}
where the subscript $I$ denotes the isospin of the channel, $q$ is the transferred momentum carried by pion, $\epsilon_2$ and $\epsilon_4^\ast$ designate the polarization vectors of initial $\bar{B}^\ast$ and final $\bar{B}^\ast$, respectively. In this section we only give the amplitudes of $I=1$ channel, and the amplitudes of $I=0$ channel are shown in Appendix \ref{Amp_BBastI0}.

For the amplitudes of the one-loop corrections to FBCI of $\bar{B}\bar{B}^\ast$ system in Fig. \ref{BBast_Contact_Corrections}, similar to $\bar{B}\bar{B}$ case, we also write them out in the following,
\begin{eqnarray}
\mathcal{Y}^{g_{2.1}}_{I=1}=\mathcal{C}^{g_{2.1}}\frac{g^2}{f^2}(\epsilon_2\cdot\epsilon_4^\ast)\left\{\mathcal{J}_{22}^g\right\}_r(m_\pi,\mathcal{E},\mathcal{E}),
\end{eqnarray}
\begin{eqnarray}
\mathcal{Y}^{g_{2.2}}_{I=1}=\frac{g^2}{f^2}(\epsilon_2\cdot\epsilon_4^\ast)\bigg\{\mathcal{C}^{g_{2.2}}\mathcal{J}_{22}^g\bigg\}_r(m_\pi,\mathcal{E}-\Delta,\mathcal{E}-\Delta),
\end{eqnarray}
\begin{eqnarray}
\mathcal{Y}^{g_{2.3}}_{I=1}=\mathcal{C}^{g_{2.3}}\frac{g^2}{f^2}(\epsilon_2\cdot\epsilon_4^\ast)\left\{\mathcal{J}_{22}^g\right\}_r(m_\pi,\mathcal{E}+\Delta,\mathcal{E}-\Delta),
\end{eqnarray}
\begin{eqnarray}
\mathcal{Y}^{g_{2.4}}_{I=1}=\mathcal{C}^{g_{2.4}}\frac{g^2}{f^2}(\epsilon_2\cdot\epsilon_4^\ast)\left\{\mathcal{J}_{22}^g\right\}_r(m_\pi,\mathcal{E},\mathcal{E}-\Delta),
\end{eqnarray}
\begin{eqnarray}
\mathcal{Y}^{g_{2.7}}_{I=1}=\mathcal{C}^{g_{2.7}}\frac{g^2}{f^2}(\epsilon_2\cdot\epsilon_4^\ast)\left\{\mathcal{J}_{22}^g\right\}_r(m_\pi,\mathcal{E}+\Delta,\mathcal{E}+\Delta),
\end{eqnarray}
\begin{eqnarray}
\mathcal{Y}^{h_{2.1}}_{I=1}=\mathcal{C}^{h_{2.1}}\frac{g^2}{f^2}(\epsilon_2\cdot\epsilon_4^\ast)\left\{\mathcal{J}_{22}^h\right\}_r(m_\pi,\mathcal{E}+\Delta,\mathcal{E}-\Delta),
\end{eqnarray}
\begin{eqnarray}
\mathcal{Y}^{h_{2.2}}_{I=1}=0,
\end{eqnarray}
\begin{eqnarray}
\mathcal{Y}^{z_{2.1}}_{I=1}=\mathcal{C}^{z_{2.1}}\frac{g^2}{f^2}(\epsilon_2\cdot\epsilon_4^\ast)\left\{\frac{\partial}{\partial x}\mathcal{J}_{22}^a\right\}_r(m_\pi,x)\Big|_{x\to\mathcal{E}},
\end{eqnarray}
\begin{eqnarray}
\mathcal{Y}^{z_{2.2}}_{I=1}=\mathcal{C}^{z_{2.2}}\frac{g^2}{f^2}(\epsilon_2\cdot\epsilon_4^\ast)\left\{\frac{\partial}{\partial x}\mathcal{J}_{22}^a\right\}_r(m_\pi,x)\Big|_{x\to\mathcal{E}+\Delta},
\end{eqnarray}
\begin{eqnarray}
\mathcal{Y}^{z_{2.3}}_{I=1}=\mathcal{C}^{z_{2.3}}\frac{g^2}{f^2}(\epsilon_2\cdot\epsilon_4^\ast)\left\{(D-1)\frac{\partial}{\partial x}\mathcal{J}_{22}^a\right\}_r(m_\pi,x)\Big|_{x\to\mathcal{E}-\Delta},
\end{eqnarray}
where the values of $\mathcal{C}^x$ are
\begin{eqnarray}
\mathcal{C}^{g_{2.1}}&=&4(3D_a-D_b-E_a-5E_b),\nonumber\\
\mathcal{C}^{g_{2.2}}&=&2\Big[D(3D_a-D_b-E_a-5E_b)-2(D_a-D_b-3E_a\nonumber\\
&&-5E_b)\Big],\nonumber\\
\mathcal{C}^{g_{2.3}}&=&-8(D_b+E_b),\quad\mathcal{C}^{g_{2.4}}=8(D_b-3E_b),\nonumber\\
\mathcal{C}^{g_{2.7}}&=&8(D_a+E_a),\quad\mathcal{C}^{h_{2.1}}=2(D_a-D_b+E_a-E_b),\nonumber\\
\mathcal{C}^{z_{2.1}}&=&12(D_a-D_b+E_a-E_b),~\mathcal{C}^{z_{2.2}}=\mathcal{C}^{z_{2.3}}=\frac{1}{2}\mathcal{C}^{z_{2.1}},\nonumber\\
\end{eqnarray}
and the amplitudes of diagrams $g_{2.5}$, $g_{2.6}$, $h_{2.3}$, and $h_{2.4}$ can be obtained by the following relations,
\begin{eqnarray}
\mathcal{Y}^{g_{2.5}}_{I=1}&=&\mathcal{Y}^{g_{2.3}}_{I=1},\quad\quad\quad\mathcal{Y}^{g_{2.6}}_{I=1}=\mathcal{Y}^{g_{2.4}}_{I=1},\nonumber\\
\mathcal{Y}^{h_{2.3}}_{I=1}&=&\mathcal{Y}^{h_{2.1}}_{I=1},\quad\quad\quad\mathcal{Y}^{h_{2.4}}_{I=1}=\mathcal{Y}^{h_{2.2}}_{I=1}.
\end{eqnarray}

The amplitudes of one-loop corrections to OPE diagram (Fig. \ref{BBast_OPE_Corrections}) are illustrated as follows,
\begin{eqnarray}
\mathcal{Y}^{p_{2.1}}_{I=1}=-\frac{g^2}{f^2}\frac{(q\cdot\epsilon_2)(q\cdot\epsilon_4^\ast)}{q^2-m_\pi^2}\Sigma(m_\pi),
\end{eqnarray}
\begin{eqnarray}
\mathcal{Y}^{c_{2.1}}_{I=1}=\frac{g^2}{3f^4}\frac{(q\cdot\epsilon_2)(q\cdot\epsilon_4^\ast)}{q^2-m_\pi^2}\left\{\mathcal{J}_0^c\right\}_r(m_\pi),
\end{eqnarray}
\begin{eqnarray}
\mathcal{Y}^{c_{2.3}}_{I=1}=\frac{-g^4}{4f^4}\frac{(q\cdot\epsilon_2)(q\cdot\epsilon_4^\ast)}{q^2-m_\pi^2}\left\{\mathcal{J}_{22}^g\right\}_r(m_\pi,\mathcal{E}+\Delta,\mathcal{E}-\Delta),
\end{eqnarray}
\begin{eqnarray}
\mathcal{Y}^{c_{2.4}}_{I=1}=\frac{g^4}{2f^4}\frac{(q\cdot\epsilon_2)(q\cdot\epsilon_4^\ast)}{q^2-m_\pi^2}\left\{\mathcal{J}_{22}^g\right\}_r(m_\pi,\mathcal{E},\mathcal{E}-\Delta),
\end{eqnarray}
\begin{eqnarray}
\mathcal{Y}^{c_{2.7}}_{I=1}=\frac{3g^4}{4f^4}\frac{(q\cdot\epsilon_2)(q\cdot\epsilon_4^\ast)}{q^2-m_\pi^2}\left\{(D-1)\frac{\partial}{\partial x}\mathcal{J}_{22}^a\right\}_r(m_\pi,x)\Big|_{x\to\mathcal{E}-\Delta},\nonumber\\
\end{eqnarray}
\begin{eqnarray}
\mathcal{Y}^{c_{2.8}}_{I=1}=\frac{3g^4}{4f^4}\frac{(q\cdot\epsilon_2)(q\cdot\epsilon_4^\ast)}{q^2-m_\pi^2}\left\{\frac{\partial}{\partial x}\mathcal{J}_{22}^a\right\}_r(m_\pi,x)\Big|_{x\to\mathcal{E}+\Delta},
\end{eqnarray}
\begin{eqnarray}
\mathcal{Y}^{c_{2.9}}_{I=1}=\frac{3g^4}{2f^4}\frac{(q\cdot\epsilon_2)(q\cdot\epsilon_4^\ast)}{q^2-m_\pi^2}\left\{\frac{\partial}{\partial x}\mathcal{J}_{22}^a\right\}_r(m_\pi,x)\Big|_{x\to\mathcal{E}},
\end{eqnarray}
\begin{eqnarray}
\mathcal{Y}^{c_{2.10}}_{I=1}=\mathcal{Y}^{c_{2.11}}_{I=1}=0,
\end{eqnarray}
where $\Sigma(m_\pi)=\left[m_\pi^2/(24\pi^2 f^2)\right]\ln(m_\pi^2/\lambda^2)$ \cite{Scherer:2002tk}, the amplitudes of diagrams $c_{2.2}$, $c_{2.5}$, and $c_{2.6}$ can be obtained by the relations
\begin{eqnarray}
\mathcal{Y}^{c_{2.2}}_{I=1}=\mathcal{Y}^{c_{2.1}}_{I=1},\quad\quad\mathcal{Y}^{c_{2.5}}_{I=1}=\mathcal{Y}^{c_{2.3}}_{I=1},\quad\quad\mathcal{Y}^{c_{2.6}}_{I=1}=\mathcal{Y}^{c_{2.4}}_{I=1}.
\end{eqnarray}

Here we list the amplitudes of the diagrams $F_{2.1}$$\sim$$B_{2.3}$ in Fig. \ref{BBast_TPE}. The amplitudes of the diagrams $R_{2.1}$$\sim$$R_{2.3}$ can be related to the ones of $B_{2.1}$$\sim$$B_{2.3}$, respectively. For conciseness, the terms that involve $q_0$ have been omitted since $q_0$ is set to be zero in our calculations. One can find the unabridged forms in Ref. \cite{Xu:2017tsr}.
\begin{widetext}\label{Amp_BBast_TPE}
\begin{eqnarray}
\mathcal{Y}^{F_{2.1}}_{I=1}&=&\frac{1}{f^4}(\epsilon_2\cdot\epsilon_4^\ast)\left\{\mathcal{J}_{22}^F\right\}_r(m_\pi,q),
\end{eqnarray}
\begin{eqnarray}
\mathcal{Y}^{T_{2.1}}_{I=1}&=&\frac{g^2}{f^4}\bigg\{(\epsilon_2\cdot\epsilon_4^\ast)\mathcal{J}_{34}^T+(q\cdot\epsilon_2)(q\cdot\epsilon_4^\ast)\left(\mathcal{J}_{24}^T+\mathcal{J}_{33}^T\right)\bigg\}_r\left(m_\pi,\mathcal{E}+\Delta,q\right),
\end{eqnarray}
\begin{eqnarray}
\mathcal{Y}^{T_{2.2}}_{I=1}&=&\frac{g^2}{f^4}\Bigg\{(\epsilon_2\cdot\epsilon_4^\ast)\bigg[2\mathcal{J}_{34}^T-\vec{q}^2\left(\mathcal{J}_{24}^T+\mathcal{J}_{33}^T\right)\bigg]-(q\cdot\epsilon_2)(q\cdot\epsilon_4^\ast)\left(\mathcal{J}_{24}^T+\mathcal{J}_{33}^T\right)\Bigg\}_r(m_\pi,\mathcal{E},q),
\end{eqnarray}
\begin{eqnarray}
\mathcal{Y}^{T_{2.3}}_{I=1}&=&\frac{g^2}{f^4}\Bigg\{(\epsilon_2\cdot\epsilon_4^\ast)\bigg[(D-1)\mathcal{J}_{34}^T-\vec{q}^2\left(\mathcal{J}_{24}^T+\mathcal{J}_{33}^T\right)\bigg]\Bigg\}_r(m_\pi,\mathcal{E}-\Delta,q),
\end{eqnarray}
\begin{eqnarray}
\mathcal{Y}^{B_{2.1}}_{I=1}&=&\frac{g^4}{4f^4}\Bigg\{(\epsilon_2\cdot\epsilon_4^\ast)\bigg[-\vec{q}^2\left[(D+5)\left(\mathcal{J}_{31}^B+\mathcal{J}_{42}^B\right)-\vec{q}^2\left(\mathcal{J}_{22}^B+2\mathcal{J}_{32}^B+\mathcal{J}_{43}^B\right)+\mathcal{J}_{21}^B\right]+2(D+1)\mathcal{J}_{41}^B\bigg]\nonumber\\
&&-(q\cdot\epsilon_2)(q\cdot\epsilon_4^\ast)\bigg[(D+3)\left(\mathcal{J}_{31}^B+\mathcal{J}_{42}^B\right)-\vec{q}^2\left(\mathcal{J}_{22}^B+2\mathcal{J}_{32}^B+\mathcal{J}_{43}^B\right)+\mathcal{J}_{21}^B\bigg]\Bigg\}_r(m_\pi,\mathcal{E}-\Delta,\mathcal{E},q),
\end{eqnarray}
\begin{eqnarray}
\mathcal{Y}^{B_{2.2}}_{I=1}&=&\frac{g^4}{4f^4}\Bigg\{(\epsilon_2\cdot\epsilon_4^\ast)\bigg[(D+1)\mathcal{J}_{41}^B-\vec{q}^2\left(\mathcal{J}_{31}^B+\mathcal{J}_{42}^B\right)\bigg]+(q\cdot\epsilon_2)(q\cdot\epsilon_4^\ast)\bigg[(D+3)\left(\mathcal{J}_{31}^B+\mathcal{J}_{42}^B\right)+\mathcal{J}_{21}^B\nonumber\\
&&-\vec{q}^2\left(\mathcal{J}_{22}^B+2\mathcal{J}_{32}^B+\mathcal{J}_{43}^B\right)\bigg]\Bigg\}_r(m_\pi,\mathcal{E}+\Delta,\mathcal{E}-\Delta,q),
\end{eqnarray}
\begin{eqnarray}
\mathcal{Y}^{B_{2.3}}_{I=1}&=&\frac{g^4}{4f^4}\bigg\{\Big[(\epsilon_2\cdot\epsilon_4^\ast)\vec{q}^2+(q\cdot\epsilon_2)(q\cdot\epsilon_4^\ast)\Big]\mathcal{J}_{21}^B\bigg\}_r(m_\pi,\mathcal{E}-\Delta,\mathcal{E},q).
\end{eqnarray}
\end{widetext}
When $q_0=0$, the amplitude structures of diagrams $R_{2.1}$$\sim$$R_{2.3}$ resemble the ones of $B_{2.1}$$\sim$$B_{2.3}$, correspondingly. Hence we can get the amplitudes of diagrams $R_{2.1}$$\sim$$R_{2.3}$ by multiplying the proper coefficients to the amplitudes of diagrams $B_{2.1}$$\sim$$B_{2.3}$, and substituting the loop integral $\mathcal{J}_{x}^B$ with $\mathcal{J}_{x}^R$ accordingly. One can readily verify the following correlations,
\begin{eqnarray}\label{JBtoJR1}
\mathcal{Y}^{R_{2.1}}_{I=1}&=&5\mathcal{Y}^{B_{2.1}}_{I=1}\Big|_{\mathcal{J}_{x}^B\to\mathcal{J}_{x}^R},\quad\mathcal{Y}^{R_{2.2}}_{I=1}=5\mathcal{Y}^{B_{2.2}}_{I=1}\Big|_{\mathcal{J}_{x}^B\to\mathcal{J}_{x}^R},\nonumber\\
\mathcal{Y}^{R_{2.3}}_{I=1}&=&-5\mathcal{Y}^{B_{2.3}}_{I=1}\Big|_{\mathcal{J}_{x}^B\to\mathcal{J}_{x}^R}.
\end{eqnarray}

In order to get the numerical results, some procedures are the same as we calculate the $\bar{B}\bar{B}$ system, i.e., Eq. (\ref{V_A}) should be used, and the 2PR part in diagrams $h_{2.1}$, $h_{2.3}$, and $B_{2.2}$ must be subtracted with the aid of Weinberg's formalism. However, some additional steps still needed, such as the pion decay constant $f$ its bare value $0.086$ GeV \cite{Xu:2017tsr} should be adopted. Since we only consider the $S$-wave interaction, the terms $(\epsilon_2\cdot\epsilon_4^\ast)$ and $(q\cdot\epsilon_2)(q\cdot\epsilon_4^\ast)$ appeared in above equations should be replaced by \cite{Sun:2012zzd,Li:2012ss,Xu:2017tsr}
\begin{eqnarray}\label{PolVecProd1}
(\epsilon_2\cdot\epsilon_4^\ast)\rightarrowtail -1,\quad\quad\quad(q\cdot\epsilon_2)(q\cdot\epsilon_4^\ast)\rightarrowtail\frac{1}{3}\vec{q}^2.
\end{eqnarray}

We first list the results of one-loop corrections to FBCI in $I=1$ and $I=0$ channel, respectively.
\begin{eqnarray}
\mathcal{Y}^{(2)\text{FBCI}}_{I=1}&=&4\left(0.086D_b-0.086E_a-0.21E_b\right)(\epsilon_2\cdot\epsilon_4^\ast),\nonumber\\
\mathcal{Y}^{(2)\text{FBCI}}_{I=0}&=&4\left(-0.21D_b+0.63E_a+0.79E_b\right)(\epsilon_2\cdot\epsilon_4^\ast).
\end{eqnarray}
By comparing with Eq. (\ref{BBast_XI1}) and Eq. (\ref{BBast_XHI0}), the results again show the convergence of chiral correction is very good.

\begin{figure*}
\begin{minipage}[t]{0.45\linewidth}
\centering
\includegraphics[width=\columnwidth]{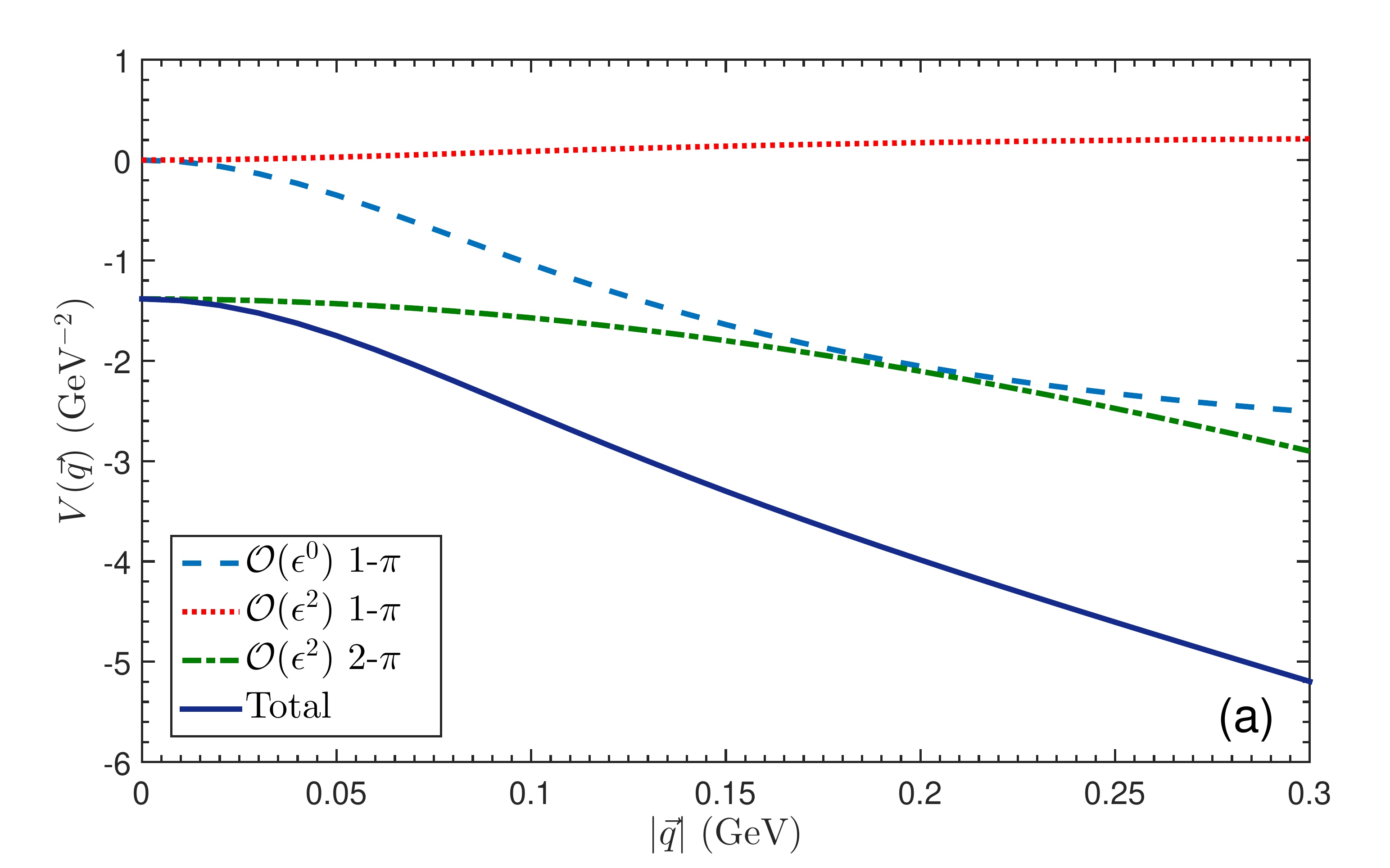}
\end{minipage}%
\begin{minipage}[t]{0.45\linewidth}
\centering
\includegraphics[width=\columnwidth]{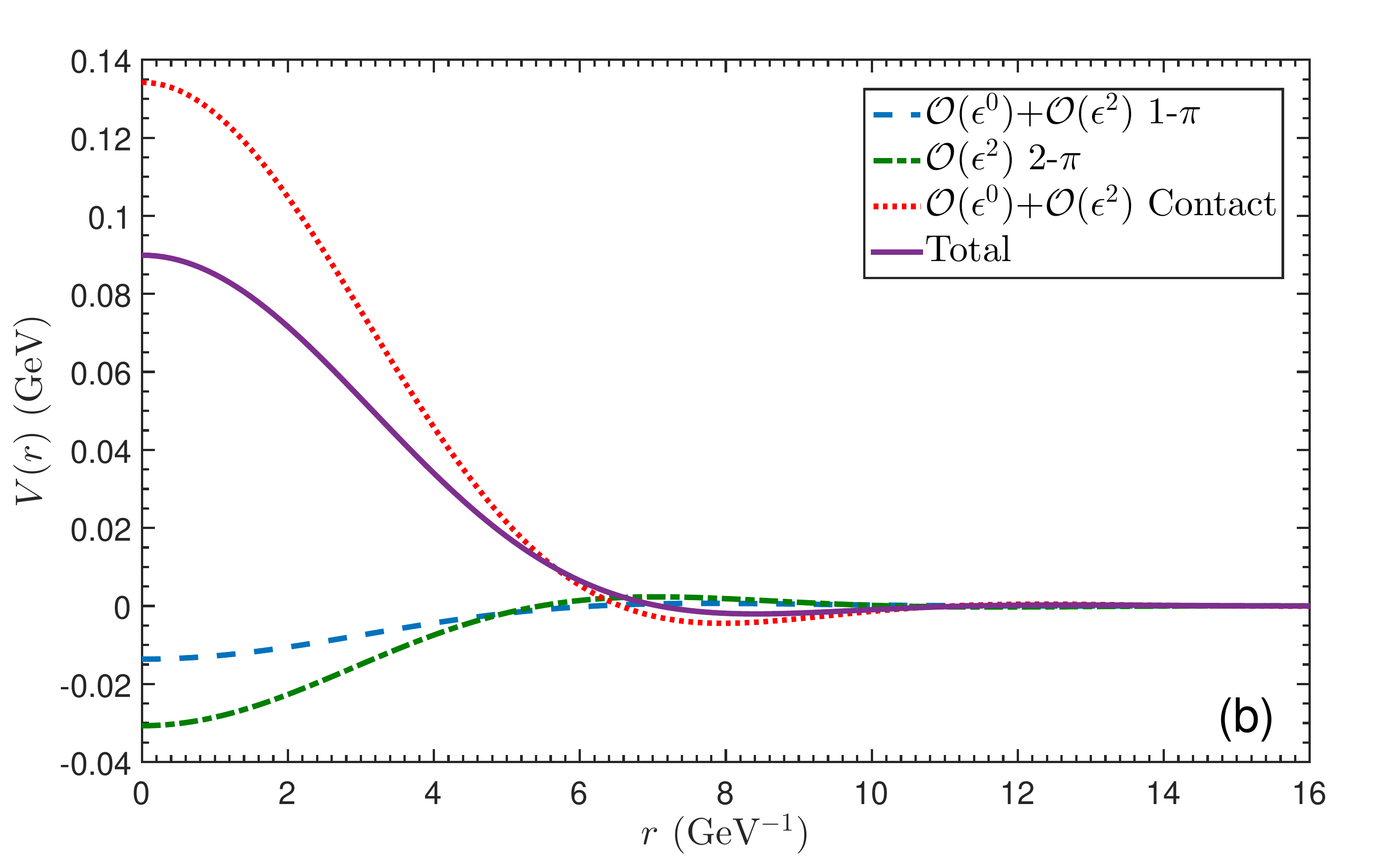}
\end{minipage}
\caption{Figure (a) represents the OPE and TPE potentials of $1(1^+)$ $\bar{B}\bar{B}^\ast$ system in momentum space, where the dashed line denotes the OPE potential at $\mathcal{O}(\epsilon^0)$, the dotted line is the OPE potential at $\mathcal{O}(\epsilon^2)$, the dot-dashed line describes the TPE potential, and the sum of these three parts are depicted by the solid line. Figure (b) shows the potentials in coordinate space, where the dashed line stands for the sum of OPE potentials at $\mathcal{O}(\epsilon^0)$ and $\mathcal{O}(\epsilon^2)$, and other basic notations are the same as in Fig. \ref{BB_Momentum_Euclid}. \label{BBastI1_Momentum_Euclid}}
\end{figure*}

\begin{figure*}
\begin{minipage}[t]{0.45\linewidth}
\centering
\includegraphics[width=\columnwidth]{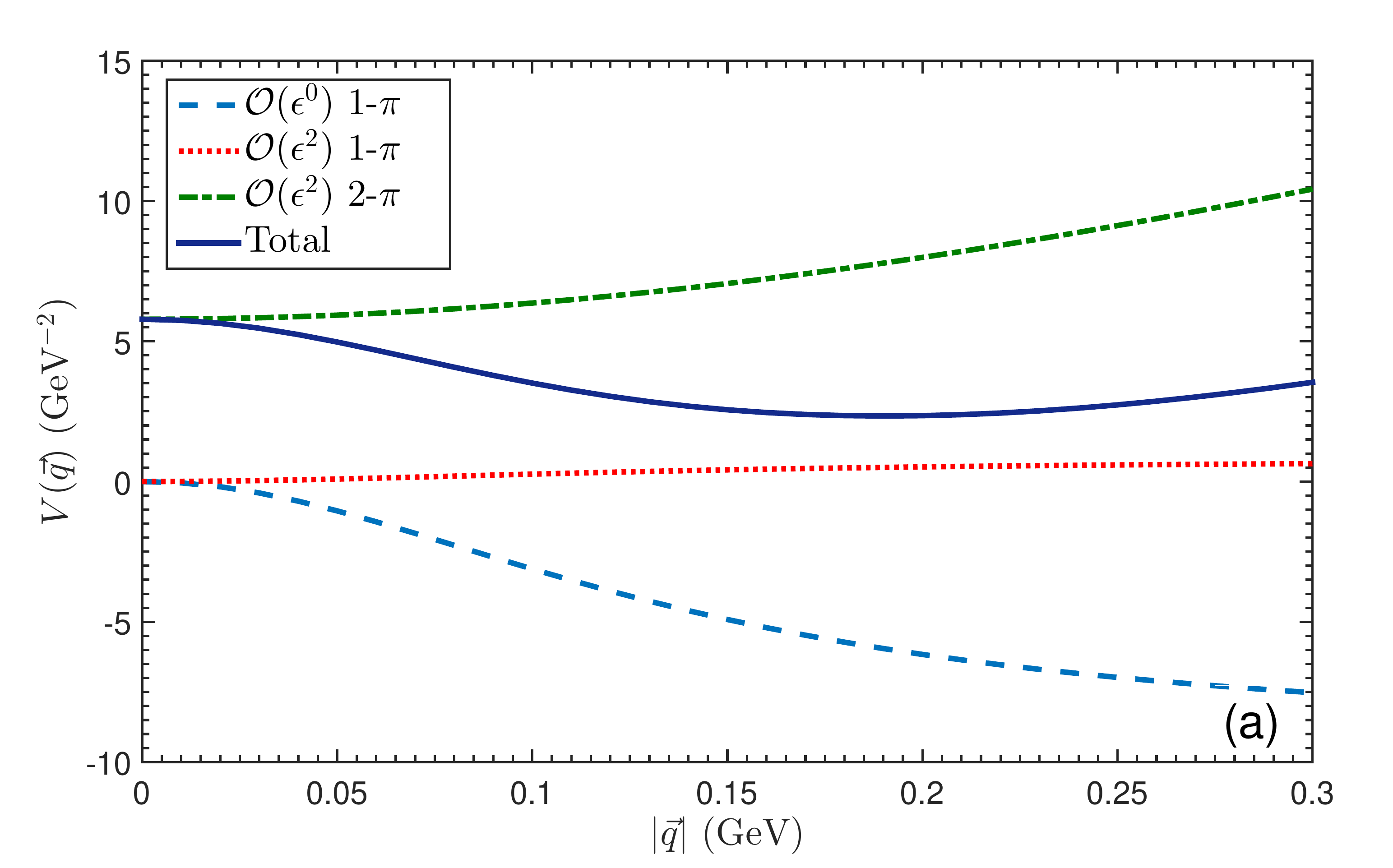}
\end{minipage}%
\begin{minipage}[t]{0.45\linewidth}
\centering
\includegraphics[width=\columnwidth]{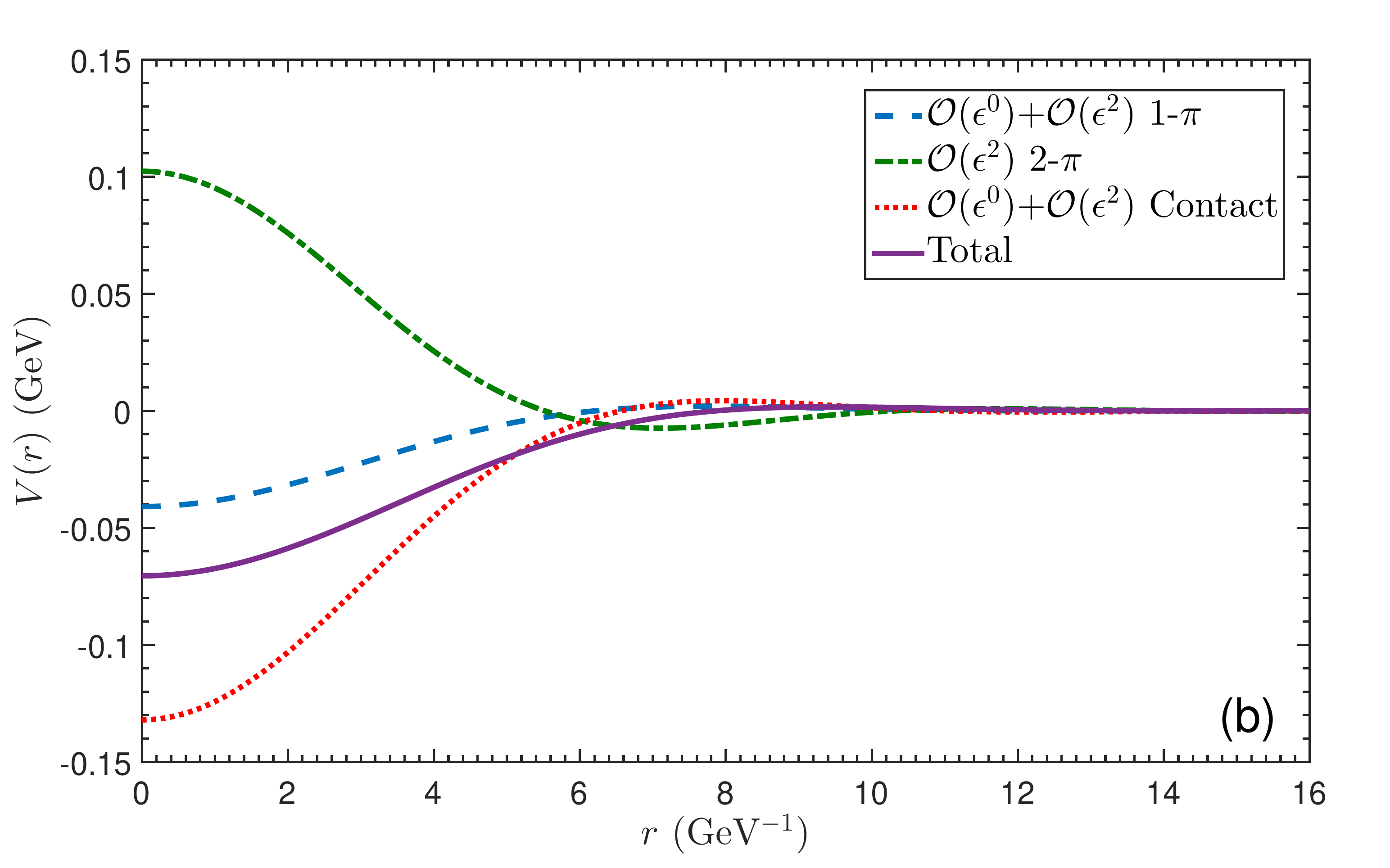}
\end{minipage}
\caption{The OPE and TPE potentials of $0(1^+)$ $\bar{B}\bar{B}^\ast$ system in momentum space (left panel) and the effective potentials in coordinate space (right panel). Notations same as in Fig. \ref{BBastI1_Momentum_Euclid}. \label{BBastI0_Momentum_Euclid}}
\end{figure*}

The effective potentials of $1(1^+)$ and $0(1^+)$ $\bar{B}\bar{B}^\ast$ systems in momentum space and coordinate space are shown in Fig. \ref{BBastI1_Momentum_Euclid} and Fig. \ref{BBastI0_Momentum_Euclid}, respectively.

From Fig. \ref{BBastI1_Momentum_Euclid}(a) and Fig. \ref{BBastI0_Momentum_Euclid}(a), we can see that the $\mathcal{O}(\epsilon^0)$ OPE potential is attractive for both $I=1$ and $I=0$ channels, and  the $\mathcal{O}(\epsilon^0)$ OPE potential in $I=0$ channel is more attractive than that of $I=1$ channel. The $\mathcal{O}(\epsilon^2)$ potential generated by one-loop corrections to OPE diagram is very small compared with the $\mathcal{O}(\epsilon^0)$ OPE potential, which demonstrates that the convergence of chiral corrections to OPE diagram is also good. The behavior of TPE potential is totally different between these two channels. In $I=1$ channel, TPE potential is attractive and it intends to be more attractive when $|\vec{q}|$ becomes larger. In $I=0$ channel, TPE potential is repulsive, and would be more repulsive when $|\vec{q}|$ becomes larger.

Because the OPE and TPE potentials are all attractive in $I=1$ channel, the total potential of $I=1$ channel is attractive without the doubt. Although OPE provides an attractive potential in $I=0$ channel, the TPE potential is repulsive and the absolute value of TPE potential is larger than OPE potential, which gives rise to a repulsive total potential in $I=0$ channel.

Although the sum of OPE and TPE potentials of $I=1$ channel in momentum space is attractive, we still can not conclude that there may exist a bound state in $I=1$ channel, because the contribution from the contact term is independent on $|\vec{q}|$, and not included in Fig. \ref{BBastI1_Momentum_Euclid}(a) and Fig. \ref{BBastI0_Momentum_Euclid}(a). The results in Fig. \ref{BBastI1_Momentum_Euclid}(b) and Fig. \ref{BBastI0_Momentum_Euclid}(b) are very interesting, because the inclusion of the contributions from FBCI totally reversed the sign of the results in momentum space. Although both the OPE and TPE potentials are attractive in $I=1$ channel, the FBCI potential is largely repulsive, and thus the total potential of $I=1$ channel in coordinate space is repulsive. For the $I=0$ channel, we notice that much of potential generated from FBCI is canceled by the one of TPE approximately, and thus the residual part of FBCI potential plus the contribution from OPE play very important role, since both of them are attractive and thus the total potential is also attractive. Moreover, we find a bound state in the $0(1^+)$ $\bar{B}\bar{B}^\ast$ system by solving the Schr\"odinger equation numerically. We obtain the binding energy is $\Delta E_{\bar{B}\bar{B}^\ast}\simeq-12.6^{+9.2}_{-12.9}$ MeV, and the corresponding root-mean-square radius is $1.0^{+0.5}_{-0.2}$ fm. Analogously, the OBE model \cite{Li:2012ss}, quark model potential \cite{Barnes:1999hs}, and Lattice QCD \cite{Bicudo:2016ooe} all demonstrated that $0(1^+)$ $\bar{B}\bar{B}^\ast$ system can form the molecular state. Our results are qualitatively consistent.

There are some other remarks on Fig. \ref{BBastI0_Momentum_Euclid}(b). One can notice that there exists a subtle cancellation between TPE and FBCI potentials in Fig. \ref{BBastI0_Momentum_Euclid}(b), and this phenomenon makes OPE alone be enough to bind $\bar{B}$ and $\bar{B}^\ast$. Although TPE is model independent, the LECs involved in the FBCI potential are determined with resonance saturation model, thus the cancellation in Fig. \ref{BBastI0_Momentum_Euclid}(b) is model dependent more or less. 
However, this result should be reasonable. $\bar{B}\bar{B}^\ast$ interactions could be related to those of $B\bar{B}^\ast$ if we ignore the annihilation effect in $B\bar{B}^\ast$ channel at very short distance \cite{Hippchen:1991rr}. The authors in Refs. \cite{Valderrama:2012jv,Nieves:2012tt,Guo:2013sya} investigated $D\bar{D}^\ast$ and $B\bar{B}^\ast$ systems with $\chi$EFT extensively. They noticed that $\mathcal{O}(\epsilon^0)$ contact-range interactions are crucial, and the inclusion of OPE only gives rise to minor modifications of the numerical results. As we have seen in Fig. \ref{BBastI0_Momentum_Euclid}(b), this is indeed the case, i.e., the attractive potential provided by the FBCI is stronger than that of OPE. The consistence tells that the values of the LECs we adopted in this work are reasonable, and the predictions given with moderate cutoff value $\Lambda=0.7$ GeV are reliable. In the following, we talk about the $\Lambda$ dependence of the total potentials.

\begin{figure}[hptb]
	\scalebox{1.0}{\includegraphics[width=\columnwidth]{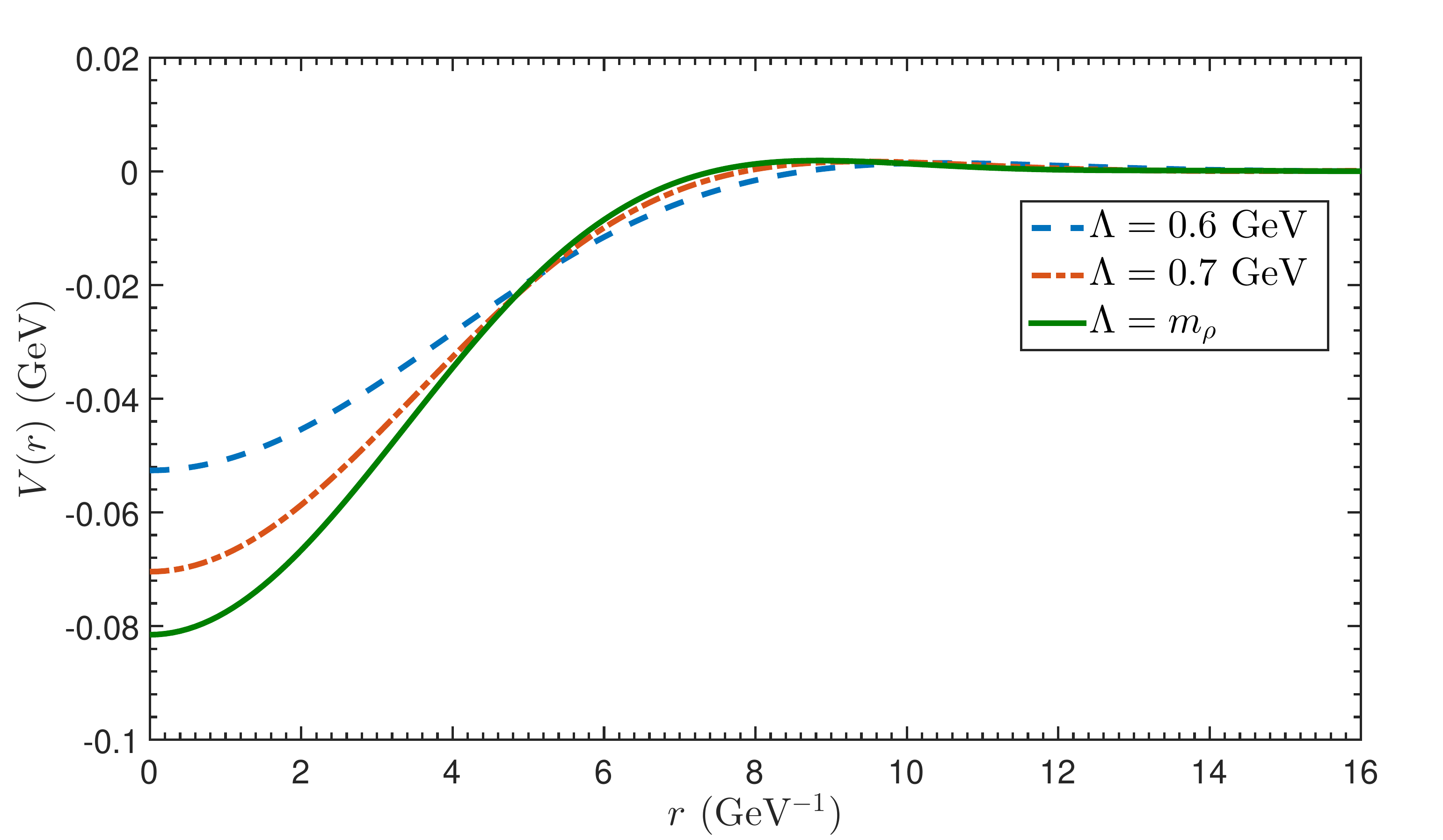}}
	\caption{The dependence of the total potentials for the $0(1^+)$ $\bar{B}\bar{B}^\ast$ system on cutoff parameter $\Lambda$, where three different results with $\Lambda=$0.6 GeV, 0.7 GeV, and 0.77 GeV are illustrated.\label{DepOnLam}}
\end{figure}

We also need to stress that in Eq. \eqref{Four_Tansf}, we introduce a Gauss regulator $\mathcal{F}(\mathbf{q})$ to regularize the potential $\mathcal{V}(\mathbf{q})$, and the function $\mathcal{F}(\mathbf{q})$ contains a cutoff parameter $\Lambda$. In theory, the binding energy should be independent of the regularization schemes adopted in Eq. \eqref{Four_Tansf}, the scale dependence in Eq. \eqref{Four_Tansf} can be compensated by that of LECs. However, in practical calculations, it is difficult to remove the influences of $\Lambda$ on observable. Therefore, we investigate the dependence of the total potentials for the $0(1^+)$ $\bar{B}\bar{B}^\ast$ system on cutoff parameter $\Lambda$, and the results are illustrated in Fig. \ref{DepOnLam}. We can read that the binding in the short range would be deeper when the value of $\Lambda$ is increased, but the results are not very sensitive to $\Lambda$. By solving the Schr\"odinger function, we find when $\Lambda=600$ MeV, 700 MeV, and 770 MeV ($m_\rho$), the corresponding binding energies are $\Delta E_{\bar{B}\bar{B}^\ast}\simeq-8.1$ MeV, $-12.6$ MeV, and $-15.6$ MeV, respectively. The result indicates that our predictions obtained at $\Lambda=700$ MeV are stable, i.e., there exists a bound state in $0(1^+)$ $\bar{B}\bar{B}^\ast$ system when the cutoff is near by $m_\rho$.

Bound $0(1^+)$ $\bar{B}\bar{B}^\ast$ means this state is stable against strong interactions, it can not decay into its components due to the constraints of phase space. However, we notice that the mass of this state lies above the threshold of electromagnetic decay mode $\bar{B}\bar{B}\gamma$, thus it can be constructed via electromagnetic interactions at experiments.

\subsection{$\bar{B}^\ast\bar{B}^\ast$ system}
\begin{figure}[hptb]
	\scalebox{0.6}{\includegraphics[width=\columnwidth]{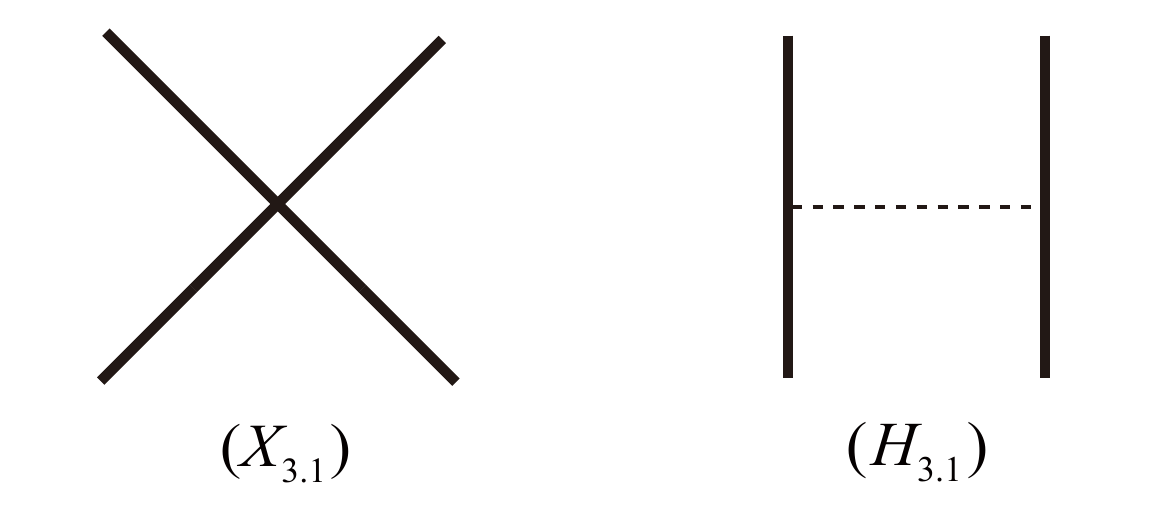}}
	\caption{The diagrams for $\bar{B}^\ast\bar{B}^\ast$ system at $\mathcal{O}(\epsilon^0)$. Notations same as in Fig. \ref{BB_Contact_Corrections}.\label{BastBast_ZeroOrder}}
\end{figure}

\begin{figure*}[hptb]
	\scalebox{0.9}{\includegraphics[width=18cm]{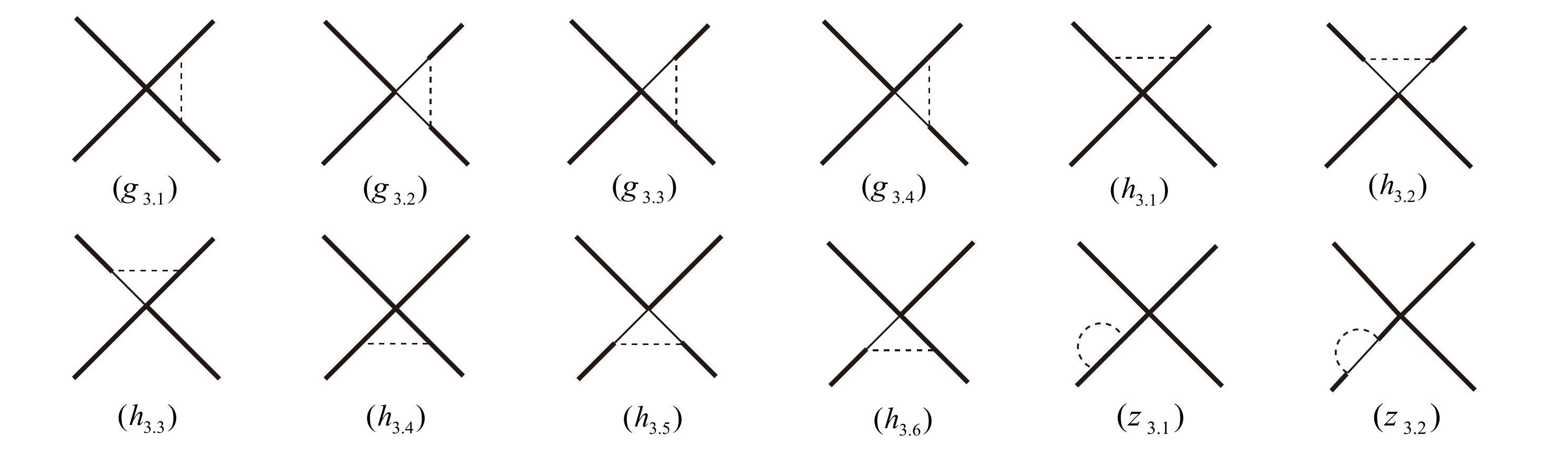}}
	\caption{One-loop corrections to the FBCI of $\bar{B}^\ast\bar{B}^\ast$ system at $\mathcal{O}(\epsilon^2)$. Notations same as in Fig. \ref{BB_Contact_Corrections}.\label{BastBast_Contact_Corrections}}
\end{figure*}

\begin{figure*}[hptb]
	\scalebox{1.0}{\includegraphics[width=18cm]{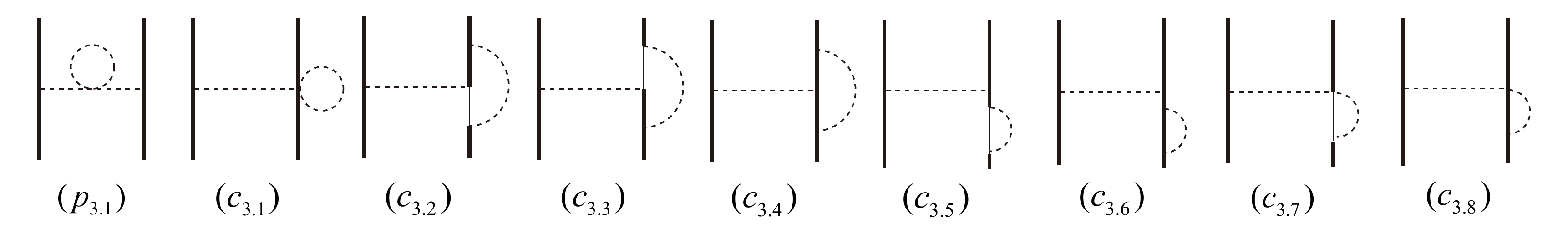}}
	\caption{One-loop corrections to the OPE diagram of $\bar{B}^\ast\bar{B}^\ast$ system at $\mathcal{O}(\epsilon^2)$. Notations same as in Fig. \ref{BB_Contact_Corrections}.\label{BastBast_OPE_Corrections}}
\end{figure*}

\begin{figure*}[hptb]
	\scalebox{1.0}{\includegraphics[width=18cm]{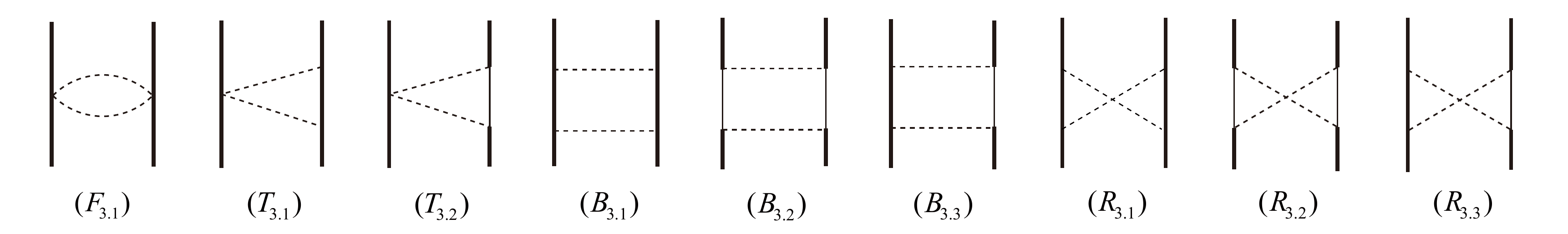}}
	\caption{Two-pion exchange diagrams of $\bar{B}^\ast\bar{B}^\ast$ system at $\mathcal{O}(\epsilon^2)$. Notations same as in Fig. \ref{BB_Contact_Corrections}.\label{BastBast_TPE}}
\end{figure*}

Like $\bar{B}\bar{B}$ system, the quantum numbers of $\bar{B}^\ast\bar{B}^\ast$ system must be constrainted by the selection rules given in Tab. \ref{allowedBB}. With $S$ wave, the physically allowed $\bar{B}^\ast\bar{B}^\ast$ states are $1(0^+)$, $1(2^+)$ and $0(1^+)$, respectively. The diagrams at the $\mathcal{O}(\epsilon^0)$ are displayed in Fig. \ref{BastBast_ZeroOrder}, and the diagrams at $\mathcal{O}(\epsilon^2)$ are illustrated in Figs. \ref{BastBast_Contact_Corrections}-\ref{BastBast_TPE}.

For the scattering process $\bar{B}^\ast(p_1)\bar{B}^\ast(p_2)\to\bar{B}^\ast(p_3)\bar{B}^\ast(p_4)$, we first list the amplitudes of the diagrams in Fig. \ref{BastBast_ZeroOrder} (we only give the amplitudes of $I=1$ channel in this section, and the amplitudes of $I=0$ channel can be found in Appendix \ref{Amp_BastBastI0}),

\begin{eqnarray}\label{BastBast_XH1I}
\mathcal{Y}^{X_{3.1}}_{I=1}&=&8\Big[(D_a-D_b+E_a-E_b)\left(\varepsilon_1+\varepsilon_2\right)\nonumber\\
&&+2(D_b+E_b)\varepsilon_3\Big],\\
\mathcal{Y}^{H_{3.1}}_{I=1}&=&-\frac{g^2}{f^2}\frac{\mathcal{G}(q,\epsilon_1,\epsilon_2,\epsilon_3^\ast,\epsilon_4^\ast)}{q^2-m_\pi^2},
\end{eqnarray}
where we define
\begin{eqnarray}
\varepsilon_1&=&(\epsilon_1\cdot\epsilon_3^\ast)(\epsilon_2\cdot\epsilon_4^\ast),\quad\varepsilon_2=(\epsilon_1\cdot\epsilon_4^\ast)(\epsilon_2\cdot\epsilon_3^\ast),\nonumber\\
\varepsilon_3&=&(\epsilon_1\cdot\epsilon_2)(\epsilon_3^\ast\cdot\epsilon_4^\ast),\quad\varepsilon_2^a=(q\cdot\epsilon_3^\ast)(q\cdot\epsilon_2)(\epsilon_1\cdot\epsilon_4^\ast),\nonumber\\
\varepsilon_2^b&=&(q\cdot\epsilon_1)(q\cdot\epsilon_4^\ast)(\epsilon_2\cdot\epsilon_3^\ast),\varepsilon_3^a=(q\cdot\epsilon_3^\ast)(q\cdot\epsilon_4^\ast)(\epsilon_1\cdot\epsilon_2),\nonumber\\
\varepsilon_3^b&=&(q\cdot\epsilon_1)(q\cdot\epsilon_2)(\epsilon_3^\ast\cdot\epsilon_4^\ast),
\end{eqnarray}
and
\begin{eqnarray}
\mathcal{G}(q,\epsilon_1,\epsilon_2,\epsilon_3^\ast,\epsilon_4^\ast)=\vec{q}^2\left(\varepsilon_2-\varepsilon_3\right)+\left(\varepsilon_2^a-\varepsilon_3^a\right)+\left(\varepsilon_2^b-\varepsilon_3^b\right).
\end{eqnarray}

The amplitudes of the diagrams in Fig. \ref{BastBast_Contact_Corrections} are given as
\begin{eqnarray}\label{BastBast_Amp_FBCII1}
\mathcal{Y}^{g_{3.1}}_{I=1}=\frac{g^2}{f^2}\left[\mathcal{C}^{g_{3.1}}_1\left(\varepsilon_1+\varepsilon_2\right)+\mathcal{C}^{g_{3.1}}_2\varepsilon_3\right]\left\{\mathcal{J}_{22}^g\right\}_r(m_\pi,\mathcal{E},\mathcal{E}),
\end{eqnarray}
\begin{eqnarray}
\mathcal{Y}^{g_{3.2}}_{I=1}=\frac{g^2}{f^2}\Big[\mathcal{C}^{g_{3.2}}\left(\varepsilon_1+\varepsilon_2\right)\Big]\left\{\mathcal{J}_{22}^g\right\}_r(m_\pi,\mathcal{E}+\Delta,\mathcal{E}+\Delta),
\end{eqnarray}
\begin{eqnarray}
\mathcal{Y}^{g_{3.3}}_{I=1}=\frac{g^2}{f^2}\Big[\mathcal{C}^{g_{3.3}}\left(\varepsilon_1+\varepsilon_2-2\varepsilon_3\right)\Big]\left\{\mathcal{J}_{22}^g\right\}_r(m_\pi,\mathcal{E},\mathcal{E}+\Delta),
\end{eqnarray}
\begin{eqnarray}
\mathcal{Y}^{h_{3.1}}_{I=1}=\frac{g^2}{f^2}\left[\mathcal{C}^{h_{3.1}}_1\left(\varepsilon_1+\varepsilon_2\right)+\mathcal{C}^{h_{3.1}}_2\varepsilon_3\right]\left\{\mathcal{J}_{22}^h\right\}_r(m_\pi,\mathcal{E},\mathcal{E}),
\end{eqnarray}
\begin{eqnarray}
\mathcal{Y}^{h_{3.2}}_{I=1}=\frac{g^2}{f^2}\left(\mathcal{C}^{h_{3.2}}\varepsilon_3\right)\left\{\mathcal{J}_{22}^h\right\}_r(m_\pi,\mathcal{E}+\Delta,\mathcal{E}+\Delta),
\end{eqnarray}
\begin{eqnarray}
\mathcal{Y}^{h_{3.3}}_{I=1}=0,
\end{eqnarray}
\begin{eqnarray}
\mathcal{Y}^{z_{3.1}}_{I=1}=\frac{g^2}{f^2}\left[\mathcal{C}^{z_{3.1}}_1\left(\varepsilon_1+\varepsilon_2\right)+\mathcal{C}^{z_{3.1}}_2\varepsilon_3\right]\left\{\frac{\partial}{\partial x}\mathcal{J}_{22}^a\right\}_r(m_\pi,x)\bigg|_{x\to \mathcal{E}},\nonumber\\
\end{eqnarray}
\begin{eqnarray}
\mathcal{Y}^{z_{3.2}}_{I=1}&=&\frac{g^2}{f^2}\left[\mathcal{C}^{z_{3.1}}_1\left(\varepsilon_1+\varepsilon_2\right)+\mathcal{C}^{z_{3.1}}_2\varepsilon_3\right]
\left\{\frac{\partial}{\partial x}\mathcal{J}_{22}^a\right\}_r(m_\pi,x)\bigg|_{x\to \mathcal{E}+\Delta},\nonumber\\
\end{eqnarray}
where the forms of the coefficients $\mathcal{C}^x$ are
\begin{eqnarray}
\mathcal{C}^{g_{3.1}}_1&=&-4(7D_a-5D_b+3E_a-9E_b),\nonumber\\
\mathcal{C}^{g_{3.1}}_2&=&8(D_a-3D_b+5E_a+E_b),\nonumber\\
\mathcal{C}^{g_{3.2}}&=&-4(3D_a-D_b-E_a-5E_b),\quad\mathcal{C}^{g_{3.3}}=-8(D_b-3E_b),\nonumber\\
\mathcal{C}^{h_{3.1}}_1&=&-2(D_a-D_b+E_a-E_b),\nonumber\\
\mathcal{C}^{h_{3.1}}_2&=&4(D_a+D_b+E_a+E_b),\nonumber\\
\mathcal{C}^{h_{3.2}}&=&4(D_b+E_b),\quad\mathcal{C}^{z_{3.1}}_1=-24(D_a-D_b+E_a-E_b),\nonumber\\
\mathcal{C}^{z_{3.1}}_2&=&-48(D_b+E_b),\quad\mathcal{C}^{z_{3.2}}_1=\frac{1}{2}\mathcal{C}^{z_{3.1}}_1,\quad\mathcal{C}^{z_{3.2}}_2=\frac{1}{2}\mathcal{C}^{z_{3.1}}_2.
\end{eqnarray}
The amplitudes of diagrams $g_{3.4}$, $h_{3.4}$, $h_{3.5}$, and $h_{3.6}$ can be obtained by the relations
\begin{eqnarray}
\mathcal{Y}^{g_{3.4}}_{I=1}&=&\mathcal{Y}^{g_{3.3}}_{I=1},\quad\quad\quad\mathcal{Y}^{h_{3.4}}_{I=1}=\mathcal{Y}^{h_{3.1}}_{I=1},\nonumber\\ \mathcal{Y}^{h_{3.5}}_{I=1}&=&\mathcal{Y}^{h_{3.2}}_{I=1},\quad\quad\quad\mathcal{Y}^{h_{3.6}}_{I=1}=\mathcal{Y}^{h_{3.3}}_{I=1}.
\end{eqnarray}

The amplitudes of the diagrams in Fig. \ref{BastBast_OPE_Corrections} are shown as follows,
\begin{eqnarray}\label{BastBast_Amp_OPEI1}
\mathcal{Y}^{p_{3.1}}_{I=1}=-\frac{g^2}{f^2}\frac{\mathcal{G}(q,\epsilon_1,\epsilon_2,\epsilon_3^\ast,\epsilon_4^\ast)}{q^2-m_\pi^2}\Sigma(m_\pi),
\end{eqnarray}
\begin{eqnarray}
\mathcal{Y}^{c_{3.1}}_{I=1}=\frac{2g^2}{3f^4}\frac{\mathcal{G}(q,\epsilon_1,\epsilon_2,\epsilon_3^\ast,\epsilon_4^\ast)}{q^2-m_\pi^2}\left\{\mathcal{J}_0^c\right\}_r(m_\pi),
\end{eqnarray}
\begin{eqnarray}
\mathcal{Y}^{c_{3.2}}_{I=1}=\frac{g^4}{2f^4}\frac{\mathcal{G}(q,\epsilon_1,\epsilon_2,\epsilon_3^\ast,\epsilon_4^\ast)}{q^2-m_\pi^2}\left\{\mathcal{J}_{22}^g\right\}_r(m_\pi,\mathcal{E}+\Delta,\mathcal{E}),
\end{eqnarray}
\begin{eqnarray}
\mathcal{Y}^{c_{3.3}}_{I=1}=\frac{g^4}{2f^4}\frac{\mathcal{G}(q,\epsilon_1,\epsilon_2,\epsilon_3^\ast,\epsilon_4^\ast)}{q^2-m_\pi^2}\left\{\mathcal{J}_{22}^g\right\}_r(m_\pi,\mathcal{E}+\Delta,\mathcal{E}),
\end{eqnarray}
\begin{eqnarray}
\mathcal{Y}^{c_{3.4}}_{I=1}=-\frac{g^4}{2f^4}\frac{\mathcal{G}(q,\epsilon_1,\epsilon_2,\epsilon_3^\ast,\epsilon_4^\ast)}{q^2-m_\pi^2}\left\{\mathcal{J}_{22}^g\right\}_r(m_\pi,\mathcal{E},\mathcal{E}),
\end{eqnarray}
\begin{eqnarray}
\mathcal{Y}^{c_{3.5}}_{I=1}=\frac{3g^4}{8f^4}\frac{\mathcal{G}(q,\epsilon_1,\epsilon_2,\epsilon_3^\ast,\epsilon_4^\ast)}{q^2-m_\pi^2}\left\{\frac{\partial}{\partial x}\mathcal{J}_{22}^a\right\}_r(m_\pi,x)\bigg|_{x\to \mathcal{E}+\Delta},
\end{eqnarray}
\begin{eqnarray}
\mathcal{Y}^{c_{3.6}}_{I=1}=\frac{3g^4}{4f^4}\frac{\mathcal{G}(q,\epsilon_1,\epsilon_2,\epsilon_3^\ast,\epsilon_4^\ast)}{q^2-m_\pi^2}\left\{\frac{\partial}{\partial x}\mathcal{J}_{22}^a\right\}_r(m_\pi,x)\bigg|_{x\to \mathcal{E}},
\end{eqnarray}
\begin{eqnarray}
\mathcal{Y}^{c_{3.7}}_{I=1}=\mathcal{Y}^{c_{3.8}}_{I=1}=0.
\end{eqnarray}

The amplitudes of the TPE diagrams in Fig. \ref{BastBast_TPE} read,
\begin{widetext}
\begin{eqnarray}
\mathcal{Y}^{F_{3.1}}_{I=1}&=&-\frac{1}{f^4}\varepsilon_1\left\{\mathcal{J}_{22}^F\right\}_r(m_\pi,q),\label{BastBast_Amp_TPEI1_F31}
\end{eqnarray}
\begin{eqnarray}
\mathcal{Y}^{T_{3.1}}_{I=1}&=&\frac{g^2}{f^4}\Bigg\{\Big[\varepsilon_1^a+2\varepsilon_1\vec{q}^2+\varepsilon_1^b\Big]\left(\mathcal{J}_{24}^T+\mathcal{J}_{33}^T\right)-4\varepsilon_1\mathcal{J}_{34}^T\Bigg\}_r(m_\pi,\mathcal{E},q),
\end{eqnarray}
\begin{eqnarray}
\mathcal{Y}^{T_{3.2}}_{I=1}&=&-\frac{g^2}{f^4}\Bigg\{\Big[\varepsilon_1^a+\varepsilon_1^b\Big]\left(\mathcal{J}_{24}^T+\mathcal{J}_{33}^T\right)+2\varepsilon_1\mathcal{J}_{34}^T\Bigg\}_r(m_\pi,\mathcal{E}+\Delta,q),
\end{eqnarray}
\begin{eqnarray}
\mathcal{Y}^{B_{3.1}}_{I=1}&=&-\frac{g^4}{4f^4}\Bigg\{\Big[\varepsilon_1\vec{q}^4+\varepsilon_1^a\vec{q}^2+\varepsilon_1^b\vec{q}^2+\varepsilon_1^c\Big]
\left(\mathcal{J}_{22}^B+2\mathcal{J}_{32}^B+\mathcal{J}_{43}^B\right)-\Big[8\varepsilon_1\vec{q}^2+6\varepsilon_1^a+6\varepsilon_1^b-\varepsilon_2^a
-\varepsilon_2^b-\varepsilon_3^a\Big]\left(\mathcal{J}_{31}^B+\mathcal{J}_{42}^B\right)\nonumber\\
&&-\Big[\varepsilon_1\vec{q}^2+\varepsilon_1^a+\varepsilon_1^b-\varepsilon_3^a\Big]\mathcal{J}_{21}^B+\Big[6\varepsilon_1+\varepsilon_2+\varepsilon_3\Big]\mathcal{J}_{41}^B
+\varepsilon_3^a\mathcal{J}_{31}^B+\varepsilon_3^b\mathcal{J}_{42}^B\Bigg\}_r(m_\pi,\mathcal{E},\mathcal{E},q),
\end{eqnarray}
\begin{eqnarray}
\mathcal{Y}^{B_{3.2}}_{I=1}&=&-\frac{g^4}{4f^4}\Bigg\{\varepsilon_3^c\left(\mathcal{J}_{22}^B+2\mathcal{J}_{32}^B+\mathcal{J}_{43}^B\right)+\Big[\varepsilon_1^a+\varepsilon_1^b+\varepsilon_2^a+\varepsilon_2^b+\varepsilon_3^b\Big]\left(\mathcal{J}_{31}^B+\mathcal{J}_{42}^B\right)+\Big[\varepsilon_1+\varepsilon_2+\varepsilon_3\Big]\mathcal{J}_{41}^B\nonumber\\
&&+\varepsilon_3^b\left(\mathcal{J}_{21}^B+\mathcal{J}_{31}^B\right)+\varepsilon_3^a\mathcal{J}_{42}^B\Bigg\}_r(m_\pi,\mathcal{E}+\Delta,\mathcal{E}+\Delta,q),
\end{eqnarray}
\begin{eqnarray}
\mathcal{Y}^{B_{3.3}}_{I=1}&=&-\frac{g^4}{4f^4}\Bigg\{\left[\varepsilon_1^a\vec{q}^2+\varepsilon_1^b\vec{q}^2+2\varepsilon_3^c\right]\left(\mathcal{J}_{22}^B+2\mathcal{J}_{32}^B+\mathcal{J}_{43}^B\right)
+\Big[-5\varepsilon_1^a-5\varepsilon_1^b+2\varepsilon_1\vec{q}^2+2\varepsilon_2^a+2\varepsilon_2^b+2\varepsilon_3^a+2\varepsilon_3^b\Big]\left(\mathcal{J}_{31}^B+\mathcal{J}_{42}^B\right)\nonumber\\
&&+\Big[-\varepsilon_1^a-\varepsilon_1^b+\varepsilon_2^a+\varepsilon_2^b\Big]\mathcal{J}_{21}^B+\Big[-8\varepsilon_1+2\varepsilon_2+2\varepsilon_3\Big]\mathcal{J}_{41}^B\Bigg\}_r(m_\pi,\mathcal{E},\mathcal{E}+\Delta,q).\label{BastBast_Amp_TPEI1_R33}
\end{eqnarray}
\end{widetext}
For succinctness, we also define
\begin{eqnarray}
\varepsilon_1^a&=&(q\cdot\epsilon_1)(q\cdot\epsilon_3^\ast)(\epsilon_2\cdot\epsilon_4^\ast),\quad\varepsilon_1^b=(q\cdot\epsilon_2)(q\cdot\epsilon_4^\ast)(\epsilon_1\cdot\epsilon_3^\ast),\nonumber\\
\varepsilon_1^c&=&(q\cdot\epsilon_1)(q\cdot\epsilon_3^\ast)(q\cdot\epsilon_2)(q\cdot\epsilon_4^\ast),\nonumber\\
\varepsilon_3^c&=&(q\cdot\epsilon_1)(q\cdot\epsilon_2)(q\cdot\epsilon_3^\ast)(q\cdot\epsilon_4^\ast).
\end{eqnarray}
Analogous to the $\bar{B}\bar{B}^\ast$ case (see Eq. \eqref{JBtoJR1}), the unlisted amplitudes of the diagrams $R_{3.1}$$\sim$$R_{3.3}$ can be achieved by the following relations,
\begin{eqnarray}\label{JBtoJR2}
\mathcal{Y}^{R_{3.1}}_{I=1}&=&5\mathcal{Y}^{B_{3.1}}_{I=1}\Big|_{\mathcal{J}_{x}^B\to\mathcal{J}_{x}^R},\quad\mathcal{Y}^{R_{3.2}}_{I=1}=5\mathcal{Y}^{B_{3.2}}_{I=1}\Big|_{\mathcal{J}_{x}^B\to\mathcal{J}_{x}^R},\nonumber\\
\mathcal{Y}^{R_{3.3}}_{I=1}&=&5\mathcal{Y}^{B_{3.3}}_{I=1}\Big|_{\mathcal{J}_{x}^B\to\mathcal{J}_{x}^R}.
\end{eqnarray}

To obtain the numerical results, some terms like $(\epsilon_i\cdot\epsilon_j)(\epsilon_k\cdot\epsilon_l)$, $(\epsilon_i\cdot\epsilon_j)(q\cdot\epsilon_k)(q\cdot\epsilon_l)$ and $(q\cdot\epsilon_i)(q\cdot\epsilon_j)(q\cdot\epsilon_k)(q\cdot\epsilon_l)$ (where $\epsilon_{i},\cdots,\epsilon_{l}$ denotes either the polarization vector of initial state or final state) appeared in Eqs. \eqref{BastBast_XH1I}-\eqref{BastBast_Amp_TPEI1_R33} must be reexpressed as a constant or a function of $\vec{q}$. Resembling the OBE model \cite{Liu:2008tn}, under $S$-wave interactions, the values of the term $(\epsilon_i\cdot\epsilon_j)(\epsilon_k\cdot\epsilon_l)$ have been given in Tab. \ref{Value_PolPro}. For the terms containing $q$, one can make the following substitutions,
\begin{eqnarray}
(\epsilon_i\cdot\epsilon_j)(q\cdot\epsilon_k)(q\cdot\epsilon_l)&\rightarrowtail& -\frac{1}{3}\vec{q}^2(\epsilon_i\cdot\epsilon_j)(\epsilon_k\cdot\epsilon_l),\nonumber\\
(q\cdot\epsilon_i)(q\cdot\epsilon_j)(q\cdot\epsilon_k)(q\cdot\epsilon_l)&\rightarrowtail& \frac{1}{9}\vec{q}^4(\epsilon_i\cdot\epsilon_j)(\epsilon_k\cdot\epsilon_l).
\end{eqnarray}
The 2PR contributions in diagrams $h_{3.1}$, $h_{3.4}$ and $B_{3.1}$ must be removed, and other parameters are the same as we calculating the $\bar{B}\bar{B}^\ast$ system.

\begin{table}
\renewcommand{\arraystretch}{1.5}
 \tabcolsep=15pt
\caption{The values of the products of polarization vectors under $S$-wave interactions \cite{Liu:2008tn}. For example, in our calculations, the term $(\epsilon_1\cdot\epsilon_2)(\epsilon_3^\ast\cdot\epsilon_4^\ast)$ should be replaced by $3$, $0$, and $0$ if the total spin $S_{\text{tot}}$ equals to $0$, $1$, and $2$, respectively.}\label{Value_PolPro}
\setlength{\tabcolsep}{4.5mm}
{
\begin{tabular}{c|ccc}
\hline\hline
Terms&$S_{\text{tot}}=0$&$S_{\text{tot}}=1$&$S_{\text{tot}}=2$\\
\hline
$(\epsilon_1\cdot\epsilon_2)(\epsilon_3^\ast\cdot\epsilon_4^\ast)$&3&0&0\\
\hline
$(\epsilon_1\cdot\epsilon_3^\ast)(\epsilon_2\cdot\epsilon_4^\ast)$&1&1&1\\
\hline
$(\epsilon_1\cdot\epsilon_4^\ast)(\epsilon_2\cdot\epsilon_3^\ast)$&1&-1&1\\
\hline\hline
\end{tabular}
}
\end{table}

\begin{figure*}
\begin{minipage}[t]{0.45\linewidth}
\centering
\includegraphics[width=\columnwidth]{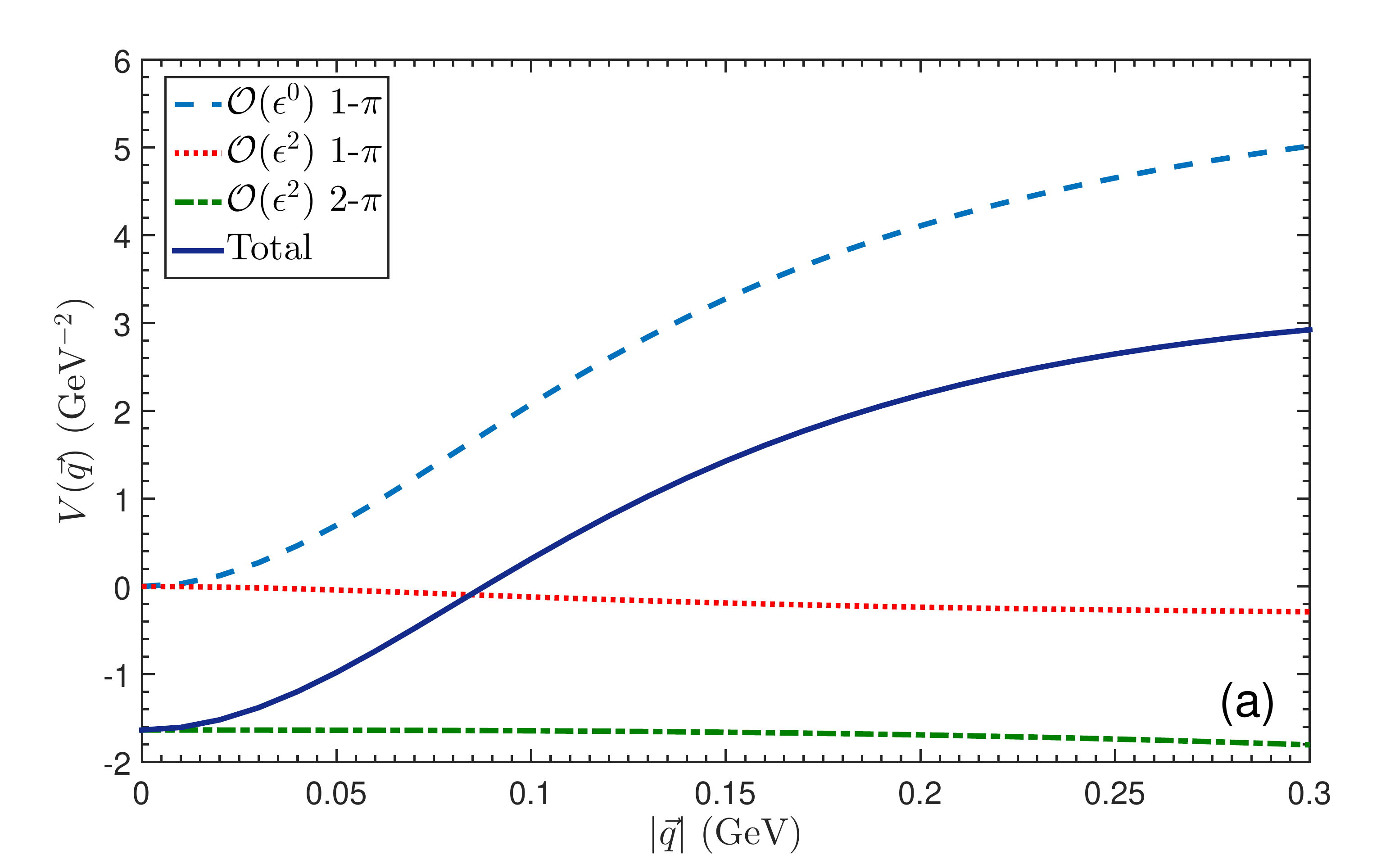}
\end{minipage}%
\begin{minipage}[t]{0.45\linewidth}
\centering
\includegraphics[width=\columnwidth]{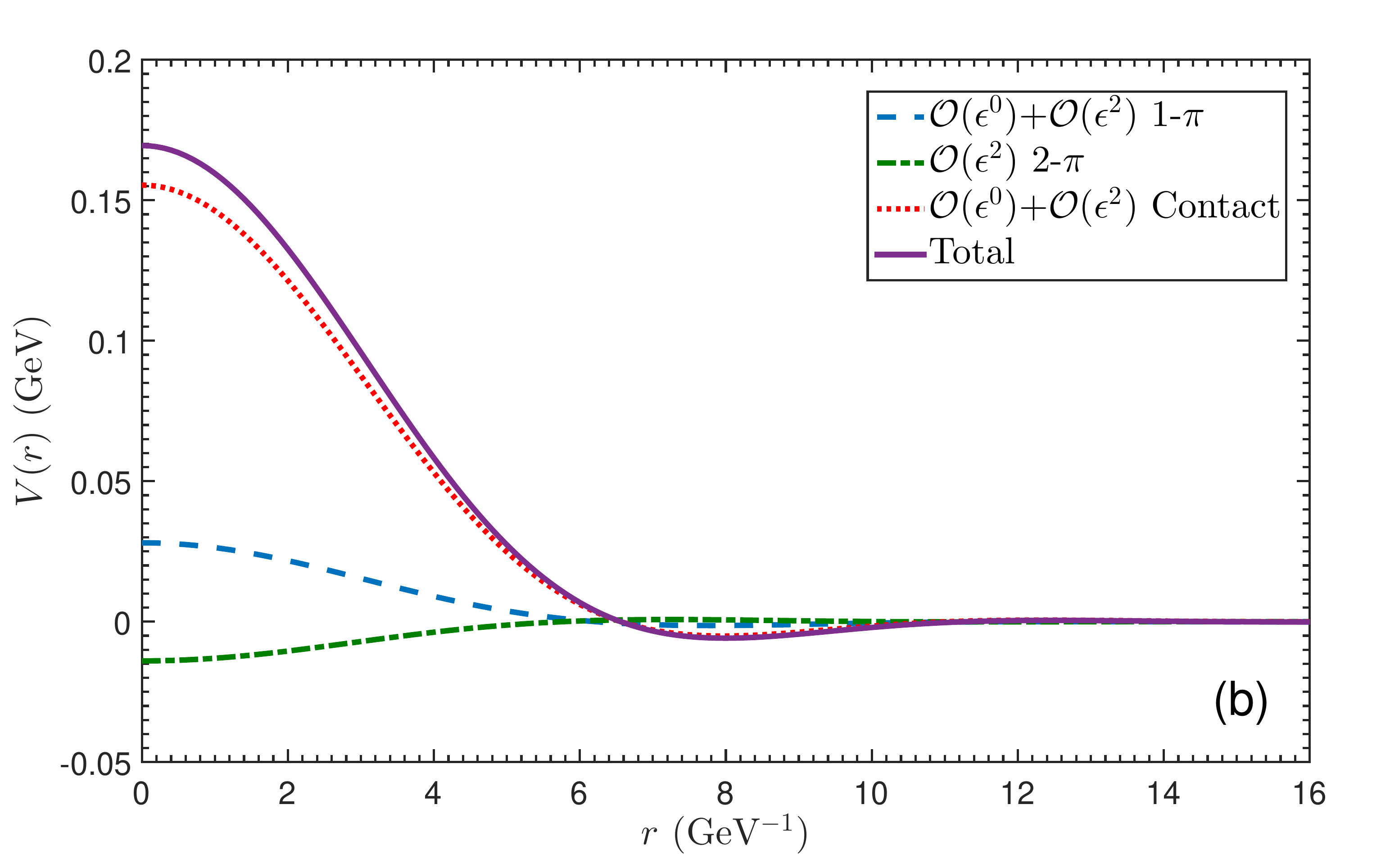}
\end{minipage}
\caption{The OPE and TPE potentials of $1(0^+)$ $\bar{B}^\ast\bar{B}^\ast$ system in momentum space (left panel) and the effective potentials in coordinate space (right panel). Notations same as in Fig. \ref{BBastI1_Momentum_Euclid}. \label{BastBastI1J0_Momentum_Euclid}}
\end{figure*}

\begin{figure*}
\begin{minipage}[t]{0.45\linewidth}
\centering
\includegraphics[width=\columnwidth]{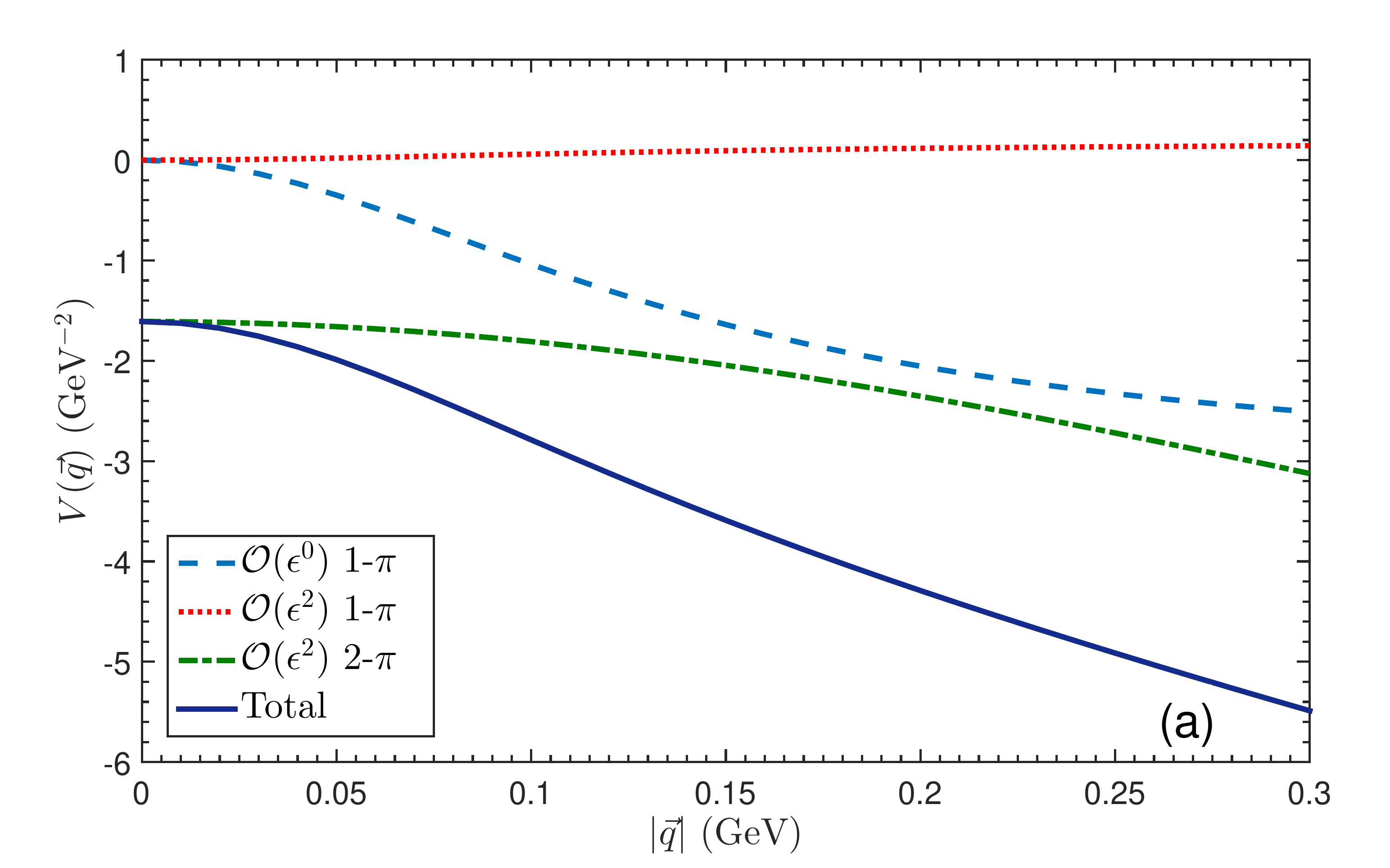}
\end{minipage}%
\begin{minipage}[t]{0.45\linewidth}
\centering
\includegraphics[width=\columnwidth]{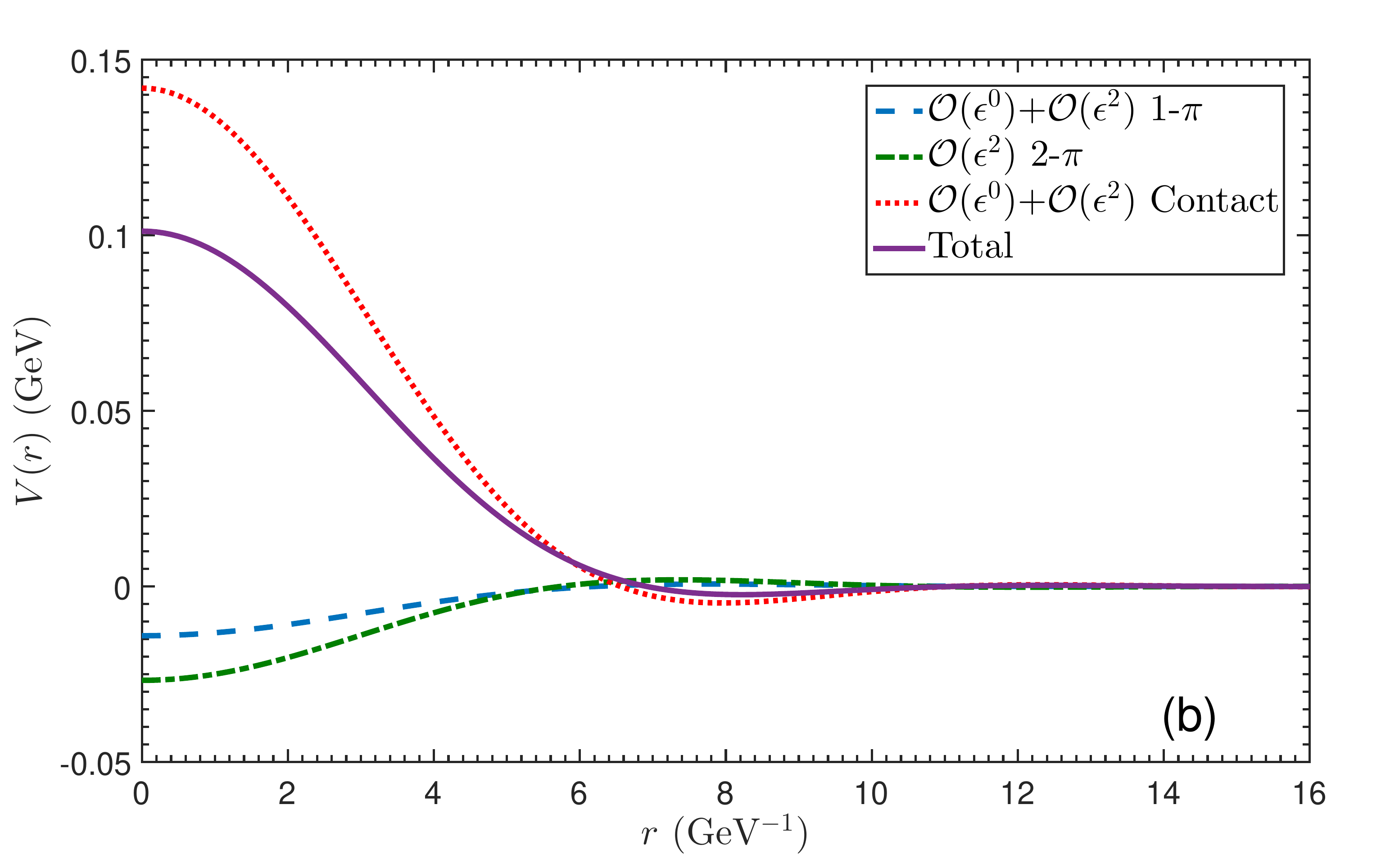}
\end{minipage}
\caption{The OPE and TPE potentials of $1(2^+)$ $\bar{B}^\ast\bar{B}^\ast$ system in momentum space (left panel) and the effective potentials in coordinate space (right panel). Notations same as in Fig. \ref{BBastI1_Momentum_Euclid}. \label{BastBastI1J2_Momentum_Euclid}}
\end{figure*}

\begin{figure*}
\begin{minipage}[t]{0.45\linewidth}
\centering
\includegraphics[width=\columnwidth]{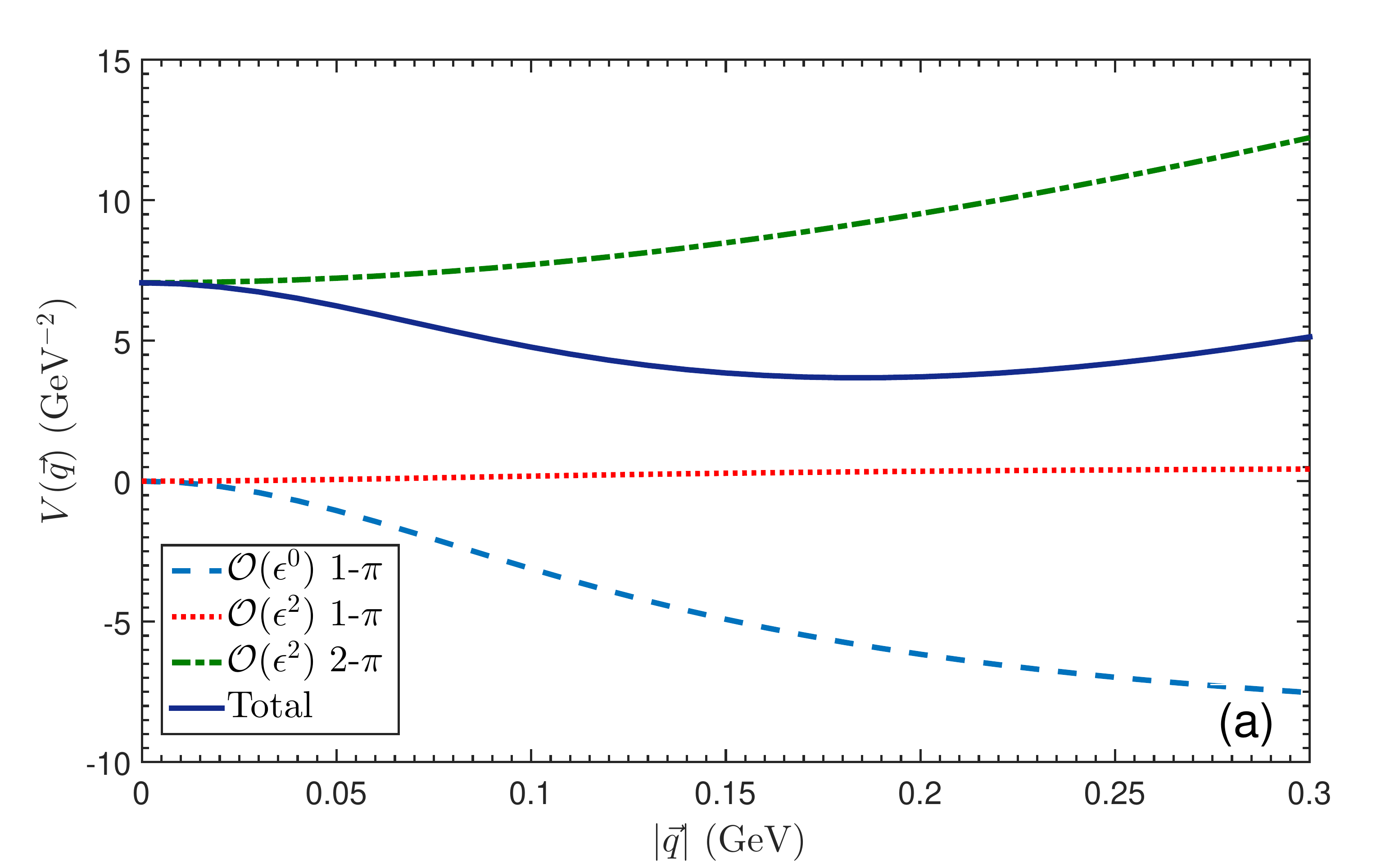}
\end{minipage}%
\begin{minipage}[t]{0.45\linewidth}
\centering
\includegraphics[width=\columnwidth]{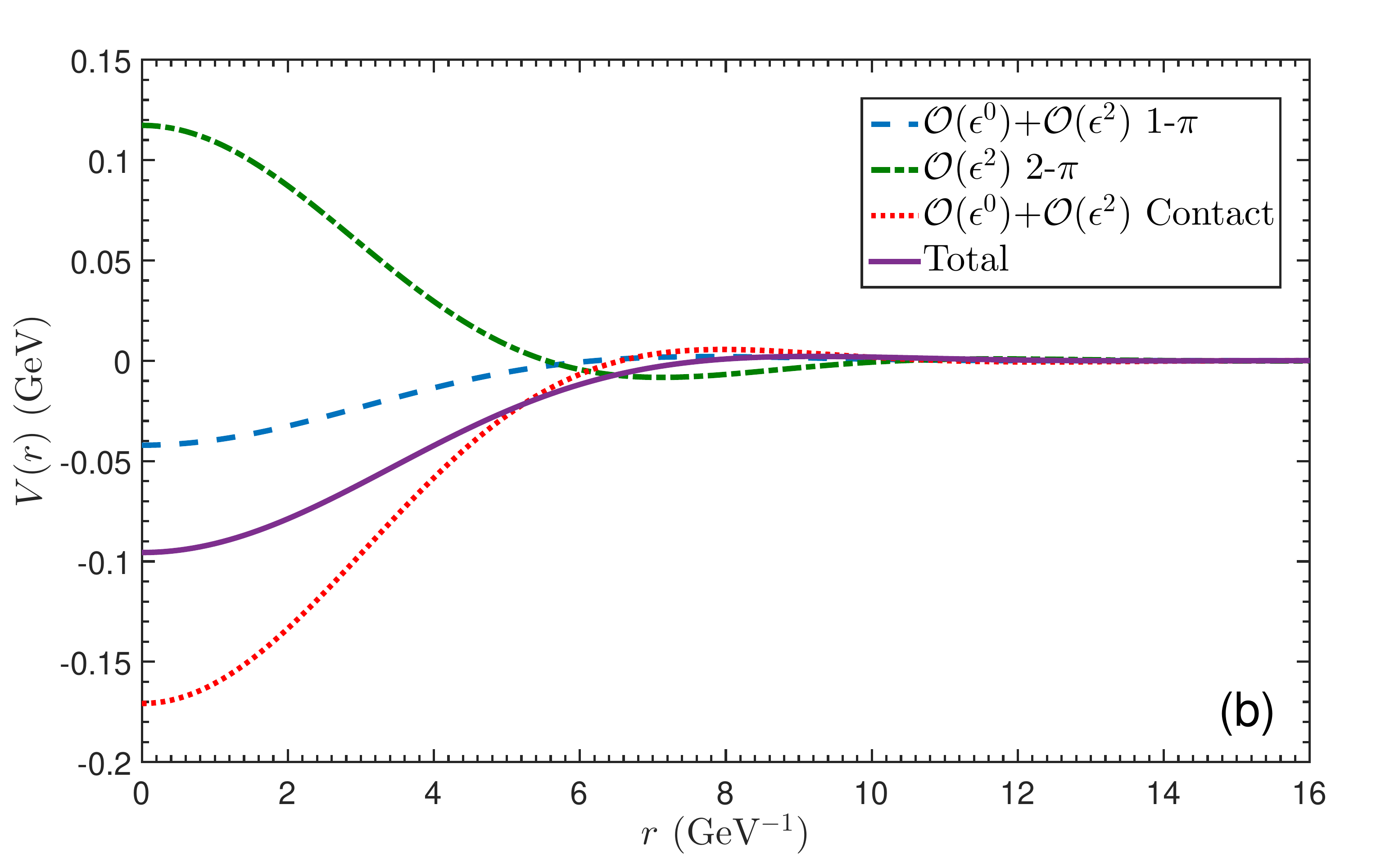}
\end{minipage}
\caption{The OPE and TPE potentials of $0(1^+)$ $\bar{B}^\ast\bar{B}^\ast$ system in momentum space (left panel) and the effective potentials in coordinate space (right panel). Notations same as in Fig. \ref{BBastI1_Momentum_Euclid}. \label{BastBastI0J1_Momentum_Euclid}}
\end{figure*}

The resulting potentials of $I=1$ and $I=0$ $\bar{B}^\ast\bar{B}^\ast$ system are depicted in Figs. \ref{BastBastI1J0_Momentum_Euclid}-\ref{BastBastI0J1_Momentum_Euclid}, respectively.

For the $1(0^+)$ $\bar{B}^\ast\bar{B}^\ast$ system, from Fig. \ref{BastBastI1J0_Momentum_Euclid} we can see that the OPE potential is repulsive and the TPE potential is slightly attractive, but the contribution from FBCI is repulsive and largely dominant, which generates a fully repulsive potential for the $1(0^+)$ $\bar{B}^\ast\bar{B}^\ast$ system. Thus no bound state can be found in the $1(0^+)$ $\bar{B}^\ast\bar{B}^\ast$ channel in our calculations.

Wh show the effective potential of $1(2^+)$ $\bar{B}^\ast\bar{B}^\ast$ system in Fig. \ref{BastBastI1J2_Momentum_Euclid}. The behavior of TPE potential is similar to the $1(0^+)$ case. The line-shape of OPE potential is totally reversed, that is, the OPE potential is shallowly attractive for $1(2^+)$ $\bar{B}^\ast\bar{B}^\ast$ system. Like the $1(0^+)$ case, the potential resulted from FBCI is also repulsive and dominant. Thus the total potential of $1(2^+)$ $\bar{B}^\ast\bar{B}^\ast$ system is also repulsive. So we can conclude that no bound state of $\bar{B}^\ast\bar{B}^\ast$ can exist in $I=1$ channel.

At last, we analyze the potential of $0(1^+)$ $\bar{B}^\ast\bar{B}^\ast$ system shown in Fig. \ref{BastBastI0J1_Momentum_Euclid}. In rigorous heavy quark limit, $\bar{B}$ and $\bar{B}^\ast$ mesons would be totally degenerate, so the potential of $\bar{B}\bar{B}^\ast$ and $\bar{B}^\ast\bar{B}^\ast$ systems should be similar if they carry the same quantum number. By comparing Fig. \ref{BastBastI0J1_Momentum_Euclid} and Fig. \ref{BBastI0_Momentum_Euclid}, we can see that the potentials of $0(1^+)$ $\bar{B}\bar{B}^\ast$ and $0(1^+)$ $\bar{B}^\ast\bar{B}^\ast$ indeed have the same behaviors. The OPE and TPE potentials are attractive and repulsive for the $0(1^+)$ $\bar{B}^\ast\bar{B}^\ast$ system, respectively, and the FBCI supplies a dominantly attractive potential, so the total potential is also attractive. By solving the schr\"odinger equation, we find a bound state in $0(1^+)$ $\bar{B}^\ast\bar{B}^\ast$ channel with the binding energy $\Delta E_{\bar{B}^\ast\bar{B}^\ast}\simeq-23.8^{+16.3}_{-21.5}$ MeV, and the corresponding root-mean-square radius is $0.81^{+0.33}_{-0.13}$ fm.

We get the mass of $0(1^+)$ $\bar{B}^\ast\bar{B}^\ast$ system also being above the threshold of $\bar{B}\bar{B}\gamma\gamma$. Therefore, it is detectable through electromagnetic interactions. In addition, we could resort to its weak decay modes as well, such as $J/\psi B^+ K^0$, with $J/\psi$ and $B^+$ being fully reconstructable from $J/\psi\to \ell^+\ell^-$ and $B^+\to\bar{D}^0\pi^+$.

\subsection{The results in strict heavy quark limit}
It is also very interesting to study the behaviors of effective potentials in strict heavy quark limit. As is discussed below Eq. \eqref{AMP_BBTPE_R11}, if $\Delta=0$, the diagrams $h_{2.1}$$\sim$$h_{2.4}$ and $B_{2.1}$$\sim$$B_{2.3}$ in Fig. \ref{BBast_Contact_Corrections} and Fig. \ref{BBast_TPE} would contribute to the 2PR parts of the $\bar{B}\bar{B}^\ast$ scattering amplitudes. Similarly, for $\bar{B}^\ast\bar{B}^\ast$ interactions, the diagrams $h_{3.1}$$\sim$$h_{3.6}$ and $B_{3.1}$$\sim$$B_{3.3}$ in Fig. \ref{BastBast_Contact_Corrections} and Fig. \ref{BastBast_TPE} also contain both 2PR and 2PI contributions. In order to get the potentials, the contributions from the 2PR parts must be eliminated by using Weinberg's formalism (see Appendix \ref{LoopIntegral_3} for more details).

\begin{figure*}
\begin{minipage}[t]{0.45\linewidth}
\centering
\includegraphics[width=\columnwidth]{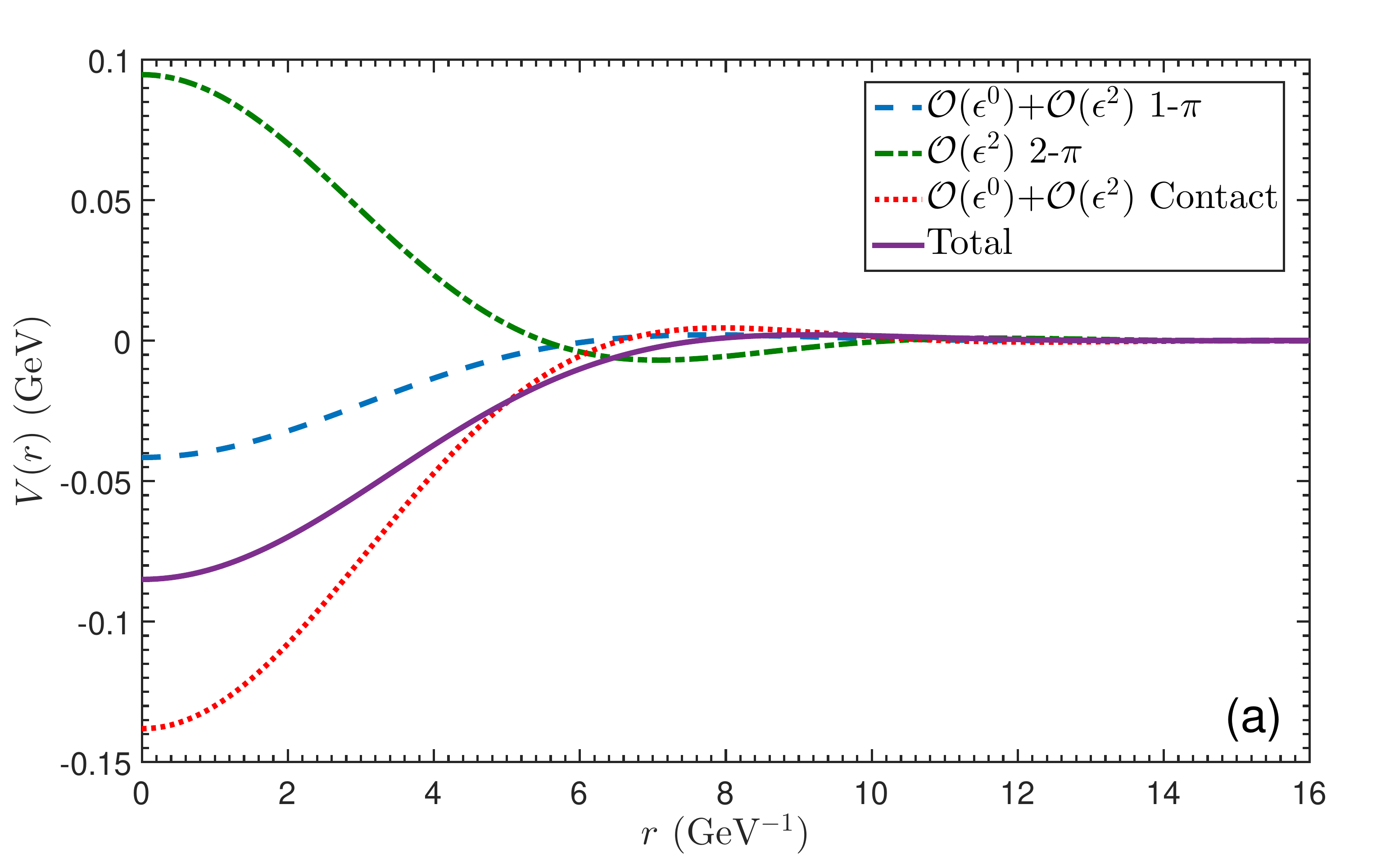}
\end{minipage}%
\begin{minipage}[t]{0.45\linewidth}
\centering
\includegraphics[width=\columnwidth]{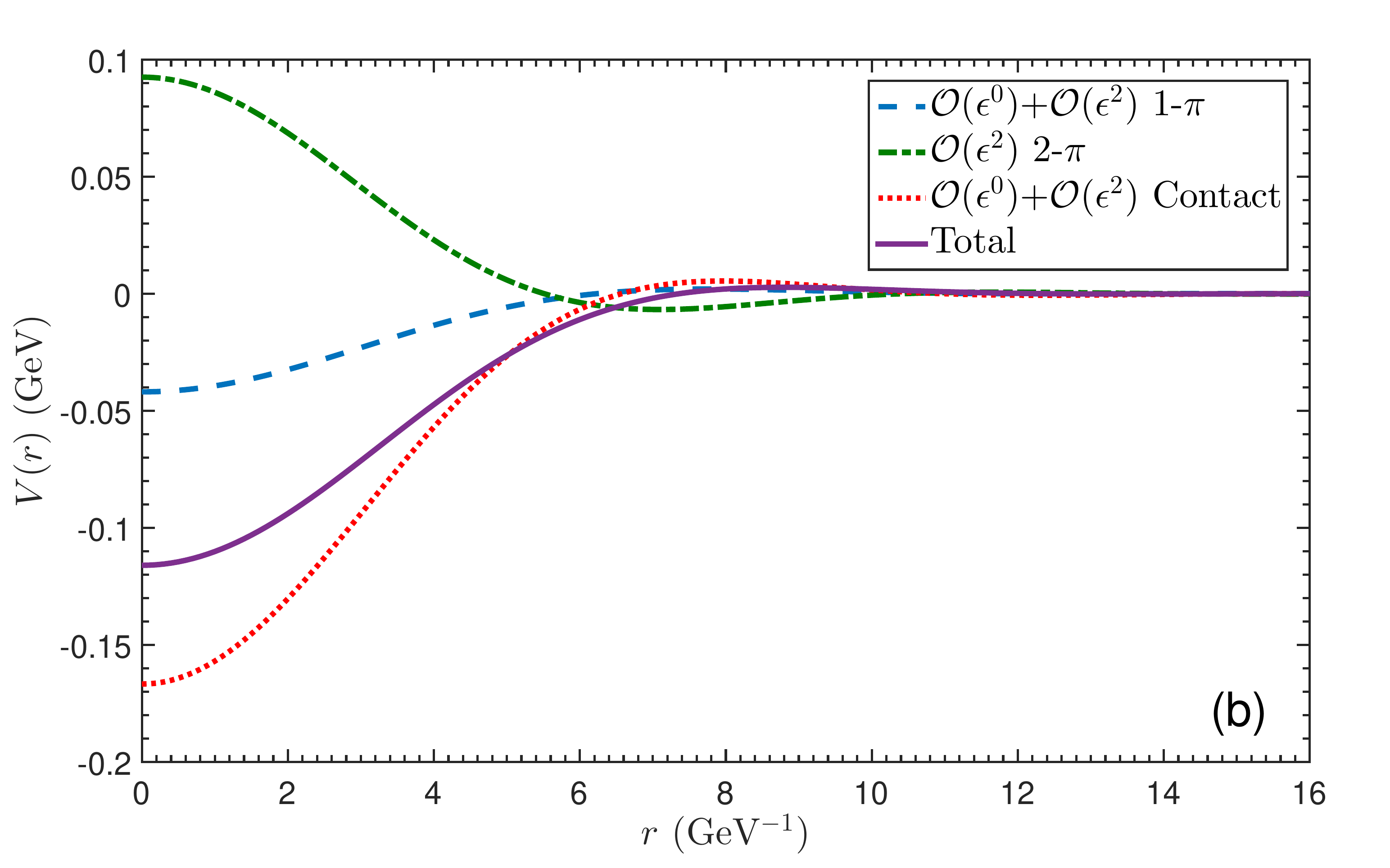}
\end{minipage}
\caption{The effective potentials of $0(1^+)$ $\bar{B}\bar{B}^\ast$ (left panel) and $\bar{B}^\ast\bar{B}^\ast$ (right panel) systems in strict heavy quark limit. Notations same as in Fig. \ref{BBastI1_Momentum_Euclid}.\label{BBI0_coordinate_New}}
\end{figure*}

Under heavy quark limit, we find the main features of the potentials in different channels remain unchanged, so for simplicity, we only give the results of two representative states, i.e., $0(1^+)$ $\bar{B}\bar{B}^\ast$ and $\bar{B}^\ast\bar{B}^\ast$. The corresponding results are plotted in Fig. \ref{BBI0_coordinate_New}. Comparing the potentials with and without $\Delta=0$ for $0(1^+)$ $\bar{B}\bar{B}^\ast$  and $\bar{B}^\ast\bar{B}^\ast$ systems, we can find the corresponding potentials are only marginally shifted, this phenomenon indicates the heavy quark symmetry holds well for $B$ mesons. Additionally, the effect caused by $\Delta$ only happens at $\mathcal{O}(\epsilon^2)$, and our calculations shown before have given the convergence of chiral corrections is very good. Furthermore, the mass difference of $\bar{B}$ and $\bar{B}^\ast$ lies far below the mass of pion $m_\pi$, thus the influence of $\Delta$ is largely suppressed. However, for $D$ and $D^\ast$ systems, the mass shift $\Delta>m_\pi$, one can expect more explicit effects of $\Delta$ in $DD^\ast$ and $D^\ast D^\ast$ systems.

 $\bar{B}$ and $\bar{B}^\ast$ would degenerate in heavy quark limit, so we anticipate the appearances of the potentials in $\bar{B}\bar{B}^\ast$  and $\bar{B}^\ast\bar{B}^\ast$ systems should be identical if they have the same quantum number. This is vividly reflected from Fig. \ref{BBI0_coordinate_New}(a) and \ref{BBI0_coordinate_New}(b). We can see that the model independent parts, namely, OPE and TPE potentials are exactly the same for the two systems. For the model dependent parts, i.e., contact interactions, at $\mathcal{O}(\epsilon^0)$ the contributions for $\bar{B}\bar{B}^\ast$ and $\bar{B}^\ast\bar{B}^\ast$ are equal, and the only difference comes from the loop corrections. Because at $\mathcal{O}(\epsilon^2)$, different contact structures with various combinations of LECs are involved into the loop diagrams.

At last, we list the binding energies and root-mean-square radii of $0(1^+)$ $\bar{B}\bar{B}^\ast$ and $\bar{B}^\ast\bar{B}^\ast$ systems in the heavy quark limit, respectively.
\begin{eqnarray}
\Delta E_{\bar{B}\bar{B}^\ast}&\simeq&-18.0^{+10.3}_{-12.9}~\text{MeV},\quad r_{\bar{B}\bar{B}^\ast}\simeq 0.87^{+0.24}_{-0.12}~\text{fm},\nonumber\\
\Delta E_{\bar{B}^\ast\bar{B}^\ast}&\simeq&-32.3^{+15.1}_{-17.4}~\text{MeV},\quad r_{\bar{B}^\ast\bar{B}^\ast}\simeq 0.73^{+0.14}_{-0.08}~\text{fm}.\nonumber
\end{eqnarray}
The result is in agreement with the patterns shown by $c\bar{c}$ and $b\bar{b}$ spectra, i.e., the binding would be deeper when the mass of the components is increased.

\section{Estimation of the $\mathcal{O}(\epsilon^2)$ LECs Contributions} \label{Two_Order_LECs}
In Sec. \ref{Sec3}, we discard the contributions of $\mathcal{O}(\epsilon^2)$ LECs since no enough data are available now to fit all of them. In this section, we pick these LECs up to see the influences on our numerical results. For this purpose, two different strategies are put to use. Although the estimations are rough and we are not trying to provide the very accurate LECs, they still can tell us some useful information about the reliability of the obtained results and how much the errors from the uncertainty of LECs at $\mathcal{O}(\epsilon^2)$ are.
\subsection{Strategy A}
We first adopt the nonanalytic dominance approximation to give an estimation of LECs contributions \cite{Liu:2012uw,Bijnens:1995yn}, This approximation is based on the fact that the scattering matrix can be decomposed into analytic and nonanalytic part in chiral perturbation theory. The nonanalytic contributions can only be obtained from loop diagrams. However, the analytic terms can originate from both tree and loop graphs, and they are the polynomials of the expanding parameter $\epsilon$. The analytic terms from loop diagrams can be absorbed by the LECs at the same order. Therefore, we can use the nonanalytic contributions to discuss the convergence of potentials since the $\mathcal{O}(\epsilon^2)$ LECs cannot be fixed now.

In the nonanalytic dominance approximation, we assume there is a large cancellation between the analytic terms of loop diagrams and the finite parts of $\mathcal{O}(\epsilon^2)$ LECs in Eqs. \eqref{4H2h}-\eqref{4H2q}. With this approximation, we redo the calculation and give the binding energies of $0(1^+)$ $\bar{B}\bar{B}^\ast$ and $\bar{B}^\ast\bar{B}^\ast$ with $\Lambda=0.7$ GeV in the third column of Tab. \ref{Energy_Comparison}.
The result given with the nonanalytic dominance approximation shows chiral expansion works well. However, one also needs to note that, in this strategy, we perhaps underestimate the effects of $\mathcal{O}(\epsilon^2)$ LECs.

\subsection{Strategy B}
The $\mathcal{O}(\epsilon^2)$ scattering amplitudes of $0(1^+)$ $\bar{B}\bar{B}^\ast$ and $\bar{B}^\ast\bar{B}^\ast$ generated from the Lagrangians in Eq. \eqref{4H2h} can be written as
\begin{eqnarray}
\mathcal{Y}^{\bar{B}\bar{B}^\ast}_{I=0}&=&32m_\pi^2\left[-D_a^h-D_b^h+3(E_a^h+E_b^h)\right](\epsilon_2\cdot\epsilon_4^\ast),\\
\mathcal{Y}^{\bar{B}^\ast\bar{B}^\ast}_{I=0}&=&32m_\pi^2\left[D_a^h+D_b^h-3(E_a^h+E_b^h)\right]\left(\varepsilon_1-\varepsilon_2\right).
\end{eqnarray}
By dimensional analysis, we can naively get
\begin{eqnarray}\label{LECs_Relations}
D_a^h\sim\frac{D_a}{\Lambda_\chi^2},\quad D_b^h\sim\frac{D_b}{\Lambda_\chi^2},\quad E_a^h\sim\frac{E_a}{\Lambda_\chi^2},\quad E_b^h\sim\frac{E_b}{\Lambda_\chi^2},
\end{eqnarray}
where $\Lambda_\chi\simeq1$ GeV.

Expanding the Lagrangians in Eq. \eqref{4H2v} and Eq. \eqref{4H2q}, we notice the contributions from these two equations will be proportional to $\mathcal{E}^2$ and $k^2$ (where $\mathcal{E}$ is the residual energy, and $k$ is the residual momentum), respectively. In our calculations, the incoming and outgoing heavy mesons are on-shell, so we can assume the contributions of $\mathcal{O}(\epsilon^2)$ tree diagrams mainly come from the ones governed by Eq. \eqref{4H2h}.

With the relations in Eq. \eqref{LECs_Relations} and the values given in Eq. \eqref{Values_LECs}, we can roughly estimate the numerical values of $D_a^h$, $\cdots$, $E_b^h$ to examine the influence of the LECs at $\mathcal{O}(\epsilon^2)$. The corresponding results are listed in the last column of Tab. \ref{Energy_Comparison}.

In Tab. \ref{Energy_Comparison}, the difference between the results of ``Strategy A" (or ``Strategy B") and ``No $\mathcal{O}(\epsilon^2)$ LECs" can be regraded as an evaluation of the errors resulted from the uncertainty of LECs at $\mathcal{O}(\epsilon^2)$. Although the two strategies used above are crude, the results in Tab. \ref{Energy_Comparison} show that the contributions of $\mathcal{O}(\epsilon^2)$ LECs should be small. Neglecting the $\mathcal{O}(\epsilon^2)$ tree diagrams is safe within the allowable range of errors at the early stage of the study. These LECs at $\mathcal{O}(\epsilon^2)$ would be determined in a complete analysis in future when more experimental data are available.
\begin{table}
\renewcommand{\arraystretch}{1.5}
 \tabcolsep=15pt
\caption{The binding energies of $0(1^+)$ $\bar{B}\bar{B}^\ast$ and $\bar{B}^\ast\bar{B}^\ast$ states obtained with different strategies in units of MeV. ``No $\mathcal{O}(\epsilon^2)$ LECs" represents the results obtained without considering the contributions from the finite part of $\mathcal{O}(\epsilon^2)$ contact terms in Eqs. \eqref{4H2h}-\eqref{4H2q}.}\label{Energy_Comparison}
\setlength{\tabcolsep}{2.3mm}
{
\begin{tabular}{c|ccc}
\hline\hline
Binding energy&No $\mathcal{O}(\epsilon^2)$ LECs&Strategy A&Strategy B\\
\hline
$\Delta E_{\bar{B}\bar{B}^\ast}$&$-12.6^{+9.2}_{-12.9}$&$-10.4^{+7.2}_{-9.7}$&$-15.9^{+9.7}_{-12.7}$\\
\hline
$\Delta E_{\bar{B}^\ast\bar{B}^\ast}$&$-23.8^{+16.3}_{-21.5}$&$-20.1^{+14.5}_{-20.0}$&$-28.2^{+18.6}_{-23.6}$\\
\hline\hline
\end{tabular}
}
\end{table}

\section{summary}\label{Sec4}
In summary, we have systematically investigated the intermeson interactions of double-beauty $\bar{B}\bar{B}$, $\bar{B}\bar{B}^\ast$, and $\bar{B}^\ast\bar{B}^\ast$ systems with HM$\chi$EFT. In addition to the $\mathcal{O}(\epsilon^0)$ FBCI and OPE diagrams, we also include the $\mathcal{O}(\epsilon^2)$ TPE diagrams and one-loop corrections to FBCI and OPE diagrams. The effective potentials are calculated with Weinberg's formalism \cite{Weinberg:1990rz,Weinberg:1991um}, i.e., we do not calculate the scattering matrix directly since the 2PR contributions will spoil the correct power counting, and instead, we only take into account the 2PI parts of Feynman diagrams to derive the effective potentials. Moreover, with the aid of simple Gauss cutoff, we make the Fourier transformation to obtain the effective potentials with more visualized form in coordinate space, and then by iterating the potentials into Schr\"odinger equation, we not only can look for the bound state solutions but also can recover the 2PR contributions that we subtract before.

We only consider the $S$-wave $\bar B^{(*)}\bar B^{(*)}$ systems in this work, and thus based on the selection rules in Tab. \ref{allowedBB}, the physically allowed states are: $1(0^+)$ $\bar{B}\bar{B}$; $1(1^+)$ and $0(1^+)$  $\bar{B}\bar{B}^\ast$; $1(0^+)$, $1(2^+)$, and $0(1^+)$ $\bar{B}^\ast \bar{B}^\ast$. The potentials of these six channels with different $I(J^P)$ quantum numbers are studied and discussed in detail. We find the convergences of chiral corrections are all very good in these six channels, and the FBCI, which depicts the short-range interaction, plays the crucial role in determining the behaviors of the total potentials.

For $1(0^+)$ $\bar{B}\bar{B}$ and $\bar{B}^\ast \bar{B}^\ast$ systems, the total potentials are both repulsive, and thus we conclude that no bound states can exist in these two channels. The same situation also holds for the $1(1^+)$ $\bar{B}\bar{B}^\ast$ and $1(2^+)$ $\bar{B}^\ast \bar{B}^\ast$ systems.

While for the $0(1^+)$ $\bar{B}\bar{B}^\ast$ and $\bar{B}^\ast \bar{B}^\ast$ systems, our results are very interesting because the total potentials of these two channels are both attractive. By solving the Schr\"odinger equation, we find two bound states in these two channels with the corresponding binding energy $\Delta E_{\bar{B}\bar{B}^\ast}\simeq-12.6^{+9.2}_{-12.9}$ MeV and $\Delta E_{\bar{B}^\ast\bar{B}^\ast}\simeq-23.8^{+16.3}_{-21.5}$ MeV, respectively. We predict their masses to be
\begin{eqnarray}
m_{\bar{B}\bar{B}^\ast}\simeq10591.4^{+9.2}_{-12.9}~\text{MeV},\quad
m_{\bar{B}^\ast\bar{B}^\ast}\simeq10625.5^{+16.3}_{-21.5}~\text{MeV}.\nonumber
\end{eqnarray}
Our result can qualitatively confirm the conclusions drawn by the OBE model \cite{Li:2012ss}, quark model potential \cite{Barnes:1999hs}, and Lattice QCD \cite{Bicudo:2016ooe}.

It will be an intriguing topic to search for the exotic double-beauty molecular states in $0(1^+)$ channels experimentally. These two molecular candidates cannot directly decompose into their components due to the constraints of phase space. However, the $\bar{B}^\ast$ meson can decay into $\bar{B}\gamma$ via electromagnetic interaction. We find the masses of $0(1^+)$ $\bar{B}\bar{B}^\ast$ and $\bar{B}^\ast\bar{B}^\ast$ states both lie above the thresholds for decaying into
$\bar{B}\bar{B}\gamma$ and $\bar{B}\bar{B}\gamma\gamma$, and thus they are reconstructable in their electromagnetic decay modes.

Our results and above typical decay modes provide important information to future search in experiments. Additionally, the analytical chiral structures of the effective potentials between $B$ mesons are very useful for the extrapolations of the $B$ meson pair interactions in Lattice QCD calculations.

\section*{Acknowledgments}
B. W is very grateful to Prof. Shi-Lin Zhu for very helpful discussions. B. W also thanks Hao Xu and Rui Chen for some useful discussions. This project is supported by the National Natural Science Foundation of China under Grants No. 11175073 and No. 11705072. This work is also supported by the Fundamental Research Funds for the Central Universities. Xiang Liu is also supported in part by the National Program for Support of Top-notch Young Professionals.
\appendix
\section{Loop integral}\label{LoopIntegral}
\subsection{Definitions of $\mathcal{J}$ functions}\label{LoopIntegral_1}
The various $\mathcal{J}$ functions that appeared in the amplitudes of the previous sections are defined in the following, and they can be obtained by calculating the loop integrals in $D$ dimensions.
\begin{widetext}
\begin{eqnarray}\label{LoopIntJc}
&&i\int\frac{d^Dl\lambda^{4-D}}{(2\pi)^D}\frac{\{1,l^\alpha,l^\alpha l^\beta\}}{l^2-m^2+i\varepsilon}\equiv\left\{\mathcal{J}_0^c,0,g^{\alpha\beta}\mathcal{J}_{21}^c\right\}(m),
\end{eqnarray}
\begin{eqnarray}\label{LoopIntJa}
&&i\int\frac{d^Dl\lambda^{4-D}}{(2\pi)^D}\frac{\{1,l^\alpha,l^\alpha l^\beta,l^\alpha l^\beta l^\gamma\}}{\left(v\cdot l+\omega+i\varepsilon\right)\left(l^2-m^2+i\varepsilon\right)}\equiv\left\{\mathcal{J}_0^a,v^\alpha \mathcal{J}_{11}^a,v^\alpha v^\beta \mathcal{J}_{21}^a+g^{\alpha\beta}\mathcal{J}_{22}^a,(g\vee v)\mathcal{J}_{31}^a+v^\alpha v^\beta v^\gamma \mathcal{J}_{32}^a\right\}(m,\omega),
\end{eqnarray}
\begin{eqnarray}\label{LoopIntJgh}
&&i\int \frac{d^Dl\lambda^{4-D}}{(2\pi)^D}\frac{\{1,l^\alpha,l^\alpha l^\beta,l^\alpha l^\beta l^\gamma\}}{\left(v\cdot l+\omega+i\varepsilon\right)\left[(+/-)v\cdot l+\delta+i\varepsilon\right]\left(l^2-m^2+i\varepsilon\right)}\nonumber\\
&&\quad\equiv\left\{\mathcal{J}_0^{g/h},v^\alpha \mathcal{J}_{11}^{g/h},v^\alpha v^\beta \mathcal{J}_{21}^{g/h}+ g^{\alpha\beta}\mathcal{J}_{22}^{g/h},(g\vee v)\mathcal{J}_{31}^{g/h}+v^\alpha v^\beta v^\gamma \mathcal{J}_{32}^{g/h}\right\}(m,\omega,\delta),
\end{eqnarray}
\begin{eqnarray}\label{LoopIntJF}
&&i\int\frac{d^Dl\lambda^{4-D}}{(2\pi)^D}\frac{\{1,l^\alpha,l^\alpha l^\beta,l^\alpha l^\beta l^\gamma\}}{\left(l^2-m^2+i\varepsilon\right)\left[(l+q)^2-m^2+i\varepsilon\right]}\equiv\left\{\mathcal{J}_0^F,q^\alpha \mathcal{J}_{11}^F,q^\alpha q^\beta \mathcal{J}_{21}^F+g^{\alpha\beta}\mathcal{J}_{22}^F,(g\vee q)\mathcal{J}_{31}^F+q^\alpha q^\beta q^\gamma \mathcal{J}_{32}^F\right\}(m,q),
\end{eqnarray}
\begin{eqnarray}
&&i\int\frac{d^Dl\lambda^{4-D}}{(2\pi)^D}\frac{\{1,l^\alpha,l^\alpha l^\beta,l^\alpha l^\beta l^\gamma,l^\alpha l^\beta l^\gamma l^\delta\}}{\left(v\cdot l+\omega+i\varepsilon\right)\left(l^2-m^2+i\varepsilon\right)\left[(l+q)^2-m^2+i\varepsilon\right]}\equiv\Big\{\mathcal{J}_0^T,q^\alpha \mathcal{J}_{11}^T+v^\alpha \mathcal{J}_{12}^T,g^{\alpha\beta}\mathcal{J}_{21}^T+q^\alpha q^\beta \mathcal{J}_{22}^T+v^\alpha v^\beta \mathcal{J}_{23}^T\nonumber\\
&&\quad+(q\vee v)\mathcal{J}_{24}^T,(g\vee q)\mathcal{J}_{31}^T+q^\alpha q^\beta q^\gamma \mathcal{J}_{32}^T+(q^2\vee v)\mathcal{J}_{33}^T+(g\vee v)\mathcal{J}_{34}^T+(q\vee v^2)\mathcal{J}_{35}^T+v^\alpha v^\beta v^\gamma \mathcal{J}_{36}^T,(g\vee g)\mathcal{J}_{41}^T\nonumber\\
&&\quad+(g\vee q^2)\mathcal{J}_{42}^T+q^\alpha q^\beta q^\gamma q^\delta \mathcal{J}_{43}^T+(g\vee v^2)\mathcal{J}_{44}^T+v^\alpha v^\beta v^\gamma v^\delta \mathcal{J}_{45}^T+(q^3\vee v)\mathcal{J}_{46}^T+(q^2\vee v^2)\mathcal{J}_{47}^T+(q\vee v^3)\mathcal{J}_{48}^T\nonumber\\
&&\quad+(g\vee q\vee v)\mathcal{J}_{49}^T\Big\}(m,\omega,q),
\end{eqnarray}
\begin{eqnarray}\label{LoopIntJRB}
&&i\int\frac{d^Dl\lambda^{4-D}}{(2\pi)^D}\frac{\{1,l^\alpha,l^\alpha l^\beta,l^\alpha l^\beta l^\gamma,l^\alpha l^\beta l^\gamma l^\delta\}}{\left(v\cdot l+\omega+i\varepsilon\right)\left[(+/-)v\cdot l+\delta+i\varepsilon\right]\left(l^2-m^2+i\varepsilon\right)\left[(l+q)^2-m^2+i\varepsilon\right]}\equiv\Big\{\mathcal{J}_0^{R/B},q^\alpha \mathcal{J}_{11}^{R/B}+v^\alpha \mathcal{J}_{12}^{R/B},g^{\alpha\beta}\mathcal{J}_{21}^{R/B}\nonumber\\
&&\quad+q^\alpha q^\beta \mathcal{J}_{22}^{R/B}+v^\alpha v^\beta \mathcal{J}_{23}^{R/B}+(q\vee v)\mathcal{J}_{24}^{R/B},(g\vee q)\mathcal{J}_{31}^{R/B}+q^\alpha q^\beta q^\gamma \mathcal{J}_{32}^{R/B}+(q^2\vee v)\mathcal{J}_{33}^{R/B}+(g\vee v)\mathcal{J}_{34}^{R/B}+(q\vee v^2)\mathcal{J}_{35}^{R/B}\nonumber\\
&&\quad+v^\alpha v^\beta v^\gamma \mathcal{J}_{36}^{R/B},(g\vee g)\mathcal{J}_{41}^{R/B}+(g\vee q^2)\mathcal{J}_{42}^{R/B}+q^\alpha q^\beta q^\gamma q^\delta \mathcal{J}_{43}^{R/B}+(g\vee v^2)\mathcal{J}_{44}^{R/B}+v^\alpha v^\beta v^\gamma v^\delta \mathcal{J}_{45}^{R/B}+(q^3\vee v)\mathcal{J}_{46}^{R/B}\nonumber\\
&&\quad+(q^2\vee v^2)\mathcal{J}_{47}^{R/B}+(q\vee v^3)\mathcal{J}_{48}^{R/B}+(g\vee q\vee v)\mathcal{J}_{49}^{R/B}\Big\}(m,\omega,\delta,q),
\end{eqnarray}
\end{widetext}
where the representation $X\vee Y\vee Z\vee\cdots$ denotes the symmetrized tensor structure of $X^\alpha Y^\beta Z^\gamma\cdots+\cdots$, and can be written as,
\begin{eqnarray}
q \vee v &\equiv& q^\alpha v^\beta+q^\beta v^\alpha, \quad g \vee q \equiv g^{\alpha\beta}q^\gamma+g^{\alpha\gamma}q^\beta+g^{\gamma\beta}q^\alpha,\nonumber
\end{eqnarray}
\begin{eqnarray}
g \vee v \equiv g^{\alpha\beta}v^\gamma+g^{\alpha\gamma}v^\beta+g^{\gamma\beta}v^\alpha,\nonumber
\end{eqnarray}
\begin{eqnarray}
q^2 \vee v &\equiv& q^{\beta} q^{\gamma}v^{\alpha}+q^{\alpha}q^{\gamma} v^{\beta}+q^{\alpha} q^{\beta} v^{\gamma},\nonumber
\end{eqnarray}
\begin{eqnarray}
q \vee v^2 \equiv q^{\gamma} v^{\alpha}v^{\beta}+q^{\beta} v^{\alpha} v^{\gamma}+q^{\alpha} v^{\beta} v^{\gamma},\nonumber
\end{eqnarray}
\begin{eqnarray}
g \vee g &\equiv& g^{\alpha\beta} g^{\gamma\delta}+g^{\alpha\delta} g^{\beta\gamma }+g^{\alpha\gamma} g^{\beta\delta},\nonumber
\end{eqnarray}
\begin{eqnarray}
g \vee q^2 &\equiv& q^{\alpha} q^{\beta} g^{\gamma
\delta}+q^{\alpha} q^{\delta} g^{\beta\gamma} +q^{\alpha}
q^{\gamma}g^{\beta\delta}+q^{\gamma}q^{\delta}g^{\alpha\beta} +q^{\beta} q^{\delta} g^{\alpha\gamma}\nonumber\\
&&+q^{\beta}q^{\gamma}g^{\alpha\delta},\nonumber
\end{eqnarray}
\begin{eqnarray}
g \vee v^2&\equiv& v^{\alpha} v^{\beta}g^{\gamma\delta}+v^{\alpha}v^{\delta} g^{\beta\gamma }+v^{\alpha} v^{\gamma} g^{\beta\delta}
+v^{\gamma} v^{\delta} g^{\alpha\beta}+v^{\beta} v^{\delta} g^{\alpha\gamma}\nonumber\\
&&+v^{\beta} v^{\gamma} g^{\alpha\delta},\nonumber
\end{eqnarray}
\begin{eqnarray}
q^3\vee v &\equiv& q^{\beta} q^{\gamma} q^{\delta}v^{\alpha}+q^{\alpha} q^{\gamma} q^{\delta} v^{\beta}
+q^{\alpha} q^{\beta} q^{\delta} v^{\gamma}+q^{\alpha}q^{\beta} q^{\gamma} v^{\delta},\nonumber
\end{eqnarray}
\begin{eqnarray}
q\vee v^3 \equiv q^{\delta} v^{\alpha} v^{\beta} v^{\gamma}+q^{\gamma}v^{\alpha} v^{\beta} v^{\delta} +q^{\beta} v^{\alpha}v^{\gamma} v^{\delta}+q^{\alpha} v^{\beta} v^{\gamma}v^{\delta },\nonumber
\end{eqnarray}
\begin{eqnarray}
q^2 \vee v^2 &\equiv& q^{\gamma}q^{\delta} v^{\alpha} v^{\beta}+q^{\beta} q^{\delta}v^{\alpha} v^{\gamma} +q^{\alpha} q^{\delta} v^{\beta}v^{\gamma}+q^{\beta} q^{\gamma} v^{\alpha} v^{\delta}\nonumber\\
&&+q^{\alpha} q^{\gamma} v^{\beta}  v^{\delta}+q^{\alpha}q^{\beta} v^{\gamma} v^{\delta},\nonumber
\end{eqnarray}
\begin{eqnarray}
g\vee q \vee v&\equiv& q^{\beta} v^{\alpha} g^{\gamma\delta}+q^{\alpha}v^{\beta} g^{\gamma\delta} +q^{\delta} v^{\alpha} g^{\beta\gamma }+q^{\gamma} v^{\alpha} g^{\beta\delta }\nonumber\\
&&+q^{\alpha}v^{\delta} g^{\beta\gamma} +q^{\alpha} v^{\gamma} g^{\beta\delta}+q^{\delta} v^{\gamma} g^{\alpha\beta}+q^{\delta}v^{\beta} g^{\alpha \gamma}\nonumber\\
&&+q^{\gamma} v^{\delta} g^{\alpha\beta}+q^{\gamma} v^{\beta}g^{\alpha\delta}+q^{\beta} v^{\delta} g^{\alpha\gamma}+q^{\beta} v^{\gamma} g^{\alpha\delta}.\nonumber
\end{eqnarray}

\subsection{Calculations of $\mathcal{J}$ functions}\label{LoopIntegral_2}
For the loop integral $\mathcal{J}_x^c$ and $\mathcal{J}_x^F$ defined in Eq. \eqref{LoopIntJc} and Eq. \eqref{LoopIntJF}, one can easily obtain their results with dimensional regularization \cite{Peskin:1995ev}. Such as
\begin{eqnarray}\label{LoopIntJ0c}
\mathcal{J}_0^c(m)&\equiv&i\int\frac{d^Dl\lambda^{4-D}}{(2\pi)^D}\frac{1}{l^2-m^2+i\varepsilon}\nonumber\\
&=&\frac{\lambda^{4-D}}{(4\pi)^{D/2}}\Gamma\left(1-\frac{D}{2}\right)\left(\frac{1}{m^2}\right)^{1-\frac{D}{2}}\nonumber\\
&=&2m^2L+\frac{m^2}{16\pi^2}\ln\frac{m^2}{\lambda^2},
\end{eqnarray}
where
\begin{eqnarray}
L=\frac{1}{16\pi^2}\left[\frac{1}{D-4}+\frac{1}{2}\left(\gamma_E-1-\ln4\pi\right)\right].
\end{eqnarray}
Here, $\gamma_E$ is the Euler-Mascheroni constant $0.5772157$.

For $\mathcal{J}_0^F$, just one more step, i.e., Feynman parameterization, is needed,
\begin{eqnarray}
\mathcal{J}_0^F(m,q)&\equiv&i\int\frac{d^Dl\lambda^{4-D}}{(2\pi)^D}\frac{1}{\left(l^2-m^2+i\varepsilon\right)\left[(l+q)^2-m^2+i\varepsilon\right]}\nonumber\\
&=&i\int_0^1dx\int\frac{d^Dl\lambda^{4-D}}{(2\pi)^D}\frac{1}{\left[(l+xq)^2-\Delta\right]^2},
\end{eqnarray}
where $\Delta=x(x-1)q^2+m^2-i\varepsilon$.
Performing a shift of integration variables $l\to l-xq$ so that there remain no terms linear in $l$ in the denominator, and repeating the same step as in Eq. \eqref{LoopIntJ0c}, we can get
\begin{eqnarray}
\mathcal{J}_0^F(m,q)=2L+\frac{1}{16\pi^2}\int_0^1dx\left(1+\ln\frac{\Delta}{\lambda^2}\right).
\end{eqnarray}

Next, we outline the deductions of $\mathcal{J}_0^a$ (Eq. \eqref{LoopIntJa}) which serves as a starting point for more complicated loop integrals, such as $\mathcal{J}_x^T$. The calculations are slightly different with the integrals containing only pion propagators. We should use the following Feynman trick to combine the denominators \cite{Scherer:2002tk},
\begin{eqnarray}\label{feytrick}
\frac{1}{AB^n}=2\int_0^\infty dy\frac{n}{(2yA+B)^{n+1}}.
\end{eqnarray}
Setting $A=v\cdot l+\omega+i\varepsilon$, $B=l^2-m^2+i\varepsilon$, and $n=1$, we get
\begin{eqnarray}
\mathcal{J}_0^a=2i\int_0^\infty dy\int\frac{d^Dl\lambda^{4-D}}{(2\pi)^D}\frac{1}{\left(l^2+2yv\cdot l+2y\omega-m^2+i\varepsilon\right)^2}.\nonumber
\end{eqnarray}
The terms linear in $l$ and $y$ in above equation can be eliminated via shifting the integration variables: (1) performing a shift $l\to l-yv$, and making use of $v^2=1$, (2) shifting the integration variable $y\to y+\omega$. Finally, we obtain
\begin{eqnarray}
\mathcal{J}_0^a=2i\int_{-\omega}^\infty dy\int\frac{d^Dl\lambda^{4-D}}{(2\pi)^D}\frac{1}{\left(l^2-y^2-m^2+\omega^2+i\varepsilon\right)^2}.\nonumber\\
\end{eqnarray}
We should notice that the range of $y$ integration now changes to $[-\omega,\infty)$, and we split this range into two parts, $[-\omega,0)$ and $[0,\infty)$. The first part can be easily obtained with dimensional regularization, which reads
\begin{eqnarray}
4\omega L+\frac{1}{8\pi^2}\int_{-\omega}^0dy\left(1+\ln\frac{\tilde{\Delta}}{\lambda^2}\right),
\end{eqnarray}
where $\tilde{\Delta}=y^2+m^2-\omega^2$.
For the second part, we first use $\beta$ function to integrate out the $y$ variables, then use dimensional regularization to integrate out $l$, and thus the result reads
\begin{eqnarray}
\frac{1}{8\pi}\sqrt{m^2-\omega^2}.
\end{eqnarray}
Finally, we can get the full form of $\mathcal{J}_0^a$,
\begin{eqnarray}
\mathcal{J}_0^a=4\omega L+\frac{1}{8\pi^2}\int_{-\omega}^0dy\left(1+\ln\frac{\tilde{\Delta}}{\lambda^2}\right)+\frac{1}{8\pi}\sqrt{m^2-\omega^2}.\nonumber\\
\end{eqnarray}
If one wants to obtain the results of $\mathcal{J}_{11}^a$, $\mathcal{J}_{21}^a$, $\mathcal{J}_{22}^a\cdots$, the shifts on integration variables $l$ and $y$ made above should be kept in mind, because the integration variables that appeared in the numerator also have to be shifted accordingly.

The calculations of $\mathcal{J}_{x}^T$ are very similar to $\mathcal{J}_{x}^a$ as illustrated above, the concrete procedures are:

(1) Using Feynman's trick to combine the denominators of two pions propagators, then shifting $l\to l-xq$, and making use of $v\cdot q=0$.

(2) Using Eq. \eqref{feytrick} to combine the denominators of heavy meson and the combined pion propagators.

(3) Then just repeating the subsequent procedures what we have done for calculating $\mathcal{J}_{x}^a$.

At last, we list the expressions of the used $\mathcal{J}$ functions in our calculations, and these functions are calculated numerically in this work.
\begin{widetext}
\begin{eqnarray}
\mathcal{J}_{22}^a(m,\omega)=2\omega\left(m^2-\frac{2}{3}\omega^2\right)L+\frac{1}{16\pi^2}\int_{-\omega}^0\tilde{\Delta}\ln\frac{\tilde{\Delta}}{\lambda^2}dy+\frac{1}{24\pi}\tilde{A}^{3/2},~\textrm{where $\tilde{\Delta}=y^2+m^2-\omega^2,~ \tilde{A}=m^2-\omega^2$}.
\end{eqnarray}
\begin{eqnarray}
\mathcal{J}_{22}^g(m,\omega,\delta)=\left\{\begin{array}{ll}
\frac{1}{\delta-\omega}\left[\mathcal{J}_{22}^a(m,\omega)-\mathcal{J}_{22}^a(m,\delta)\right] & \textrm{if $\omega\neq\delta$}\\
-\frac{\partial}{\partial x}\mathcal{J}_{22}^a(m,x)\Big|_{x\to\omega(\textrm{or}\ \delta)} & \textrm{if $\omega=\delta$}
\end{array} \right..
\end{eqnarray}
\begin{eqnarray}
\mathcal{J}_{22}^F(m,q)=\left(m^2-\frac{q^2}{6}\right)L+\frac{1}{32\pi^2}\int_0^1\bar{\Delta}\ln\frac{\bar{\Delta}}{\lambda^2}dx,~\textrm{where $\bar{\Delta}=x(x-1)q^2+m^2$}.
\end{eqnarray}
\begin{eqnarray}
\mathcal{J}_{21}^T(m,\omega,q)=2\omega L+\frac{1}{16\pi^2}\int_0^1dx\int_{-\omega}^0\left(1+\ln\frac{\Delta}{\lambda^2}\right)dy+\frac{1}{16\pi}\int_0^1A^{1/2}dx,
\end{eqnarray}
\begin{eqnarray}
\mathcal{J}_{22}^T(m,\omega,q)=\frac{1}{8\pi^2}\int_0^1dx\int_{-\omega}^0\frac{x^2}{\Delta}dy+\frac{1}{16\pi}\int_0^1x^2A^{-1/2}dx,
\end{eqnarray}
\begin{eqnarray}
\mathcal{J}_{24}^T(m,\omega,q)=-L+\frac{1}{8\pi^2}\int_0^1dx\int_{-\omega}^0\frac{x(y+\omega)}{\Delta}dy-\frac{1}{16\pi^2}\int_0^1x\left(1+\ln\frac{A}{\lambda^2}\right)dx+\frac{\omega}{16\pi}\int_0^1x A^{-1/2}dx,
\end{eqnarray}
\begin{eqnarray}
\mathcal{J}_{31}^T(m,\omega,q)=-\omega L-\frac{1}{16\pi^2}\int_0^1dx\int_{-\omega}^0x\left(1+\ln\frac{\Delta}{\lambda^2}\right)dy-\frac{1}{16\pi}\int_0^1xA^{1/2}dx,
\end{eqnarray}
\begin{eqnarray}
\mathcal{J}_{32}^T(m,\omega,q)=-\frac{1}{8\pi^2}\int_0^1dx\int_{-\omega}^0\frac{x^3}{\Delta}dy-\frac{1}{16\pi}\int_0^1x^3A^{-1/2}dx,
\end{eqnarray}
\begin{eqnarray}
\mathcal{J}_{33}^T(m,\omega,q)=\frac{2}{3}L-\frac{1}{8\pi^2}\int_0^1dx\int_{-\omega}^0\frac{x^2(y+\omega)}{\Delta}dy+\frac{1}{16\pi^2}\int_0^1x^2\left(1+\ln\frac{A}{\lambda^2}\right)dx-\frac{\omega}{16\pi}\int_0^1x^2 A^{-1/2}dx,
\end{eqnarray}
\begin{eqnarray}
\mathcal{J}_{34}^T(m,\omega,q)=\left(m^2-\frac{q^2}{6}-2\omega^2\right)L-\frac{1}{16\pi^2}\int_0^1dx\int_{-\omega}^0(y+\omega)\left(1+\ln\frac{\Delta}{\lambda^2}\right)dy-\frac{\omega}{16\pi}\int_0^1 A^{1/2}dx+\frac{1}{32\pi^2}\int_0^1A\ln\frac{A}{\lambda^2}dx,
\end{eqnarray}
\begin{eqnarray}
\mathcal{J}_{36}^T(m,\omega,q)&=&\left(-2m^2+\frac{q^2}{3}+8\omega^2\right)L-\frac{1}{8\pi^2}\int_0^1dx\int_{-\omega}^0\frac{(y+\omega)^3}{\Delta}dy-\frac{\omega^3}{16\pi}\int_0^1xA^{-1/2}dx\nonumber\\
&&-\frac{1}{16\pi^2}\int_0^1xA\ln\frac{A}{\lambda^2}dx+\frac{3\omega}{16\pi}\int_0^1x A^{1/2}
+\frac{3\omega^2}{16\pi^2}\int_0^1x\left(1+\ln\frac{A}{\lambda^2}\right)dx,
\end{eqnarray}
\begin{eqnarray}
\mathcal{J}_{41}^T(m,\omega,q)=\omega\left(m^2-\frac{q^2}{6}-\frac{2}{3}\omega^2\right)L+\frac{1}{32\pi^2}\int_0^1dx\int_{-\omega}^0\Delta\ln\frac{\Delta}{\lambda^2}dy+\frac{1}{48\pi}\int_0^1A^{3/2}dx,
\end{eqnarray}
\begin{eqnarray}
\mathcal{J}_{42}^T(m,\omega,q)=\frac{2}{3}\omega L+\frac{1}{16\pi^2}\int_0^1dx\int_{-\omega}^0x^2\left(1+\ln\frac{\Delta}{\lambda^2}\right)dy+\frac{1}{16\pi}\int_0^1x^2A^{1/2}dx,
\end{eqnarray}
\begin{eqnarray}
\mathcal{J}_{43}^T(m,\omega,q)=\frac{1}{8\pi^2}\int_0^1dx\int_{-\omega}^0\frac{x^4}{\Delta}dy+\frac{1}{16\pi}\int_0^1x^4A^{-1/2}dx,
\end{eqnarray}
where $\Delta=y^2+A$, $A=x(x-1)q^2+m^2-\omega^2$.
\begin{eqnarray}
\mathcal{J}_{x}^R(m,\omega,\delta,q)=\left\{\begin{array}{ll}
\frac{1}{\delta-\omega}\left[\mathcal{J}_{x}^T(m,\omega,q)-\mathcal{J}_{x}^T(m,\delta,q)\right] & \textrm{if $\omega\neq\delta$}\\
-\frac{\partial}{\partial x}\mathcal{J}_{x}^T(m,x,q)\Big|_{x\to\omega(\textrm{or}\ \delta)} & \textrm{if $\omega=\delta$}
\end{array} \right..
\end{eqnarray}
\end{widetext}
\subsection{Removing the 2PR contributions from $\mathcal{J}_x^h$ and $\mathcal{J}_x^B$ functions}\label{LoopIntegral_3}
We can not directly calculate the full expressions of $\mathcal{J}_x^h$ and $\mathcal{J}_x^B$ functions as defined in Eq. \eqref{LoopIntJgh} and Eq. \eqref{LoopIntJRB}, respectively, because they contain both 2PR and 2PI contributions. The 2PR part will be recovered when the effective potential is inserted into Schr\"odinger equation, and thus this part has to be removed when we calculate the potentials. In the following part, we illustrate how to subtract the 2PR contributions by utilizing the Weinberg's formalism (one can find a more convenient approach in Ref. \cite{Zhu:2004vw}), and take the $\mathcal{J}_{21}^B$ as an example.
\begin{widetext}
\begin{eqnarray}\label{J22B_1}
&&i\int\frac{d^Dl\lambda^{4-D}}{(2\pi)^D}\frac{l^\alpha l^\beta}{\left(v\cdot l+\frac{\vec{p}^{2}-\vec{l}^{2}}{2M_1}+\omega+i\varepsilon\right)\left(-v\cdot l+\frac{\vec{p}^2-\vec{l}^2}{2M_2}+\delta+i\varepsilon\right)\left(l^2-m^2+i\varepsilon\right)\left[(l+q)^2-m^2+i\varepsilon\right]}\nonumber\\
&&=i\int_0^1dx\int\frac{d^Dl\lambda^{4-D}}{(2\pi)^D}\frac{(l-xq)^\alpha (l-xq)^\beta}{\left(v\cdot l+\frac{\vec{p}^2-\vec{l}^2}{2M_1}+\omega+i\varepsilon\right)\left(-v\cdot l+\frac{\vec{p}^2-\vec{l}^2}{2M_2}+\delta+i\varepsilon\right)\left(l^2-\mathcal{M}^2+i\varepsilon\right)^2},
\end{eqnarray}
\end{widetext}
where $\mathcal{M}^2=x(x-1)q^2+m^2$, and we have used $v\cdot q=0$. The kinetic term $(\vec{p}^2-\vec{l}^2)/2M_i$ ($M_i$ is the mass of heavy meson) is included in the denominators of heavy mesons propagators for avoiding the pinch singularity when $\omega=\delta=0$  (cf. discussions in Ref. \cite{Weinberg:1991um}). In Eq. \eqref{J22B_1}, the pion poles contribute, and there is also a contribution from the double heavy meson poles. The contribution from the latter one is enhanced in the $1/M_B$ expansion and gives the result of the iterated $\mathcal{O}(\epsilon^0)$ OPE diagram, which is just the 2PR part that we need to eliminate. In other words, only the contribution from the pion poles accounts for the effective potential. The pion poles are located at
\begin{eqnarray}
l_1^0=-\sqrt{\vec{l}^2+\mathcal{M}^2}+i\varepsilon,\quad l_2^0=\sqrt{\vec{l}^2+\mathcal{M}^2}-i\varepsilon.\nonumber
\end{eqnarray}

Picking out $l^\alpha l^\beta$ term in the second line of Eq. \eqref{J22B_1} and substituting it with $g^{\alpha\beta}(l_0^2-\vec{l}^2)/D$, then we can get the scalar function $\mathcal{J}_{21}^B$ defined in Eq. \eqref{LoopIntJRB}. Closing the $l_0$ contour integral in the lower half-plane with the pole of interest located at $l_0=\sqrt{\vec{l}^2+\mathcal{M}^2}-i\varepsilon$, the 2PI part of $\mathcal{J}_{21}^B$ can be easily obtained by using residue theorem and dimensional regularization in $D-1$ dimensions, the corresponding result is written as
\begin{eqnarray}\label{J21B_2}
\left[\mathcal{J}_{21}^B\right]_{\text{2PI}}&=&\left(\frac{1}{4D}\right)\int_0^1dx\int\frac{d^{D-1}l\lambda^{4-D}}{(2\pi)^{D-1}}\Bigg[\frac{1}{(\boldsymbol{l}^2+\mathcal{M}^2-i\varepsilon)^{3/2}}\nonumber\\
&&-\frac{3\boldsymbol{l}^2}{(\boldsymbol{l}^2+\mathcal{M}^2-i\varepsilon)^{5/2}}\Bigg]\nonumber\\
&=&L+\frac{1}{64\pi^2}\int_0^1dx\left(3+2\ln\frac{\mathcal{M}^2}{\lambda^2}\right).
\end{eqnarray}

In the following, we write out the used 2PI parts of $\mathcal{J}_x^h$ and $\mathcal{J}_x^B$ functions in this calculations.
\begin{eqnarray}
\left[\mathcal{J}_{22}^h\right]_{\text{2PI}}=m^2L+\frac{m^2}{64\pi^2}\left(1+2\ln\frac{m^2}{\lambda^2}\right),
\end{eqnarray}
\begin{eqnarray}
\left[\mathcal{J}_{22}^B\right]_{\text{2PI}}=\frac{1}{8\pi^2}\int_0^1\frac{x^2}{\mathcal{M}^2}dx,
\end{eqnarray}
\begin{eqnarray}
\left[\mathcal{J}_{31}^B\right]_{\text{2PI}}=-\frac{1}{2}L-\frac{1}{64\pi^2}\int_0^1x\left(3+2\ln\frac{\mathcal{M}^2}{\lambda^2}\right)dx,
\end{eqnarray}
\begin{eqnarray}
\left[\mathcal{J}_{32}^B\right]_{\text{2PI}}=-\frac{1}{8\pi^2}\int_0^1\frac{x^3}{\mathcal{M}^2}dx,
\end{eqnarray}
\begin{eqnarray}
\left[\mathcal{J}_{41}^B\right]_{\text{2PI}}&=&\frac{1}{3}\left(m^2-\frac{q^2}{6}\right)L+\frac{1}{288\pi^2}\int_0^1\mathcal{M}^2\Bigg(2\nonumber\\
&&+3\ln\frac{\mathcal{M}^2}{\lambda^2}\Bigg)dx,
\end{eqnarray}
\begin{eqnarray}
\left[\mathcal{J}_{42}^B\right]_{\text{2PI}}=\frac{1}{3}L+\frac{1}{64\pi^2}\int_0^1x^2\left(3+2\ln\frac{\Delta}{\lambda^2}\right)dx,
\end{eqnarray}
\begin{eqnarray}
\left[\mathcal{J}_{43}^B\right]_{\text{2PI}}=\frac{1}{8\pi^2}\int_0^1\frac{x^4}{\mathcal{M}^2}dx.
\end{eqnarray}

\section{The amplitudes of $0(1^+)$ $\bar{B}\bar{B}^\ast$ state} \label{Amp_BBastI0}

The $I=0$ amplitudes that come from Fig. \ref{BBast_ZeroOrder} are:
\begin{eqnarray}\label{BBast_XHI0}
\mathcal{Y}^{X_{2.1}}_{I=0}&=&8\left[-D_a-D_b+3(E_a+E_b)\right](\epsilon_2\cdot\epsilon_4^\ast),\\
\mathcal{Y}^{H_{2.1}}_{I=0}&=&-\frac{3g^2}{f^2}\frac{(q\cdot\epsilon_2)(q\cdot\epsilon_4^\ast)}{q^2-m_\pi^2}.
\end{eqnarray}

The $I=0$ amplitudes that come from Fig. \ref{BBast_Contact_Corrections} are:
\begin{eqnarray}
\mathcal{Y}^{g_{2.1}}_{I=0}&=&\mathcal{C}^{g_{2.1}}\frac{g^2}{f^2}(\epsilon_2\cdot\epsilon_4^\ast)\left\{\mathcal{J}_{22}^g\right\}_r(m_\pi,\mathcal{E},\mathcal{E}),
\end{eqnarray}
\begin{eqnarray}
\mathcal{Y}^{g_{2.2}}_{I=0}=\frac{g^2}{f^2}(\epsilon_2\cdot\epsilon_4^\ast)\bigg\{\mathcal{C}^{g_{2.2}}\mathcal{J}_{22}^g\bigg\}_r(m_\pi,\mathcal{E}-\Delta,\mathcal{E}-\Delta),
\end{eqnarray}
\begin{eqnarray}
\mathcal{Y}^{g_{2.3}}_{I=0}&=&\mathcal{C}^{g_{2.3}}\frac{g^2}{f^2}(\epsilon_2\cdot\epsilon_4^\ast)\left\{\mathcal{J}_{22}^g\right\}_r(m_\pi,\mathcal{E}+\Delta,\mathcal{E}-\Delta),
\end{eqnarray}
\begin{eqnarray}
\mathcal{Y}^{g_{2.4}}_{I=0}&=&0,
\end{eqnarray}
\begin{eqnarray}
\mathcal{Y}^{g_{2.7}}_{I=0}&=&\mathcal{C}^{g_{2.7}}\frac{g^2}{f^2}(\epsilon_2\cdot\epsilon_4^\ast)\left\{\mathcal{J}_{22}^g\right\}_r(m_\pi,\mathcal{E}+\Delta,\mathcal{E}+\Delta),
\end{eqnarray}
\begin{eqnarray}
\mathcal{Y}^{h_{2.1}}_{I=0}&=&\mathcal{C}^{h_{2.1}}\frac{g^2}{f^2}(\epsilon_2\cdot\epsilon_4^\ast)\left\{\mathcal{J}_{22}^h\right\}_r(m_\pi,\mathcal{E}+\Delta,\mathcal{E}-\Delta),
\end{eqnarray}
\begin{eqnarray}
\mathcal{Y}^{h_{2.2}}_{I=0}&=&\mathcal{C}^{h_{2.2}}\frac{g^2}{f^2}(\epsilon_2\cdot\epsilon_4^\ast)\left\{\mathcal{J}_{22}^h\right\}_r(m_\pi,\mathcal{E},\mathcal{E}-\Delta),
\end{eqnarray}
\begin{eqnarray}
\mathcal{Y}^{z_{2.1}}_{I=0}&=&\mathcal{C}^{z_{2.1}}\frac{g^2}{f^2}(\epsilon_2\cdot\epsilon_4^\ast)\left\{\frac{\partial}{\partial x}\mathcal{J}_{22}^a\right\}_r(m_\pi,x)\bigg|_{x\to \mathcal{E}},
\end{eqnarray}
\begin{eqnarray}
\mathcal{Y}^{z_{2.2}}_{I=0}&=&\mathcal{C}^{z_{2.2}}\frac{g^2}{f^2}(\epsilon_2\cdot\epsilon_4^\ast)\left\{\frac{\partial}{\partial x}\mathcal{J}_{22}^a\right\}_r(m_\pi,x)\bigg|_{x\to\mathcal{E}+\Delta},
\end{eqnarray}
\begin{eqnarray}
\mathcal{Y}^{z_{2.3}}_{I=0}=\mathcal{C}^{z_{2.3}}\frac{g^2}{f^2}(\epsilon_2\cdot\epsilon_4^\ast)\left\{(D-1)\frac{\partial}{\partial x}\mathcal{J}_{22}^a\right\}_r(m_\pi,x)\bigg|_{x\to\mathcal{E}-\Delta},\nonumber\\
\end{eqnarray}
where
\begin{eqnarray}
\mathcal{C}^{g_{2.1}}&=&12(D_a-D_b+E_a-E_b),\nonumber\\
\mathcal{C}^{g_{2.2}}&=&6\Big[D(D_a-D_b+E_a-E_b)+2(D_b+E_b)\Big],\nonumber\\
\mathcal{C}^{g_{2.3}}&=&12(D_b+E_b),\quad\mathcal{C}^{g_{2.7}}=12(D_a+E_a),\nonumber\\
\mathcal{C}^{h_{2.1}}&=&6(D_a+D_b-3E_a-3E_b),\quad\mathcal{C}^{h_{2.2}}=24(D_b-3E_b),\nonumber\\
\mathcal{C}^{z_{2.1}}&=&12(D_a+D_b-3E_a-3E_b),\nonumber\\
\mathcal{C}^{z_{2.2}}&=&\mathcal{C}^{z_{2.3}}=\frac{1}{2}\mathcal{C}^{z_{2.1}}.
\end{eqnarray}
Note that the coefficients $\mathcal{C}^{x}$ appeared here should not be confused with those for the $I=1$ channels.

The amplitudes of diagrams $g_{2.5}$, $g_{2.6}$, $h_{2.3}$, and $h_{2.4}$ can be obtained by the relations
\begin{eqnarray}
\mathcal{Y}^{g_{2.5}}_{I=0}&=&\mathcal{Y}^{g_{2.3}}_{I=0},~~\mathcal{Y}^{g_{2.6}}_{I=0}=\mathcal{Y}^{g_{2.4}}_{I=0},~~
\mathcal{Y}^{h_{2.3}}_{I=0}=\mathcal{Y}^{h_{2.1}}_{I=0},~~\mathcal{Y}^{h_{2.4}}_{I=0}=\mathcal{Y}^{h_{2.2}}_{I=0}.\nonumber
\end{eqnarray}

The $I=0$ amplitudes that come from Fig. \ref{BBast_OPE_Corrections} are:
\begin{eqnarray}
\mathcal{Y}^{p_{2.1}}_{I=0}=-\frac{3g^2}{f^2}\frac{(q\cdot\epsilon_2)(q\cdot\epsilon_4^\ast)}{q^2-m_\pi^2}\Sigma(m_\pi),
\end{eqnarray}
\begin{eqnarray}
\mathcal{Y}^{c_{2.1}}_{I=0}=\frac{g^2}{f^4}\frac{(q\cdot\epsilon_2)(q\cdot\epsilon_4^\ast)}{q^2-m_\pi^2}\left\{\mathcal{J}_0^c\right\}_r(m_\pi),
\end{eqnarray}
\begin{eqnarray}
\mathcal{Y}^{c_{2.3}}_{I=0}=-\frac{3g^4}{4f^4}\frac{(q\cdot\epsilon_2)(q\cdot\epsilon_4^\ast)}{q^2-m_\pi^2}\left\{\mathcal{J}_{22}^g\right\}_r(m_\pi,\mathcal{E}+\Delta,\mathcal{E}-\Delta),
\end{eqnarray}
\begin{eqnarray}
\mathcal{Y}^{c_{2.4}}_{I=0}=\frac{3g^4}{2f^4}\frac{(q\cdot\epsilon_2)(q\cdot\epsilon_4^\ast)}{q^2-m_\pi^2}\left\{\mathcal{J}_{22}^g\right\}_r(m_\pi,\mathcal{E},\mathcal{E}-\Delta),
\end{eqnarray}

\begin{eqnarray}
\mathcal{Y}^{c_{2.7}}_{I=0}&=&\frac{9g^4}{4f^4}\frac{(q\cdot\epsilon_2)(q\cdot\epsilon_4^\ast)}{q^2-m_\pi^2}\left\{(D-1)\frac{\partial}{\partial x}\mathcal{J}_{22}^a\right\}_r(m_\pi,x)\bigg|_{x\to\mathcal{E}-\Delta},\nonumber\\
\end{eqnarray}
\begin{eqnarray}
\mathcal{Y}^{c_{2.8}}_{I=0}=\frac{9g^4}{4f^4}\frac{(q\cdot\epsilon_2)(q\cdot\epsilon_4^\ast)}{q^2-m_\pi^2}\left\{\frac{\partial}{\partial x}\mathcal{J}_{22}^a\right\}_r(m_\pi,x)\bigg|_{x\to\mathcal{E}+\Delta},
\end{eqnarray}
\begin{eqnarray}
\mathcal{Y}^{c_{2.9}}_{I=0}=\frac{9g^4}{2f^4}\frac{(q\cdot\epsilon_2)(q\cdot\epsilon_4^\ast)}{q^2-m_\pi^2}\left\{\frac{\partial}{\partial x}\mathcal{J}_{22}^a\right\}_r(m_\pi,x)\bigg|_{x\to\mathcal{E}},
\end{eqnarray}
\begin{eqnarray}
\mathcal{Y}^{c_{2.10}}_{I=0}=\mathcal{Y}^{c_{2.11}}_{I=0}=0.
\end{eqnarray}
The amplitudes of diagrams $c_{2.2}$, $c_{2.5}$, and $c_{2.6}$ can be obtained by the relations
$$\mathcal{Y}^{c_{2.2}}_{I=0}=\mathcal{Y}^{c_{2.1}}_{I=0},~~~~~\mathcal{Y}^{c_{2.5}}_{I=0}=\mathcal{Y}^{c_{2.3}}_{I=0},~~~~~\mathcal{Y}^{c_{2.6}}_{I=0}=\mathcal{Y}^{c_{2.4}}_{I=0}.$$

The $I=0$ amplitudes that come from Fig. \ref{BBast_TPE} are:
\begin{widetext}
\begin{eqnarray}
\mathcal{Y}^{F_{2.1}}_{I=0}&=&-\frac{3}{f^4}(\epsilon_2\cdot\epsilon_4^\ast)\left\{\mathcal{J}_{22}^F\right\}_r(m_\pi,q),
\end{eqnarray}
\begin{eqnarray}
\mathcal{Y}^{T_{2.1}}_{I=0}&=&-\frac{3g^2}{f^4}\bigg\{(\epsilon_2\cdot\epsilon_4^\ast)\mathcal{J}_{34}^T+(q\cdot\epsilon_2)(q\cdot\epsilon_4^\ast)\left(\mathcal{J}_{24}^T+\mathcal{J}_{33}^T\right)\bigg\}_r\left(m_\pi,\mathcal{E}+\Delta,q\right),
\end{eqnarray}
\begin{eqnarray}
\mathcal{Y}^{T_{2.2}}_{I=0}&=&-\frac{3g^2}{f^4}\Bigg\{(\epsilon_2\cdot\epsilon_4^\ast)\bigg[2\mathcal{J}_{34}^T-\vec{q}^2\left(\mathcal{J}_{24}^T+\mathcal{J}_{33}^T\right)\bigg]-(q\cdot\epsilon_2)(q\cdot\epsilon_4^\ast)\left(\mathcal{J}_{24}^T+\mathcal{J}_{33}^T\right)\Bigg\}_r(m_\pi,\mathcal{E},q),
\end{eqnarray}
\begin{eqnarray}
\mathcal{Y}^{T_{2.3}}_{I=0}&=&-\frac{3g^2}{f^4}\Bigg\{(\epsilon_2\cdot\epsilon_4^\ast)\bigg[(D-1)\mathcal{J}_{34}^T-\vec{q}^2\left(\mathcal{J}_{24}^T+\mathcal{J}_{33}^T\right)\bigg]\Bigg\}_r(m_\pi,\mathcal{E}-\Delta,q),
\end{eqnarray}
\begin{eqnarray}
\mathcal{Y}^{B_{2.1}}_{I=0}&=&\frac{9g^4}{4f^4}\Bigg\{(\epsilon_2\cdot\epsilon_4^\ast)\bigg[-\vec{q}^2\left[(D+5)\left(\mathcal{J}_{31}^B+\mathcal{J}_{42}^B\right)-\vec{q}^2\left(\mathcal{J}_{22}^B+2\mathcal{J}_{32}^B+\mathcal{J}_{43}^B\right)+\mathcal{J}_{21}^B\right]+2(D+1)\mathcal{J}_{41}^B\bigg]\nonumber\\
&&-(q\cdot\epsilon_2)(q\cdot\epsilon_4^\ast)\bigg[(D+3)\left(\mathcal{J}_{31}^B+\mathcal{J}_{42}^B\right)-\vec{q}^2\left(\mathcal{J}_{22}^B+2\mathcal{J}_{32}^B+\mathcal{J}_{43}^B\right)+\mathcal{J}_{21}^B\bigg]\Bigg\}_r(m_\pi,\mathcal{E}-\Delta,\mathcal{E},q),
\end{eqnarray}
\begin{eqnarray}
\mathcal{Y}^{B_{2.2}}_{I=0}&=&\frac{9g^4}{4f^4}\Bigg\{(\epsilon_2\cdot\epsilon_4^\ast)\bigg[(D+1)\mathcal{J}_{41}^B-\vec{q}^2\left(\mathcal{J}_{31}^B+\mathcal{J}_{42}^B\right)\bigg]+(q\cdot\epsilon_2)(q\cdot\epsilon_4^\ast)\bigg[(D+3)\left(\mathcal{J}_{31}^B+\mathcal{J}_{42}^B\right)+\mathcal{J}_{21}^B\nonumber\\
&&-\vec{q}^2\left(\mathcal{J}_{22}^B+2\mathcal{J}_{32}^B+\mathcal{J}_{43}^B\right)\bigg]\Bigg\}_r(m_\pi,\mathcal{E}+\Delta,\mathcal{E}-\Delta,q),
\end{eqnarray}
\begin{eqnarray}
\mathcal{Y}^{B_{2.3}}_{I=0}&=&-\frac{9g^4}{4f^4}\bigg\{\Big[(\epsilon_2\cdot\epsilon_4^\ast)\vec{q}^2+(q\cdot\epsilon_2)(q\cdot\epsilon_4^\ast)\Big]\mathcal{J}_{21}^B\bigg\}_r(m_\pi,\mathcal{E}-\Delta,\mathcal{E},q).
\end{eqnarray}
When $q_0=0$, the amplitudes of the diagrams $R_{2.1}$$\sim$$R_{2.3}$ can be obtained by the relations
\begin{eqnarray}
\mathcal{Y}^{R_{2.1}}_{I=0}&=&-\frac{1}{3}\mathcal{Y}^{B_{2.1}}_{I=0}\Big|_{\mathcal{J}_{x}^B\to\mathcal{J}_{x}^R},\quad\quad\mathcal{Y}^{R_{2.2}}_{I=0}=-\frac{1}{3}\mathcal{Y}^{B_{2.2}}_{I=0}\Big|_{\mathcal{J}_{x}^B\to\mathcal{J}_{x}^R},\quad\quad
\mathcal{Y}^{R_{2.3}}_{I=0}=\frac{1}{3}\mathcal{Y}^{B_{2.3}}_{I=0}\Big|_{\mathcal{J}_{x}^B\to\mathcal{J}_{x}^R}.
\end{eqnarray}
\end{widetext}

\section{The amplitudes of $0(1^+)$ $\bar{B}^\ast\bar{B}^\ast$ state} \label{Amp_BastBastI0}
The $I=0$ amplitudes that come from Fig. \ref{BastBast_ZeroOrder} are:
\begin{eqnarray}\label{BastBast_XHI0}
\mathcal{Y}^{X_{3.1}}_{I=0}&=&4\left[D_a+D_b-3(E_a+E_b)\right]\left(\varepsilon_1-\varepsilon_2\right),\\
\mathcal{Y}^{H_{3.1}}_{I=0}&=&\frac{3g^2}{f^2}\frac{\mathcal{G}(q,\epsilon_1,\epsilon_2,\epsilon_3^\ast,\epsilon_4^\ast)}{q^2-m_\pi^2}.
\end{eqnarray}

The $I=0$ amplitudes that come from Fig. \ref{BastBast_Contact_Corrections} are:
\begin{eqnarray}
\mathcal{Y}^{g_{3.1}}_{I=0}=\mathcal{C}^{g_{3.1}}\frac{g^2}{f^2}\left(\varepsilon_1-\varepsilon_2\right)\left\{\mathcal{J}_{22}^g\right\}_r(m_\pi,\mathcal{E},\mathcal{E}),
\end{eqnarray}
\begin{eqnarray}
\mathcal{Y}^{g_{3.2}}_{I=0}=\mathcal{C}^{g_{3.2}}\frac{g^2}{f^2}\left(\varepsilon_1-\varepsilon_2\right)\left\{\mathcal{J}_{22}^g\right\}_r(m_\pi,\mathcal{E}+\Delta,\mathcal{E}+\Delta),
\end{eqnarray}
\begin{eqnarray}
\mathcal{Y}^{g_{3.3}}_{I=0}=0,
\end{eqnarray}
\begin{eqnarray}
\mathcal{Y}^{h_{3.1}}_{I=0}=\mathcal{C}^{h_{3.1}}\frac{g^2}{f^2}\left(\varepsilon_1-\varepsilon_2\right)\left\{\mathcal{J}_{22}^h\right\}_r(m_\pi,\mathcal{E},\mathcal{E}),
\end{eqnarray}
\begin{eqnarray}
\mathcal{Y}^{h_{3.2}}_{I=0}=0,
\end{eqnarray}
\begin{eqnarray}
\mathcal{Y}^{h_{3.3}}_{I=0}=\mathcal{C}^{h_{3.3}}\frac{g^2}{f^2}\left(\varepsilon_1-\varepsilon_2\right)\left\{\mathcal{J}_{22}^h\right\}_r(m_\pi,\mathcal{E}+\Delta,\mathcal{E}),
\end{eqnarray}
\begin{eqnarray}
\mathcal{Y}^{z_{3.1}}_{I=0}=\mathcal{C}^{z_{3.1}}\frac{g^2}{f^2}\left(\varepsilon_1-\varepsilon_2\right)\left\{\frac{\partial}{\partial x}\mathcal{J}_{22}^a\right\}_r(m_\pi,x)\bigg|_{x\to\mathcal{E}},
\end{eqnarray}
\begin{eqnarray}
\mathcal{Y}^{z_{3.2}}_{I=0}=\mathcal{C}^{z_{3.2}}\frac{g^2}{f^2}\left(\varepsilon_1-\varepsilon_2\right)\left\{\frac{\partial}{\partial x}\mathcal{J}_{22}^a\right\}_r(m_\pi,x)\bigg|_{x\to \mathcal{E}+\Delta},
\end{eqnarray}
where
\begin{eqnarray}
\mathcal{C}^{g_{3.1}}&=&-12(3D_a+D_b+3E_a+E_b),\nonumber\\
\mathcal{C}^{g_{3.2}}&=&-12(D_a-D_b+E_a-E_b),\nonumber\\
\mathcal{C}^{h_{3.1}}&=&-6(D_a+D_b-3E_a-3E_b),~\mathcal{C}^{h_{3.3}}=-24(D_b-3E_b),\nonumber\\
\mathcal{C}^{z_{3.1}}&=&-24(D_a+D_b-3E_a-3E_b),\quad\mathcal{C}^{z_{3.2}}=\frac{1}{2}\mathcal{C}^{z_{3.1}},
\end{eqnarray}
and the amplitudes of diagrams $g_{3.4}$, $h_{3.4}$, $h_{3.5}$, and $h_{3.6}$ can be obtained by the relations
\begin{eqnarray}
\mathcal{Y}^{g_{3.4}}_{I=0}=\mathcal{Y}^{g_{3.3}}_{I=0},~~\mathcal{Y}^{h_{3.4}}_{I=0}=\mathcal{Y}^{h_{3.1}}_{I=0},~~ \mathcal{Y}^{h_{3.5}}_{I=0}=\mathcal{Y}^{h_{3.2}}_{I=0},~~\mathcal{Y}^{h_{3.6}}_{I=0}=\mathcal{Y}^{h_{3.3}}_{I=0}.\nonumber
\end{eqnarray}

The $I=0$ amplitudes that come from Fig. \ref{BastBast_OPE_Corrections} are:
\begin{eqnarray}
\mathcal{Y}^{p_{3.1}}_{I=0}&=&\frac{3g^2}{f^2}\frac{\mathcal{G}(q,\epsilon_1,\epsilon_2,\epsilon_3^\ast,\epsilon_4^\ast)}{q^2-m_\pi^2}\Sigma(m_\pi),
\end{eqnarray}
\begin{eqnarray}
\mathcal{Y}^{c_{3.1}}_{I=0}=-\frac{2g^2}{f^4}\frac{\mathcal{G}(q,\epsilon_1,\epsilon_2,\epsilon_3^\ast,\epsilon_4^\ast)}{q^2-m_\pi^2}\left\{\mathcal{J}_0^c\right\}_r(m_\pi),
\end{eqnarray}
\begin{eqnarray}
\mathcal{Y}^{c_{3.2}}_{I=0}=-\frac{3g^4}{2f^4}\frac{\mathcal{G}(q,\epsilon_1,\epsilon_2,\epsilon_3^\ast,\epsilon_4^\ast)}{q^2-m_\pi^2}\left\{\mathcal{J}_{22}^g\right\}_r(m_\pi,\mathcal{E}+\Delta,\mathcal{E}),
\end{eqnarray}
\begin{eqnarray}
\mathcal{Y}^{c_{3.3}}_{I=0}=-\frac{3g^4}{2f^4}\frac{\mathcal{G}(q,\epsilon_1,\epsilon_2,\epsilon_3^\ast,\epsilon_4^\ast)}{q^2-m_\pi^2}\left\{\mathcal{J}_{22}^g\right\}_r(m_\pi,\mathcal{E}+\Delta,\mathcal{E}),
\end{eqnarray}
\begin{eqnarray}
\mathcal{Y}^{c_{3.4}}_{I=0}&=&\frac{3g^4}{2f^4}\frac{\mathcal{G}(q,\epsilon_1,\epsilon_2,\epsilon_3^\ast,\epsilon_4^\ast)}{q^2-m_\pi^2}\left\{\mathcal{J}_{22}^g\right\}_r(m_\pi,\mathcal{E},\mathcal{E}),
\end{eqnarray}
\begin{eqnarray}
\mathcal{Y}^{c_{3.5}}_{I=0}&=&-\frac{9g^4}{8f^4}\frac{\mathcal{G}(q,\epsilon_1,\epsilon_2,\epsilon_3^\ast,\epsilon_4^\ast)}{q^2-m_\pi^2}\left\{\frac{\partial}{\partial x}\mathcal{J}_{22}^a\right\}_r(m_\pi,x)\bigg|_{x\to \mathcal{E}+\Delta},\nonumber\\
\end{eqnarray}
\begin{eqnarray}
\mathcal{Y}^{c_{3.6}}_{I=0}=-\frac{9g^4}{4f^4}\frac{\mathcal{G}(q,\epsilon_1,\epsilon_2,\epsilon_3^\ast,\epsilon_4^\ast)}{q^2-m_\pi^2}\left\{\frac{\partial}{\partial x}\mathcal{J}_{22}^a\right\}_r(m_\pi,x)\bigg|_{x\to \mathcal{E}},\nonumber\\
\end{eqnarray}
\begin{eqnarray}
\mathcal{Y}^{c_{3.7}}_{I=0}=\mathcal{Y}^{c_{3.8}}_{I=0}=0,
\end{eqnarray}

The $I=0$ amplitudes that come from Fig. \ref{BastBast_TPE} are:
\begin{widetext}
\begin{eqnarray}
\mathcal{Y}^{F_{3.1}}_{I=0}&=&\frac{3}{f^4}\varepsilon_1\left\{\mathcal{J}_{22}^F\right\}_r(m_\pi,q),
\end{eqnarray}
\begin{eqnarray}
\mathcal{Y}^{T_{3.1}}_{I=0}&=&-\frac{3g^2}{f^4}\Bigg\{\Big[\varepsilon_1^a+2\varepsilon_1\vec{q}^2+\varepsilon_1^b\Big]\left(\mathcal{J}_{24}^T+\mathcal{J}_{33}^T\right)-4\varepsilon_1\mathcal{J}_{34}^T\Bigg\}_r(m_\pi,\mathcal{E},q),
\end{eqnarray}
\begin{eqnarray}
\mathcal{Y}^{T_{3.2}}_{I=0}&=&\frac{3g^2}{f^4}\Bigg\{\Big[\varepsilon_1^a+\varepsilon_1^b\Big]\left(\mathcal{J}_{24}^T+\mathcal{J}_{33}^T\right)+2\varepsilon_1\mathcal{J}_{34}^T\Bigg\}_r(m_\pi,\mathcal{E}+\Delta,q),
\end{eqnarray}
\begin{eqnarray}
\mathcal{Y}^{B_{3.1}}_{I=0}&=&-\frac{9g^4}{4f^4}\Bigg\{\Big[\varepsilon_1\vec{q}^4+\varepsilon_1^a\vec{q}^2+\varepsilon_1^b\vec{q}^2+\varepsilon_1^c\Big]
\left(\mathcal{J}_{22}^B+2\mathcal{J}_{32}^B+\mathcal{J}_{43}^B\right)-\Big[8\varepsilon_1\vec{q}^2+6\varepsilon_1^a+6\varepsilon_1^b-\varepsilon_2^a
-\varepsilon_2^b-\varepsilon_3^a\Big]\left(\mathcal{J}_{31}^B+\mathcal{J}_{42}^B\right)\nonumber\\
&&-\Big[\varepsilon_1\vec{q}^2+\varepsilon_1^a+\varepsilon_1^b-\varepsilon_3^a\Big]\mathcal{J}_{21}^B+\Big[6\varepsilon_1+\varepsilon_2+\varepsilon_3\Big]\mathcal{J}_{41}^B
+\varepsilon_3^a\mathcal{J}_{31}^B+\varepsilon_3^b\mathcal{J}_{42}^B\Bigg\}_r(m_\pi,\mathcal{E},\mathcal{E},q),
\end{eqnarray}
\begin{eqnarray}
\mathcal{Y}^{B_{3.2}}_{I=0}&=&-\frac{9g^4}{4f^4}\Bigg\{\varepsilon_3^c\left(\mathcal{J}_{22}^B+2\mathcal{J}_{32}^B+\mathcal{J}_{43}^B\right)+\Big[\varepsilon_1^a+\varepsilon_1^b+\varepsilon_2^a+\varepsilon_2^b+\varepsilon_3^b\Big]\left(\mathcal{J}_{31}^B+\mathcal{J}_{42}^B\right)+\Big[\varepsilon_1+\varepsilon_2+\varepsilon_3\Big]\mathcal{J}_{41}^B\nonumber\\
&&+\varepsilon_3^b\left(\mathcal{J}_{21}^B+\mathcal{J}_{31}^B\right)+\varepsilon_3^a\mathcal{J}_{42}^B\Bigg\}_r(m_\pi,\mathcal{E}+\Delta,\mathcal{E}+\Delta,q),
\end{eqnarray}
\begin{eqnarray}
\mathcal{Y}^{B_{3.3}}_{I=0}&=&-\frac{9g^4}{4f^4}\Bigg\{\left[\varepsilon_1^a\vec{q}^2+\varepsilon_1^b\vec{q}^2+2\varepsilon_3^c\right]\left(\mathcal{J}_{22}^B+2\mathcal{J}_{32}^B+\mathcal{J}_{43}^B\right)
+\Big[-5\varepsilon_1^a-5\varepsilon_1^b+2\varepsilon_1\vec{q}^2+2\varepsilon_2^a+2\varepsilon_2^b+2\varepsilon_3^a+2\varepsilon_3^b\Big]\left(\mathcal{J}_{31}^B+\mathcal{J}_{42}^B\right)\nonumber\\
&&+\Big[-\varepsilon_1^a-\varepsilon_1^b+\varepsilon_2^a+\varepsilon_2^b\Big]\mathcal{J}_{21}^B+\Big[-8\varepsilon_1+2\varepsilon_2+2\varepsilon_3\Big]\mathcal{J}_{41}^B\Bigg\}_r(m_\pi,\mathcal{E},\mathcal{E}+\Delta,q),
\end{eqnarray}
\begin{eqnarray}
\mathcal{Y}^{R_{3.1}}_{I=0}&=&-\frac{1}{3}\mathcal{Y}^{B_{3.1}}_{I=0}\Big|_{\mathcal{J}_{x}^B\to\mathcal{J}_{x}^R},\quad\quad\mathcal{Y}^{R_{3.2}}_{I=0}=-\frac{1}{3}\mathcal{Y}^{B_{3.2}}_{I=0}\Big|_{\mathcal{J}_{x}^B\to\mathcal{J}_{x}^R},\quad\quad
\mathcal{Y}^{R_{3.3}}_{I=0}=-\frac{1}{3}\mathcal{Y}^{B_{3.3}}_{I=0}\Big|_{\mathcal{J}_{x}^B\to\mathcal{J}_{x}^R}.
\end{eqnarray}
\end{widetext}


\begin{thebibliography}{199}
\bibitem{Zweig:19xx}
  G. Zweig,
   An SU(3) model for strong interaction symmetry and its breaking. Version 1.
\end{thebibliography}
\end{document}